\documentclass[aps,prd,showpacs,showkeys,twoside,eqsecnum,twocolumn]{revtex4-1}
\usepackage{amsfonts,amsmath,amssymb,bm,graphicx}

\begin{document}


\title{Field theory description of neutrino oscillations}

\author{Maxim Dvornikov}
\email{maxdvo@izmiran.ru}
\affiliation{N.~V.~Pushkov Institute of Terrestrial Magnetism,
Ionosphere and Radiowave Propagation \\ 142190, Troitsk, Moscow
Region, Russia}

\date{\today}

\begin{abstract}
We review various field theory approaches to the description of
neutrino oscillations in vacuum and external fields. First we
discuss a relativistic quantum mechanics based approach which
involves the temporal evolution of massive neutrinos. To describe
the dynamics of the neutrinos system we use exact solutions of
wave equations in presence of an external field. It allows one to
exactly take into account both the characteristics of neutrinos
and the properties of an external field. In particular, we examine
flavor oscillations an vacuum and in background matter as well as
spin flavor oscillations in matter under the influence of an
external electromagnetic field. Moreover we consider the situation
of hypothetical nonstandard neutrino interactions with background
fermions. In the case of ultrarelativistic particles we reproduce
an effective Hamiltonian which is used in the standard quantum
mechanical approach for the description of neutrino oscillations.
The corrections to the quantum mechanical Hamiltonian are also
discussed. Note that within the relativistic quantum mechanics
method one can study the evolution of both Dirac and Majorana
neutrinos. We also consider several applications of this formalism
to the description of oscillations of astrophysical neutrinos
emitted by a supernova and compare the behavior of Dirac and
Majorana neutrinos. Then we study a spatial evolution of mixed
massive neutrinos emitted by classical sources. This method seems
to be more realistic since it predicts neutrino oscillations in
space. Besides oscillations among different neutrino flavors, we
also study transitions between particle and antiparticle states.
Finally we use the quantum field theory method, which involves
virtual neutrinos propagating between production and detection
points, to describe particle-antiparticle transitions of Majorana
neutrinos in presence of background matter.
\end{abstract}



\maketitle

\tableofcontents

\section{Introduction}

Neutrino physics is one of the most rapidly developing area of
high energy physics especially after the great success in
experimental studies of neutrino properties. In the first place we
should mention the investigation of astrophysical neutrinos since
they play an important role in the evolution of various
astronomical objects like stars, supernovas, quasars etc. In the
course of recent experiments for the detection of solar
neutrinos~\cite{Abe10} it was revealed that transitions between
different neutrino flavors, neutrino oscillations, are the most
plausible theoretical explanation of the solar neutrinos
deficit~\cite{BahPin04}. Flavor oscillations were also observed in
the experimental studies of atmospheric neutrinos~\cite{Wen10}. It
is worth noticing that there are numerous attempts to detect
neutrinos from outside the solar system (see, e.g.,
Ref.~\cite{Abb10}), however, presently only SN1987A supernova
neutrinos were observed~\cite{Hir87}.

Besides natural sources, one can use neutrinos produced by an
accelerator or a nuclear reactor to study oscillations of these
particles~\cite{Gan10}. In this case we control the flavor content
of both initial and final fluxes, i.e. it is the best strategy to
examine neutrino oscillations. Besides the studies of neutrino
oscillations, some accelerator based experiments are dedicated to
the investigation of neutrino interactions (see, e.g.,
Ref.~\cite{Kur10}).

The existence of neutrino oscillations is a direct indication to
the facts that neutrinos are massive particles and there is a
mixing between different neutrino generations. There are multiple
theoretical mechanisms to generate masses and mixing of
neutrinos~\cite{MohSmi06} to fit the data of the aforementioned
neutrino oscillations experiments.

It was found that besides the possibility of flavor conversions in
vacuum, various external fields, like interaction with background
matter~\cite{Wol78} or with an external magnetic
field~\cite{Cis71}, can significantly influence neutrino
oscillations. For example, the resonant enhancement of neutrino
oscillations in matter, Mikheev-Smirnov-Wolfenstein (MSW) effect,
plays an important role in the solution of the solar neutrino
problem~\cite{HolSmi03}. It should be noted that the standard
model of electroweak interactions does not imply the mixing
between different neutrino flavors when particles propagate in
background matter. However the possibility of nonstandard neutrino
interactions, which can cause transitions between neutrino flavors
even at the absence of vacuum mixing, was recently
discussed~\cite{Big09}.

Since neutrinos are unlikely to have a nonzero electric
charge~\cite{IgnJos95}, the interaction with an external
electromagnetic field may be implemented due to the presence of
anomalous magnetic moments. Note that, unlike an electron which
has a vacuum magnetic moment, electromagnetic moments of a
neutrino always arise from the loop corrections. The contributions
to neutrino electromagnetic moments in various extensions of the
standard model are reviewed in Ref.~\cite{GiuStu09}.

It should be noted that neutrino electromagnetic properties have
rather complicated structure. First, in the system of mixed
neutrinos there can be both diagonal and transition magnetic
moments. In presence of an external electromagnetic field the
former are responsible for the helicity flip within one neutrino
generation and the latter cause the change of neutrino flavor
(spin flavor oscillations). Second, electromagnetic properties of
Dirac and Majorana neutrinos are completely different. In general
case Dirac neutrinos can have all kinds of magnetic moments,
whereas Majorana neutrinos possess only transition magnetic
moments which are antisymmetric in neutrino
flavors~\cite{FukYan03p461}. Nowadays there is no universally
recognized experimental confirmation of the nature of
neutrinos~\cite{AviEllEng08}. Thus it is important to study the
evolution of both Dirac and Majorana neutrinos in an external
electromagnetic field.

The indication to the Majorana nature of neutrinos should be
experimental confirmation of the existence of neutrinoless double
$\beta$-decay ($0 \nu 2 \beta$), which is the manifestation of
oscillations between neutrinos and antineutrinos~\cite{Bil10}.
This kind of transitions is possible only if neutrinos are
Majorana particles. Despite numerous attempts to detect ($0 \nu 2
\beta$)-decay in a laboratory were made~\cite{Arg09}, no confirmed
events are known presently.

The great variety of evidences of neutrino oscillations requires a
rigorous theoretical explanation of this phenomenon. The
conventional approach to the description of neutrino oscillations
is based on the quantum mechanical evolution of neutrino flavor
eigenstates governed by an effective Hamiltonian~\cite{BilPon78}.
This intuitive approach may be easily extended to the description
of neutrino oscillations in background matter~\cite{Wol78} and
spin flavor oscillations in an external magnetic
field~\cite{LimMar88} giving one a reasonable description of
neutrino oscillations in these external fields.

Neutrinos are supposed to propagate as plane waves in the quantum
mechanical approach. However, due to uncertainties in production
and detection processes, neutrinos seem to have some distribution
of their momenta, i.e. they propagate like wave
packets~\cite{Kay81}. The Gaussian distribution of the neutrino
momentum is typically discussed~\cite{Giu02}. However the actual
form of the distribution strongly depends on the production and
detection processes (see, e.g., Ref.~\cite{Dol05}). The necessity
of the wave packets treatment of neutrino oscillations is
discussed in Ref.~\cite{Sto98}. The details of this approach
including its extension to the Dirac theory were considered in
Ref.~\cite{BerGuzNis10}.

The attempts to reproduce the quantum mechanical transition
probability formula in vacuum were made in
Refs.~\cite{Kob82,GiuKimLeeLee93} treating massive neutrinos as
virtual particles propagating between production and detection
points. Using this quantum field theory method one deals with the
observables of charged leptons rather than with the
characteristics of mixed neutrinos. The special Gaussian form of
the source and detector within this quantum field theory approach
was discussed in Ref.~\cite{IoaPil99}. The analysis of
relativistic effects in neutrino oscillations using this approach
was made in Ref.~\cite{NauNau10}.

In Ref.~\cite{BlaVit95} neutrino flavor oscillations in vacuum are
described on the basis of time evolution of the Fock states of
flavor neutrinos. The time dependent transition probability
formula, obtained within the $S$-matrix approach, was studied in
Ref.~\cite{Wu10}. The quantum mechanics analysis of the neutrino
detection process and the possibility to obtain the nonstandard
transition probability formulas is considered in
Ref.~\cite{AnaSav10}.

We contributed to the field theoretical description of neutrino
oscillations in
Refs.~\cite{Dvo05,Dvo06,DvoMaa07,Dvo08JPCS,Dvo08JPG,Dvo09,DvoMaa09,Dvo10}.
We developed an approach based on relativistic quantum mechanics
and applied it for the description of neutrino flavor oscillations
in vacuum~\cite{Dvo05}, in background
matter~\cite{Dvo06,Dvo08JPCS}, and to spin flavor oscillations in
an external electromagnetic
field~\cite{DvoMaa07,Dvo08JPG,DvoMaa09,Dvo10}. This method is
based on the first quantized neutrino wave packets which was also
studied in Ref.~\cite{BerLeo04} to study neutrino oscillations in
vacuum and in an external electromagnetic field.

In frames of the relativistic quantum mechanics method we
formulate the initial condition problem for the system of flavor
neutrinos and study the subsequent time evolution of neutrino wave
packets. When we discuss neutrino propagation in an external
field, we use an exact solution of the corresponding wave equation
in presence of this external field. With help of this method we
could reproduce the Schr\"{o}dinger type evolution equation, which
is used in the standard quantum mechanical description of Dirac
and Majorana neutrino oscillations, and discuss the correction to
the standard approach.

In Ref.~\cite{Dvo09} we studied neutrino flavor oscillation in
vacuum within the external classical sources method. Note that the
analogous approach to the description of neutrino oscillations was
also discussed in Ref.~\cite{KieWei98}. Within this approach we
studied the spatial evolution of flavor neutrino waves emitted by
external classical sources. We could describe the evolution of
both Dirac and Majorana neutrinos in vacuum.

In the present work we review our recent achievements in
theoretical description of neutrino flavor oscillations. This work
is organized as follows. In Secs.~\ref{VACUUM}-\ref{MATTERB} we
discuss oscillations of Dirac neutrinos in frames of relativistic
quantum mechanics method. We start with the description of
neutrino flavor oscillations in vacuum (Sec.~\ref{VACUUM}). Then
we consider oscillations in background matter (Sec.~\ref{MATTER}),
where we also study oscillations in case of nonstandard neutrino
interactions with matter. In Sec.~\ref{B} we apply the
relativistic quantum mechanics method to the description of spin
flavor oscillations of Dirac neutrinos in an external magnetic
field. Finally, in Sec.~\ref{MATTERB}, we discuss the most general
situation of spin flavor oscillations in matter under the
influence of an external magnetic field. In Secs.~\ref{SPECMATT}
and~\ref{STERILE} we examine applications of the relativistic
quantum mechanics method to the description of propagation and
oscillations of astrophysical neutrinos in a magnetized envelope
after the supernova explosion.

In Secs.~\ref{MAJVACUUM} and~\ref{MAJNUMATTB} we analyze the
evolution of massive mixed Majorana neutrinos in vacuum as well as
in matter and magnetic field in frames of the relativistic quantum
mechanics approach. Besides oscillations among different neutrino
flavors we also consider transitions between neutrino and
antineutrino states which are allowed if neutrinos are Majorana
particles. In Sec.~\ref{MAJSPECMATT} we apply the general results
to the studies of oscillations of astrophysical neutrinos in the
supernova envelope supposing that neutrinos are Majorana particles
and compare them with the Dirac neutrino case studied in
Sec.~\ref{SPECMATT}.

Then, in Sec.~\ref{INHOMOGEN}, we formulate an alternative
formalism for the description of neutrino flavor oscillations
which is based on the mixed massive neutrinos emission by
classical sources. We examine the spatial distribution of flavor
neutrino wave functions in case of localized sources. Note that
vacuum oscillations of both Dirac (Sec.~\ref{DMT}) and Majorana
(Sec.~\ref{GMT}) neutrinos are studied. When we discuss Majorana
neutrinos case, we also consider the possibility of transitions
between particles and antiparticles.

Finally, in Sec.~\ref{NU2ANTINU}, we consider the transitions
among neutrinos and antineutrinos in frames of the quantum field
theory, treating neutrinos as virtual particles propagating
between macroscopically separated production and detection points
(see also Refs.~\cite{Kob82,GiuKimLeeLee93,IoaPil99,NauNau10}). In
particular we are interested in the influence of background matter
on the oscillations process since this kind of transitions
typically happens inside a nucleus (see, e.g., Ref.~\cite{Bil10})
and one cannot neglect the presence of dense nuclear matter in
neutrino oscillations.

In Sec.~\ref{CONCL} we summarize our results. Several technical
issues are considered in
Appendices~\ref{DETAILSABSOL}-\ref{NU2ANTINUINT} in order not to
encumber the description of the main results in
Secs.~\ref{VACUUM}-\ref{NU2ANTINU}.

\section{Dirac neutrinos in vacuum\label{VACUUM}}

In this section we discuss the evolution of mixed flavor neutrinos
in vacuum, i.e. at the absence of external fields, using the
relativistic quantum mechanics approach~\cite{Dvo05}. First we
formulate the initial condition problem for the system of flavor
neutrinos. We exactly solve this problem and find the time
dependent wave functions and the transition probability. We also
discuss the validity of the developed formalism.

Without loss of generality we study the situation of only two
particles $\nu = (\nu_\alpha,\nu_\beta)$, where $\alpha$ and
$\beta$ can stay for electron, muon or $\tau$-neutrinos. In
various theoretical models (see, e.g., Ref.~\cite{Moh04}) bigger
number of flavor neutrino fields $N_\nu > 3$ is proposed. The case
of arbitrary number of neutrino fields can be considered in our
formalism by straightforward generalization. The Lorentz invariant
Lagrangian for this system has the following form:
\begin{equation}\label{Lagrnuvac}
  \mathcal{L} =
  \sum_{\lambda = \alpha, \beta}
  \bar{\nu}_\lambda
  \mathrm{i}\gamma^\mu\partial_\mu
  \nu_\lambda -
  \sum_{\lambda \lambda' = \alpha, \beta}
  m_{\lambda\lambda'}
  \bar{\nu}_\lambda \nu_{\lambda'},
\end{equation}
where $\gamma^\mu = (\gamma^0,\bm{\gamma})$ are Dirac matrices and
$(m_{\lambda \lambda'})$ is the mass matrix which is not diagonal
in the flavor neutrinos basis. The nondiagonal elements of this
matrix $m_{\alpha\beta} = m_{\beta\alpha}$ correspond to the
mixing between different neutrino flavors.

To describe the time evolution of the system~\eqref{Lagrnuvac} we
formulate the initial condition problem for flavor neutrinos
$\nu_\lambda$,
\begin{equation}\label{inicondnulambda}
  \nu_\lambda(\mathbf{r}, t = 0) = \nu_\lambda^{(0)}(\mathbf{r}),
  \quad
  \lambda = \alpha,\beta,
\end{equation}
where $\nu_\lambda^{(0)}(\mathbf{r})$ are known functions.
Eq.~\eqref{inicondnulambda} means that initial field distributions
of flavor neutrinos are known and we will search for their wave
functions at subsequent moments of time:
$\nu_\lambda(\mathbf{r},t>0)$. The situation when only one of the
flavor neutrinos, e.g. belonging to the type ``$\beta$", is
present initially corresponds to a typical neutrino oscillations
experiment: one looks for the initially absent neutrino flavor
``$\alpha$" in a beam consisting of neutrinos of the flavor
``$\beta$".

To solve the initial condition problem~\eqref{inicondnulambda} for
the system~\eqref{Lagrnuvac} we introduce the mass eigenstates
neutrinos $\psi_a$, $a=1,2$,
\begin{equation}\label{matrtranslambda}
  \nu_{\lambda} = \sum_{a=1,2}U_{\lambda a} \psi_a,
\end{equation}
where the matrix $(U_{\lambda a})$ is chosen in such a way to
diagonalize the mass matrix $(m_{\lambda \lambda'})$. The
eigenvalues of the matrix $(m_{\lambda \lambda'})$, which are real
and positive, have the meaning of the masses of the fields
$\psi_a$. We define them as $m_a$.

The Lagrangian formulated in terms of the flavor neutrino fields
does not provide any information about the nature of neutrinos,
i.e. whether neutrinos are Dirac or Majorana particles, since in
general case it is written using the two component left and right
handed spinors~\cite{Kob80,SchVal80}. Only when we introduce the
mass eigenstates and analyze the structure of resulting mass
matrices, we can reveal the nature of neutrinos. We suppose that
in our case the fields $\psi_a$ are Dirac particles.

For the case of only two neutrinos system the mixing matrix
$(U_{\lambda a})$ in Eq.~\eqref{matrtranslambda} has the form,
\begin{equation}\label{Ulambdaa}
  (U_{\lambda a}) =
  \begin{pmatrix}
    \cos \theta & -\sin \theta \\
    \sin \theta & \cos \theta \
  \end{pmatrix},
\end{equation}
where $\theta$ is the vacuum mixing angle.

The Lagrangian~\eqref{Lagrnuvac} expressed in terms of the fields
$\psi_a$ reads
\begin{equation}\label{Lagrpsivac}
  \mathcal{L} = \sum_{a = 1,2}
  \bar{\psi}_a
  (\mathrm{i}\gamma^\mu\partial_\mu - m_a)
  \psi_a.
\end{equation}
The Lagrangian~\eqref{Lagrpsivac} should be supplied with the
initial condition,
\begin{equation}\label{inicondpsi}
  \psi_a(\mathbf{r}, t = 0) = \psi_a^{(0)}(\mathbf{r}),
  \quad
  \psi_a^{(0)} = (U_{a\lambda}^{-1}) \nu_\lambda^{(0)},
\end{equation}
which follows from Eqs.~\eqref{inicondnulambda}
and~\eqref{matrtranslambda}.

The Dirac equations,
\begin{equation}\label{Direqpsivac}
  \mathrm{i}\dot{\psi}_a = (\bm{\alpha}\mathbf{p} + \beta m_a) \psi_a,
\end{equation}
which result from the Lagrangian~\eqref{Lagrpsivac}, reveal that
the fields $\psi_a$ decouple. In Eq.~\eqref{Direqpsivac} we use
the standard definitions for Dirac matrices $\bm{\alpha} =
\gamma^0\bm{\gamma}$ and $\beta = \gamma^0$. The solution of
Eq.~\eqref{Direqpsivac} can be found in the following way:
\begin{align}\label{GenSolDirpsivac}
  \psi_{a}(\mathbf{r},t) = &
  \int \frac{\mathrm{d}^3 \mathbf{p}}{(2\pi)^{3/2}}
  e^{\mathrm{i} \mathbf{p} \mathbf{r}}
  \notag
  \\
  & \times
  \sum_{\zeta=\pm 1}
  \left[
    a_a^{(\zeta)} u_a^{(\zeta)}e^{-\mathrm{i} E_a t}+
    b_a^{(\zeta)} v_a^{(\zeta)}e^{\mathrm{i} E_a t}
  \right],
\end{align}
where $E_a = \sqrt{|\mathbf{p}|^2 + m_a^2}$ is the energy of a
massive neutrino in vacuum, $\zeta = \pm 1$ is the helicity of
massive neutrinos, and $u_a^{(\zeta)}$ and $v_a^{(\zeta)}$ are the
basis spinors corresponding to a definite helicity.

In the relativistic quantum mechanics approach to the description
of neutrinos evolution~\cite{Dvo05} the coefficients
$a_a^{(\zeta)}$ and $b_a^{(\zeta)}$ are $c$-number quantities
rather than operators acting in the Fock space. Our task is to
find these coefficients. Since the fields $\psi_a$ are
independent, the values of the coefficients $a_a^{(\zeta)}$ and
$b_a^{(\zeta)}$ depends only on the initial
condition~\eqref{inicondpsi}.

Using the orthonormality of the basis spinors,
\begin{equation}\label{normspinorsvac}
  u_a^{(\zeta')\dag} u_a^{(\zeta)} =
  v_a^{(\zeta')\dag} v_a^{(\zeta)} = \delta_{\zeta'\zeta},
  \quad
  u_a^{(\zeta')\dag} v_a^{(\zeta)} = 0,
\end{equation}
we find the coefficients $a_a^{(\zeta)}$ and $b_a^{(\zeta)}$ in
the form,
\begin{align}\label{abvac}
  a_a^{(\zeta)} = &
  \frac{1}{(2\pi)^{3/2}} u_a^{(\zeta)\dag} \psi_a^{(0)}(\mathbf{p}),
  \notag
  \\
  b_a^{(\zeta)} = &
  \frac{1}{(2\pi)^{3/2}} v_a^{(\zeta)\dag} \psi_a^{(0)}(\mathbf{p}),
\end{align}
where
\begin{equation}\label{Fourierpsi0}
  \psi_a^{(0)}(\mathbf{p}) = \int \mathrm{d}^3 \mathbf{r}
  e^{- \mathrm{i} \mathbf{p} \mathbf{r}} \psi_a^{(0)}(\mathbf{r}),
\end{equation}
is the Fourier transform of the initial
condition~\eqref{inicondpsi}.

With help of Eqs.~\eqref{GenSolDirpsivac}-\eqref{Fourierpsi0} we
arrive to the expression for the wave function of the neutrino
mass eigenstates,
\begin{equation}\label{SpecSolpsirvac}
  \psi_{a}(\mathbf{r},t) =
  \int \mathrm{d}^3 \mathbf{r}'
  S_a(\mathbf{r}' - \mathbf{r},t)
  (-\mathrm{i}\gamma^0) \psi_a^{(0)}(\mathbf{r}'),
\end{equation}
where
\begin{equation}\label{PJFpsirvac}
  S_a(\mathbf{r},t) =
  (\mathrm{i} \gamma^\mu \partial_\mu + m_a) D_a(\mathbf{r},t),
\end{equation}
is the Pauli-Jourdan function for a spinor particle and
\begin{equation}\label{PJFscalvac}
  D_a(\mathbf{r},t) =
  \int \frac{\mathrm{d}^3 \mathbf{p}}{(2\pi)^3}
  e^{\mathrm{i} \mathbf{p} \mathbf{r}} \frac{\sin E_a t}{E_a},
\end{equation}
is the Pauli-Jourdan function for a scalar particle.

It is convenient to rewrite Eq.~\eqref{SpecSolpsirvac} in the
form,
\begin{equation}\label{SpecSolpsipvac}
  \psi_{a}(\mathbf{r},t) =
  \int \frac{\mathrm{d}^3 \mathbf{p}}{(2\pi)^3}
  e^{\mathrm{i} \mathbf{p} \mathbf{r}}
  S_a(-\mathbf{p},t)
  (-\mathrm{i}\gamma^0) \psi_a^{(0)}(\mathbf{p}),
\end{equation}
where
\begin{align}\label{PJFpsipvac}
  S_a(-\mathbf{p},t) = &
  \sum_{\zeta = \pm 1}
  \big(
    u_a^{(\zeta)} \otimes u_a^{(\zeta)\dag} e^{-\mathrm{i} E_a t}
    \notag
    \\
    & +
    v_a^{(\zeta)} \otimes v_a^{(\zeta)\dag} e^{\mathrm{i} E_a t}
  \big)(\mathrm{i}\gamma^0)
  \notag
  \\
  & =
  \left[
    \cos E_a t - \mathrm{i} \frac{\sin E_a t}{E_a}
    (\bm{\alpha}\mathbf{p}+ \beta m_a)
  \right]
  \notag
  \\
  & \times
  (\mathrm{i}\gamma^0),
\end{align}
is the Fourier transform of the Pauli-Jourdan
function~\eqref{PJFpsirvac}. To derive Eq.~\eqref{PJFpsipvac} we
use the summation over the helicity index
formulas~\cite{BogShi80},
\begin{align}
  \sum_{\zeta = \pm 1} u_a^{(\zeta)} \otimes u_a^{(\zeta)\dag} = &
  \frac{1}{2} + \frac{1}{2 E_a} (\bm{\alpha}\mathbf{p}+ \beta m_a),
  \notag
  \\
  \sum_{\zeta = \pm 1} v_a^{(\zeta)} \otimes v_a^{(\zeta)\dag} = &
  \frac{1}{2} - \frac{1}{2 E_a} (\bm{\alpha}\mathbf{p}+ \beta m_a),
\end{align}
which are consistent with the normalization of the basis
spinors~\eqref{normspinorsvac}.

Now let us specify the initial condition~\eqref{inicondnulambda}.
We suggest that initially very broad wave packet is present, i.e.
the coordinate dependence of the wave functions
$\nu_\lambda^{(0)}(\mathbf{r})$ is close to a plane wave
corresponding to the initial momentum $\mathbf{k}$. Moreover we
choose the situation when only one flavor neutrino is present,
\begin{equation}\label{inicondnulambdaexp}
  \nu_\alpha^{(0)}(\mathbf{r}) = 0,
  \quad
  \nu_\beta^{(0)}(\mathbf{r}) =
  e^{\mathrm{i}\mathbf{k}\mathbf{r}} \nu_\beta^{(0)}(\mathbf{k}),
\end{equation}
where $\nu_\alpha^{(0)}(\mathbf{k})$ is the coordinate independent
normalization spinor, $|\nu_\beta^{(0)}(\mathbf{k})|^2 = 1$.

Using Eqs.~\eqref{matrtranslambda}, \eqref{Ulambdaa}, and
\eqref{PJFscalvac}-\eqref{inicondnulambdaexp} we obtain the wave
function of the initially absent flavor neutrino $\nu_\alpha$ as
\begin{align}\label{nugenvac}
  \nu_\alpha(\mathbf{r},t)= &
  e^{\mathrm{i}\mathbf{k}\mathbf{r}} \sin\theta\cos\theta
  \bigg\{
    \cos E_1 t -
    \cos E_2 t
    \notag
    \\
    & -
    \mathrm{i}\frac{\sin E_1 t}{E_1}
    (\bm{\alpha}\mathbf{k}+\beta m_1)
    \notag
    \\
    & +
    \mathrm{i}\frac{\sin E_2 t}{E_2}
    (\bm{\alpha}\mathbf{k}+\beta m_2)
  \bigg\} \nu_\beta^{(0)}(\mathbf{k}),
\end{align}
where the energies are the functions of the initial momentum, $E_a
= \sqrt{|\mathbf{k}|^2 + m_a^2}$.

With help of Eq.~\eqref{nugenvac} we get the transition
probability $P_{\nu_\beta \to \nu_\alpha}(t) =
|\nu_\alpha(\mathbf{r},t)|^2$
as
\begin{align}\label{PtrPsurvac}
  P_{\nu_\beta \to \nu_\alpha}(t) = &
  \sin^2 (2\theta)
  \bigg[
    \sin^2(\Phi t)
    -
    \sin(\Phi t) \cos(\sigma t)
    \notag
    \\
    & \times
    \frac{1}{2}
    \left(
      \frac{m_1^2}{k^2} \sin E_1 t -
      \frac{m_2^2}{k^2} \sin E_2 t
    \right)
  \bigg]
  \notag
  \\
  &
  + \mathcal{O}
  \left(
    \frac{m_a^4}{k^4}
  \right),
\end{align}
where
\begin{align}\label{Phisigma}
  \Phi = & \frac{E_1 - E_2}{2} \approx \frac{\delta m^2}{4k} + \dotsb,
  \notag
  \\
  \sigma = & \frac{E_1 + E_2}{2} \approx k + \frac{m_1^2+m_2^2}{4k} + \dotsb,
\end{align}
and $\delta m^2 = m_1^2 - m_2^2$. The quantity $\Phi$ has the
meaning of the phase of neutrino oscillations in vacuum. Note that
the coordinate dependence is washed out from
Eq.~\eqref{PtrPsurvac} since we study a very broad initial wave
packet.

In the untrarelativistic limit~\eqref{Phisigma} the main term in
Eq.~\eqref{PtrPsurvac} resembles the usual transition probability
for neutrino oscillations in vacuum. The correction to the main
result is suppressed by the factor $m_a^2/k^2 \ll 1$. Using
Eq.~\eqref{Phisigma} we can represent Eq.~\eqref{PtrPsurvac} in
the following form:
\begin{align}\label{Ptr1vac}
  P_{\nu_\beta \to \nu_\alpha}(t) = &
  \sin^2 (2\theta)
  \bigg[
    \sin^2
    \left(
      \frac{\delta m^2}{4k} t
    \right)
    -
    \frac{\delta m^2}{4k^2}
    \notag
    \\
    & \times
    \sin
    \left(
      \frac{\delta m^2}{4k} t
    \right)
    \sin (2 k t)
  \bigg] + \dotsb,
\end{align}
where we drop small terms $\sim m_a^4/k^4$.

Now the leading term in Eq.~\eqref{Ptr1vac} reproduces the well
known transition probability for neutrino oscillations in vacuum
derived in frames of the quantum mechanical
approach~\cite{GriPon69}. The correction to the leading term,
which is a rapidly oscillating function on the frequency $\sim k$,
was first studied in Ref.~\cite{BlaVit95} in frames of the quantum
field theory approach to neutrino oscillations. We obtained
analogous result using the relativistic quantum mechanics method
(see also Ref.~\cite{Dvo05}). This correction to the leading term
in the transition probability results from the accurate account of
the Lorentz invariance.

Note that Eq.~\eqref{Ptr1vac} is invariant under the $m_1
\leftrightarrow m_2$ transformation. It means that one cannot
obtain the information about neutrino mass hierarchy studying
neutrino oscillations in vacuum even taking into account the
correction to the leading term in the transition probability.

Now let us discuss the possibility of initial conditions which
differ from the plane wave~\eqref{inicondnulambdaexp}. If the
initial wave function is localized in a spatial region with a
typical size $L_0$ and we measure a signal in the wave zone
$|\mathbf{r}| \gg L_0$, the dependence on the particle masses
$m_a$ is washed out from the Pauli-Jourdan
functions~\eqref{PJFscalvac} and~\eqref{PJFpsirvac} (see
Refs.~\cite{Dvo05,MorFes53}). Thus neutrinos with spatially
localized initial wave packets evolve in the wave zone like
massless particles, which are known not to reveal flavor
oscillations. Therefore the initial wave packet should be
sufficiently broad.

\section{Dirac neutrinos in background matter\label{MATTER}}

In this section we use the formalism developed in
Sec.~\ref{VACUUM} to study the evolution of the system of mixed
flavor neutrinos $\nu = (\nu_\alpha,\nu_\beta)$ in background
matter~\cite{Dvo06,Dvo08JPCS}. We formulate the initial condition
problem for this system and solve it for ultrarelativistic
neutrinos. The case of the standard model neutrino interactions is
studied in details. Then we also analyze the dynamics of neutrino
oscillations in presence of the nonstandard interactions which mix
neutrino flavors.

The neutrino interaction with matter can be represented in the
form of an external axial-vector field
$f_{\lambda\lambda'}^\mu$~\cite{DvoStu02,LobStu01}. As in
Sec.~\ref{VACUUM}, we start from the Lorentz invariant Lagrangian,
\begin{align}\label{Lagrnumatt}
  \mathcal{L} = &
  \sum_{\lambda = \alpha,\beta}
  \bar{\nu}_\lambda \mathrm{i}\gamma^\mu\partial_\mu \nu_\lambda
  \notag
  \\
  & -
  \sum_{\lambda,\lambda' = \alpha,\beta}
  \bar{\nu}_\lambda
  \left(
    m_{\lambda\lambda'} +
    f_{\lambda\lambda'}^{\mu}
    \gamma_\mu^\mathrm{L}
  \right)
  \nu_{\lambda'},
\end{align}
where $\gamma_\mu^\mathrm{L} = \gamma_\mu (1 - \gamma^5)/2$,
$\gamma^5 = \mathrm{i} \gamma^0 \gamma^1 \gamma^2 \gamma^3$, and
keep the same notation for the mass matrix as in
Sec.~\ref{VACUUM}.

Note that in general case the axial vector field
$f_{\lambda\lambda'}^\mu$ can be nondiagonal in the neutrino
flavor basis. The appearance of nondiagonal elements
$(f_{\alpha\beta}^\mu)$ of this matrix is the indication to the
presence of nonstandard neutrino interactions which mix neutrino
flavors since in the standard model of electroweak interactions
only diagonal elements of this matrix can appear. The time
component of diagonal elements of this matrix,
$f_{\lambda\lambda}^0$, is proportional to the density of
background matter and spatial component,
$\mathbf{f}_{\lambda\lambda}$, to the mean velocity and
polarization of background fermions. The details of the averaging
over the background fermions are presented in Ref.~\cite{LobStu01}

To describe the evolution of the system~\eqref{Lagrnumatt} we
formulate the initial condition problem for flavor neutrinos, with
the initial wave functions having an analogous form as in
Eq.~\eqref{inicondnulambda}. To solve the initial condition
problem we introduce the set of neutrino mass eigenstates $\psi_a$
[see Eqs.~\eqref{matrtranslambda} and~\eqref{Ulambdaa}] to
diagonalize the mass matrix $(m_{\lambda\lambda'})$. As in
Sec.~\ref{VACUUM} we suggest that mass eigenstates $\psi_a$ are
Dirac particles.

The Lagrangian~\eqref{Lagrnumatt} expressed in terms of the fields
$\psi_a$ has the form,
\begin{equation}\label{Lagrpsimatt}
  \mathcal{L} =
  \sum_{a=1,2}
  \bar{\psi}_a (\mathrm{i}\gamma^\mu \partial_\mu-m_a) \psi_a -
  \sum_{a,b=1,2}
  g_{ab}^{\mu} \bar{\psi}_a \gamma_\mu^\mathrm{L} \psi_{b},
\end{equation}
where
\begin{equation}\label{gmatrmatt}
  (g_{ab}^{\mu}) = U^\dag(f_{\lambda\lambda'}^{\mu})U =
  \begin{pmatrix}
    g_1^\mu & g^\mu \\
    g^\mu & g_2^\mu \
  \end{pmatrix},
\end{equation}
is the external axial-vector field expressed in the mass
eigenstates basis.

One can derive Dirac equations for the neutrino mass eigenstates,
\begin{align}\label{Direqpsimatt}
  \mathrm{i}\dot{\psi}_a = & \mathcal{H}_a \psi_a +
  \mathcal{V} \psi_b,
  \quad
  a,b=1,2,
  \quad
  a \neq b,
  \notag
  \\
  \mathcal{H}_a = & \bm{\alpha}\mathbf{p} + \beta m_a +
  \beta\gamma_\mu^\mathrm{L} g_a^\mu,
  \quad
  \mathcal{V} = \beta\gamma_\mu^\mathrm{L} g^\mu,
\end{align}
directly from the Lagrangian~\eqref{Lagrpsimatt}. It should be
noted that Dirac equations for different neutrino mass eigenstates
are coupled because of the presence of the interaction
$\mathcal{V}$. Studying an exact solution of the Dirac equation
for a massive neutrino in background matter we can exactly take
into account the contribution of the term
$\beta\gamma_\mu^\mathrm{L} g_a^\mu$ into the dynamics of a
particle. On the contrary, the term proportional to $\mathcal{V}$,
which mixes different mass eigenstates, should be studied
perturbatively. Nevertheless one can account for all terms in the
perturbative expansion for ultrarelativistic neutrinos.

We will study the case of nonmoving and unpolarized matter which
corresponds to $\mathbf{g}_{ab} = 0$. The matrix $(g_{ab}^\mu)$
has only time component now,
\begin{equation}\label{gab0}
  (g_{ab}) \equiv (g_{ab}^0) =
  \begin{pmatrix}
    g_1 & g \\
    g & g_2 \
  \end{pmatrix},
\end{equation}
where we introduce the new notations, $g_a \equiv g_{aa}^0$ and $g
\equiv g_{12}^0 = g_{21}^0$. If the background matter is nonmoving
and unpolarized, the Hamiltonian $\mathcal{H}_a$ commutes with the
helicity operator $(\bm{\Sigma}\mathbf{p})/|\mathbf{p}|$, where
$\bm{\Sigma}=\gamma^0\gamma^5\bm{\gamma}$, and we can classify the
states of massive neutrinos with help of the eigenvalues of the
helicity operator $\zeta = \pm 1$.

The general solution of Eq.~\eqref{Direqpsimatt} can be presented
in the following way:
\begin{align}\label{GenSolDirpsimatt}
  \psi_{a}(\mathbf{r},t) = &
  e^{- \mathrm{i} g_a t/2}
  \int \frac{\mathrm{d}^3\mathbf{p}}{(2\pi)^{3/2}}
  e^{\mathrm{i}\mathbf{p}\mathbf{r}}
  \notag
  \\
  & \times
  \sum_{\zeta=\pm 1}
  \big[
    a_a^{(\zeta)}(t)u_a^{(\zeta)}\exp{(-\mathrm{i}E_a^{(\zeta)} t)}
    \notag
    \\
    & +
    b_a^{(\zeta)}(t)v_a^{(\zeta)}\exp{(+\mathrm{i}E_a^{(\zeta)} t)}
  \big],
\end{align}
where $a_a^{(\zeta)}$ and $b_a^{(\zeta)}$ are the undetermined
nonoperator coefficients [see Eq.~\eqref{GenSolDirpsivac}], which
are, however, time dependent now because of the presence of the
term $\mathcal{V}$ in Eq.~\eqref{Direqpsimatt}.

The energy spectrum $E_a^{(\zeta)}$ in
Eq.~\eqref{GenSolDirpsimatt} was found in Ref.~\cite{StuTer05},
\begin{equation}\label{energymatt}
  E_a^{(\zeta)}=
  \sqrt{
  \left(
    |\mathbf{p}|-\zeta g_a/2
  \right)^2+m_a^2},
\end{equation}
for the case of nonmoving and unpolarized neutrinos. The basis
spinors $u_a^{(\zeta)}$ and $v_a^{(\zeta)}$ in
Eq.~\eqref{GenSolDirpsimatt} are the eigenvectors of the helicity
operator $(\bf{\Sigma}\mathbf{p})/|\mathbf{p}|$, with the
eigenvalues $\zeta$. As an example, we present here the basis
spinors which correspond to an ultrarelativistic particle
propagating along the $z$-axis,
\begin{align}\label{spinorsmatt}
  u^{-{}} = &
  \frac{1}{\sqrt{2}}
  \begin{pmatrix}
     0 \\
     -1 \\
     0 \\
     1 \
  \end{pmatrix},
  \quad
  u^{+{}} =
  \frac{1}{\sqrt{2}}
  \begin{pmatrix}
     1 \\
     0 \\
     1 \\
     0 \
  \end{pmatrix},
  \notag
  \\
  v^{-{}} = &
  \frac{1}{\sqrt{2}}
  \begin{pmatrix}
     0 \\
     1 \\
     0 \\
     1 \
  \end{pmatrix}
  \quad
  v^{+{}} =
  \frac{1}{\sqrt{2}}
  \begin{pmatrix}
     1 \\
     0 \\
     -1 \\
     0 \
  \end{pmatrix},
\end{align}
where we omit the subscript ``$a$" since we neglect the neutrino
mass in Eq.~\eqref{spinorsmatt}. Basis spinors corresponding to
arbitrary energies were also found in the explicit form in
Ref.~\cite{StuTer05}.

Now we should specify the initial condition. We can choose it as
in Eq.~\eqref{inicondnulambdaexp}, with $\mathbf{k}=(0, 0, k)$ and
$k \gg m_a$. It is also convenient to take $\nu_\beta^{(0)} =
u^{-{}}$ [see Eq.~\eqref{spinorsmatt}]. Such an initial wave
function corresponds to a neutrino propagating along the $z$-axis,
with the spin directed opposite to the particle momentum, i.e. it
describes a left polarized neutrino.

If we put the \textit{ansatz}~\eqref{GenSolDirpsimatt} in the wave
equations~\eqref{Direqpsimatt}, we get the following ordinary
differential equations for the functions $a_a^{(\zeta)}(t)$ and
$b_a^{(\zeta)}(t)$:
\begin{align}\label{ODEDirmatt}
  \mathrm{i}\dot{a}_a^{(\zeta)}= &
  e^{\mathrm{i}(g_a-g_b)t/2}
  \exp\left( \mathrm{i}E_a^{(\zeta)}t \right) u^{(\zeta)\dag}
  \mathcal{V}
  \notag
  \\
  & \times
  \sum_{\zeta'=\pm 1}
  \Big[
    a_b^{(\zeta')}u^{(\zeta')} \exp\left( -\mathrm{i}E_b^{(\zeta')} t \right)
    \notag
    \\
    & +
    b_b^{(\zeta')}v^{(\zeta')} \exp\left( \mathrm{i}E_b^{(\zeta')} t \right)
  \Big],
  \notag
  \\
  \mathrm{i}\dot{b}_a^{(\zeta)}= &
  e^{\mathrm{i}(g_a-g_b)t/2}
  \exp\left( -\mathrm{i}E_a^{(\zeta)}t \right) v^{(\zeta)\dag}
  \mathcal{V}
  \notag
  \\
  & \times
  \sum_{\zeta'=\pm 1}
  \Big[
    a_b^{(\zeta')}u^{(\zeta')} \exp\left( -\mathrm{i}E_b^{(\zeta')} t \right)
    \notag
    \\
    & +
    b_b^{(\zeta')}v^{(\zeta')} \exp\left( \mathrm{i}E_b^{(\zeta')} t \right)
  \Big].
\end{align}
To obtain Eq.~\eqref{ODEDirmatt} we use the orthonormality of the
basis spinors~\eqref{spinorsmatt} [see
Eq.~\eqref{normspinorsvac}]. We should supply the
Eq.~\eqref{ODEDirmatt} with the initial condition,
\begin{align}\label{inicondamatt}
  a_1^{(\zeta)}(0) = & \frac{\sin\theta}{(2\pi)^{3/2}}
  u^{(\zeta)\dag}\nu_\beta^{(0)},
  \notag
  \\
  a_2^{(\zeta)}(0) = & \frac{\cos\theta}{(2\pi)^{3/2}}
  u^{(\zeta)\dag}\nu_\beta^{(0)},
\end{align}
that result from Eqs.~\eqref{inicondpsi}
and~\eqref{GenSolDirpsimatt}. If we study an arbitrary wave packet
initial condition rather than the plane wave distribution for
$\nu_\beta^{(0)}(\mathbf{r})$~\eqref{inicondnulambdaexp}, we have
to replace $\nu_\beta^{(0)}$ in Eq.~\eqref{inicondamatt} with the
Fourier transform of the initial wave function
$\nu_\beta^{(0)}(\mathbf{r})$.

Taking into account the fact that $\langle u^{(\zeta)} |
\mathcal{V} | v^{(\zeta')} \rangle = 0$, we get that equations for
$a_a^{(\zeta)}(t)$ and $b_a^{(\zeta)}(t)$ decouple, i.e. the
interaction $\mathcal{V}$ does not mix positive and negative
energy eigenstates. In the following we will consider the
evolution of only $a_a^{(\zeta)}(t)$ since the dynamics of
$b_a^{(\zeta)}(t)$ is studied analogously.

The only nonzero matrix elements of the potential $\mathcal{V}$ in
Eq.~\eqref{ODEDirmatt} are $\langle u^{-{}} | \mathcal{V} | u^{-}
\rangle = \langle v^{+{}} | \mathcal{V} | v^{+{}} \rangle = g$,
which result from Eq.~\eqref{spinorsmatt}. Finally
Eq.~\eqref{ODEDirmatt} are reduced to the ordinary differential
equations only for the functions $a^{-{}}_a(t)$,
\begin{align}\label{aeqmatt}
  \mathrm{i}\dot{a}_a^{-{}} = &
  a_b^{-{}} g
  \exp{[\mathrm{i} \{ E_a^{-{}}-E_b^{-{}} +(g_a - g_b)/2 \} t]},
  \notag
  \\
  &
  a,b=1,2,
  \quad
  a \neq b.
\end{align}
It follows from Eq.~\eqref{aeqmatt} that equations for the
functions $a_a^{-{}}$ and $a_a^{+{}}$ (not shown here), decouple
since the interaction with background matter conserves the
particle helicity and we have chosen the initial condition
corresponding to a left polarized particle. Indeed one can obtain
from Eq.~\eqref{inicondamatt} that the functions $a_a^{+{}}(t)$
are equal to zero at $t=0$.

The solution of Eq.~\eqref{aeqmatt} can be expressed in the form
(see Appendix~\ref{DETAILSABSOL}),
\begin{align}\label{aeqsolmatt}
  a_1^{-{}}(t)= &
  F a_1^{-{}}(0)+G a_2^{-{}}(0),
  \notag
  \\
  a_2^{-{}}(t)= &
  F^{*{}}a_2^{-{}}(0)-G^{*{}}a_1^{-{}}(0),
  \notag
  \\
  F = &
  \left[
    \cos\Omega_{-{}} t-
    \mathrm{i}\frac{\omega_{-{}}}{2\Omega_{-{}}}\sin\Omega t
  \right]
  \exp{(\mathrm{i}\omega_{-{}} t/2)},
  \notag
  \\
  G = &
  -\mathrm{i}\frac{g}{\Omega_{-{}}}\sin\Omega_{-{}} t
  \exp{(\mathrm{i}\omega_{-{}} t/2)},
\end{align}
where $\Omega_{-{}} = \sqrt{g^2+(\omega_{-{}}/2)^2}$ and
$\omega_{-{}} = E_1^{-{}} - E_2^{-{}} + (g_1 - g_2)/2$.

Using the identity $\left(v^{+{}} \otimes
v^{+{}\dag}\right)\nu^{(0)}_\beta = 0$ [see
Eq.~\eqref{spinorsmatt}] as well as Eqs.~\eqref{matrtranslambda},
\eqref{Ulambdaa}, \eqref{GenSolDirpsimatt}, and~\eqref{aeqsolmatt}
we arrive to the wave function of the flavor neutrino
$\nu_\alpha$,
\begin{align}\label{nualphamatt}
  \nu_\alpha(z,t) = &
  -\mathrm{i}
  \exp{(- \mathrm{i} \sigma_{-{}} t + \mathrm{i} k z)}
  \frac{\sin\Omega_{-{}} t}{\Omega_{-{}}}
  \notag
  \\
  & \times
  [g\cos 2\theta+(\omega_{-{}}/2)\sin 2\theta] \nu^{(0)}_\beta
  \notag
  \\
  & +
  \mathcal{O}
  \left(
    \frac{m_a}{k}
  \right),
\end{align}
where $\sigma_{-{}} = (E_1^{-{}} + E_2^{-{}})/2 + (g_1 + g_2)/2$.
Note that Eq.~\eqref{nualphamatt} is the most general one which
takes into account the nonstandard interactions of relativistic
neutrinos with nonmoving and unpolarized matter of arbitrary
density.

Now let us discuss the standard model neutrino interactions with
background matter. In this case the matrix
$(f_{\lambda\lambda'}^\mu)$ is diagonal: $f_{\lambda\lambda'}^\mu
= f_\lambda^{\mu} \delta_{\lambda\lambda'}$. Since we study the
nonmoving and unpolarized matter the spatial components of the
four vector $f_\lambda^{\mu}$ are equal zero. If we study the
matter composed of electrons, neutrons and protons, the zero-th
component $f_\lambda^0 \equiv f_\lambda$ is (see, e.g.,
Ref.~\cite{DvoStu02})
\begin{align}\label{fdefinition}
  f_\lambda = & \sqrt{2} G_\mathrm{F} \sum_{f=e,p,n} n_f q_f^{(\lambda)},
  \notag
  \\
  q_f^{(\lambda)} = &
  \left(
    I^{(f)}_{3\mathrm{L}} - 2 Q^{(f)} \sin^2\theta_W +
    \delta_{f e}\delta_{\lambda \nu_e}
  \right),
\end{align}
where $n_f$ is the number density of background particles,
$I^{(f)}_{3\mathrm{L}}$ is the third isospin component of the
matter fermion $f$, $Q^{(f)}$ is its electric charge, $\theta_W$
is the Weinberg angle and $G_\mathrm{F}$ is the Fermi constant.

Using Eqs.~\eqref{Ulambdaa}, \eqref{gmatrmatt}, and~\eqref{gab0}
and we get that matrix $(g_{ab})$ has the form,
\begin{multline}\label{flambda}
  (g_{ab}) =
  \\
  \begin{pmatrix}
    f_\alpha\cos^2 \theta + f_\beta\sin^2 \theta & \sin 2\theta \Delta V_\mathrm{eff}/2 \\
    \sin 2\theta \Delta V_\mathrm{eff}/2 & f_\alpha \sin^2\theta + f_\beta \cos^2\theta \
  \end{pmatrix},
\end{multline}
where $\Delta V_\mathrm{eff} = f_\beta - f_\alpha$ is the
difference between the effective potentials of the flavor
neutrinos interaction with background matter. With help of
Eq.~\eqref{fdefinition} we present $\Delta V_\mathrm{eff}$ in the
form,
\begin{equation}\label{Veffcases}
  \Delta V_\mathrm{eff} =
  \sqrt{2} G_\mathrm{F} \times
  \begin{cases}
    n_e, & \text{for $\nu_e \to \nu_{\mu,\tau}$}, \\
    0, & \text{for $\nu_{\mu,\tau} \to \nu_{\tau,\mu}$},
  \end{cases}
\end{equation}
for various oscillations channels.

In the following we will discuss the low density matter limit,
$g_a \ll k$, which is fulfilled for all realistic neutrino momenta
and densities of background matter. Indeed, even for the
background matter in the center of a neutron star where $n_n =
10^{38}\thinspace\text{cm}^{-3}$, using Eq.~\eqref{fdefinition} we
get $g_a \sim 10\thinspace\text{eV}$, which is much less than any
reasonable neutrino energy. With help of Eq.~\eqref{energymatt} we
get that in this approximation $\omega_{-{}}/2 \approx \Phi - \cos
2\theta \Delta V_\mathrm{eff}/2$, where $\Phi$ and $\delta m^2$
are defined in Eq.~\eqref{Phisigma}.

The transition probability for the process $\nu_\beta \to
\nu_\alpha$ can be calculated on the basis of
Eq.~\eqref{nualphamatt} as
\begin{equation}\label{Ptrmatt}
  P_{\nu_\beta \to \nu_\alpha}(t) =
  |\nu_\alpha(z,t)|^2 \approx
  P_\mathrm{max} \sin^2
  \left(
    \frac{\pi}{L_\mathrm{osc}}t
  \right),
\end{equation}
where
\begin{align}\label{parametersmatt}
  P_\mathrm{max} = &
  \frac{\Phi^2 \sin^2(2\theta)}
  {(\Phi \cos2\theta - \Delta V_\mathrm{eff}/2)^2 +
  \Phi^2\sin^2(2\theta)},
  \notag
  \\
  \frac{\pi}{L_\mathrm{osc}} = &
  \sqrt{(\Phi\cos2\theta - \Delta V_\mathrm{eff}/2)^2 +
  \Phi^2 \sin^2(2\theta)},
\end{align}
are the maximal transition probability and the oscillations
length.

One can see that Eqs.~\eqref{Ptrmatt} and~\eqref{parametersmatt}
reproduce the well known formula for the neutrino oscillations
probability in the background matter (see Ref.~\cite{Wol78}). If
the background density has the resonance value determined by,
$\Phi \cos2\theta = \Delta V_\mathrm{eff}^{(\mathrm{res})}/2$, the
maximal transition probability reaches big values $\sim 1$. This
resonance enhancement of neutrino oscillations in matter is known
as the MSW effect~\cite{Wol78}.

Now let us discuss the modification of Eqs.~\eqref{Ptrmatt}
and~\eqref{parametersmatt} which include a hypothetical
nonstandard neutrino interaction. One of the possibilities to
account for this kind of interactions is to study the nondiagonal
element of the matrix $(f_{\lambda\lambda'}^\mu)$. This
interaction can produce the neutrino flavor conversion in presence
of background matter. If still we discuss the nonmoving and
unpolarized matter, we define the additional nonzero element as $f
\equiv f_{\alpha\beta}^0 \neq 0$. We can express the nonstandard
interaction as $f = \epsilon_{\alpha\beta}(f_\alpha + f_\beta)/2$
although the exact form of the nonstandard interaction dependence
on the densities of background fermions is still open. The
experimental constraint on the $\epsilon_{\alpha\beta}$ parameters
reads $|\epsilon_{\alpha\beta}| \lesssim 0.4$~\cite{Den10}.

Using the same technique as to get Eqs.~\eqref{Ptrmatt}
and~\eqref{parametersmatt} we arrive to the modified maximal
transition probability and the oscillations length,
\begin{widetext}
\begin{equation}\label{nsimatt}
  P_\mathrm{max} =
  \frac{(\Phi \sin 2\theta + f)^2}
  {(\Phi \cos2\theta - \Delta V_\mathrm{eff}/2)^2+
  (\Phi \sin 2\theta + f)^2},
  \quad
  \frac{\pi}{L_\mathrm{osc}} =
  \sqrt{(\Phi\cos2\theta - \Delta V_\mathrm{eff}/2)^2+
  (\Phi \sin 2\theta + f)^2}.
\end{equation}
\end{widetext}
Eq.~\eqref{nsimatt} is exact and valid for arbitrary magnitude of
the nonstandard interaction $f$ in contrast to the perturbative
formulas derived in Ref.~\cite{KikMinUch09}.

As one can see in Eq.~\eqref{nsimatt} that in the majority of
cases the nonstandard neutrino interaction of the considered type
does not generate any additional resonances in neutrino
oscillations. The small new interaction can just slightly change
the shape of the transition probability.

We can however notice that the new interaction produces flavor
oscillations even for massless neutrinos. Indeed, if we suggest
that $m_a = 0$ (or $\Phi = 0$), we get that the parameters of
transition probability formula, given in Eq.~\eqref{nsimatt},
formally coincide with that in Eq.~\eqref{parametersmatt}, derived
for the massive neutrinos, if replace $\Phi \sin 2\theta \to f$
there. However we cannot expect the appearance of the usual MSW
resonance in this model since $\Phi = 0$. The amplification of
neutrino oscillation can happen only if $\Delta V_\mathrm{eff} =
0$. For example, it is the case for $\nu_\mu \leftrightarrow
\nu_\tau$ oscillations [see Eq.~\eqref{Veffcases}].

Note that we have chosen the plane wave initial condition
corresponding to ultrarelativistic particles to study neutrino
oscillations in background matter. It allowed us to exactly take
into account the contribution of the field $g$ in
Eq.~\eqref{ODEDirmatt}. It is, however, possible to study the
neutrino evolution with arbitrary initial condition in low density
matter~\cite{Dvo06}. It was shown in Ref.~\cite{Dvo06} that the
dynamics of neutrino oscillations is consistent with the results
of Ref.~\cite{Wol78}.

\section{Dirac neutrinos in an external magnetic field\label{B}}

In this section we apply the formalism developed in
Secs.~\ref{VACUUM} and~\ref{MATTER} for the description of
neutrino evolution in an external electromagnetic
field~\cite{DvoMaa07}. In contrast to the previous sections we
examine the situation when the helicity of a neutrino changes
together with its flavor, i.e. we study so called neutrino spin
flavor oscillations, $\nu_\beta^\mathrm{L,R} \leftrightarrow
\nu_\alpha^\mathrm{R,L}$. We derive the new transition probability
formulas which account for arbitrary magnetic moments matrix.

Neutrinos are known to be uncharged particles. The constraint on
the neutrino electric charge is at the level of
$10^{-13}\thinspace e$~\cite{IgnJos95}. Nevertheless there is a
possibility for them to interact with an external electromagnetic
field $F_{\mu\nu} = (\mathbf{E},\mathbf{B})$ via anomalous
magnetic moments. The experimental constraint on the neutrino
magnetic moments is $\sim 10^{-10}
\mu_\mathrm{B}$~\cite{Den10,Bed07}, where $\mu_\mathrm{B}$ is the
Bohr magneton. Despite the smallness of magnetic moments, its
interaction with strong electromagnetic fields can produces
sizeable effects (see, e.g., Secs.~\ref{SPECMATT}
and~\ref{STERILE} below).

The Lagrangian for the considered system of two flavor neutrinos
$\nu = (\nu_\alpha, \nu_\beta)$ is expressed in the following way:
\begin{align}\label{LagrnuB}
  \mathcal{L} = &
  \sum_{\lambda = \alpha, \beta}
  \bar{\nu}_\lambda
  \mathrm{i}\gamma^\mu\partial_\mu
  \nu_\lambda
  \notag
  \\
  & -
  \sum_{\lambda \lambda' = \alpha, \beta}
  \bar{\nu}_{\lambda}
  \left(
    m_{\lambda\lambda'} +
    \frac{1}{2} M_{\lambda\lambda'} \sigma_{\mu\nu} F^{\mu\nu}
  \right)
  \nu_{\lambda'},
\end{align}
where $\sigma_{\mu\nu} = (\mathrm{i}/2) (\gamma_\mu\gamma_\nu -
\gamma_\nu\gamma_\mu)$. The magnetic moments matrix
$(M_{\lambda\lambda'})$ in Eq.~\eqref{LagrnuB} is defined in the
flavor eigenstates basis. In general case this matrix is
independent from the mass matrix $(m_{\lambda\lambda'})$, i.e. the
diagonalization of the mass matrix does not necessarily imply the
diagonal form of the magnetic moments matrix.

To analyze the dynamics of the system~\eqref{LagrnuB} we formulate
the initial condition problem [see Eqs.~\eqref{inicondnulambda},
\eqref{inicondpsi}, and~\eqref{inicondnulambdaexp}] and introduce
the mass eigenstates $\psi_a$ [see Eqs.~\eqref{matrtranslambda}
and~\eqref{Ulambdaa}]. However, in contrast to the previous
sections we should choose the normalization spinor
$\nu_\beta^{(0)}$ in Eq.~\eqref{inicondnulambdaexp} in a specific
form.

When a neutrino with anomalous magnetic moment propagate in an
external electromagnetic field its helicity changes. Therefore we
can impose the additional constraint on the initial spinor,
\begin{equation}\label{inicondhel}
  P_{\pm{}} \nu_\beta^{(0)} =
  \nu_\beta^{(0)},
  \quad
  P_{\pm{}} =
  \left(
    1 \pm \frac{(\bm{\Sigma}\cdot\mathbf{k})}{|\mathbf{k}|}
  \right),
\end{equation}
which means that one has neutrinos of the specific helicity
initially. Here $\bm{\Sigma} = \gamma^5\bm{\alpha}$ is the Dirac
matrix. If we act with the operator $P_{\mp{}}$ on the final state
$\nu_\alpha(\mathbf{r},t)$, we can study the appearance of the
opposite helicity eigenstates among neutrinos of the flavor
``$\alpha$", i.e this situation corresponds to the neutrino spin
flavor oscillations $\nu_\beta^\mathrm{L,R} \leftrightarrow
\nu_\alpha^\mathrm{R,L}$. For the sake of definiteness we choose
the initial wave function corresponding to a left polarized
neutrino and the final one to a right polarized particle.

Now we express the Lagrangian~\eqref{LagrnuB} using the mass
eigenstates $\psi_a$, which diagonalize the mass matrix,
\begin{align}\label{LagrpsiB}
  \mathcal{L} = &
  \sum_{a=1,2}
  \bar{\psi}_a (\mathrm{i}\gamma^\mu \partial_\mu - m_a) \psi_a
  \notag
  \\
  & -
  \frac{1}{2}
  \sum_{ab=1,2}
  \mu_{ab} \bar{\psi}_a \sigma_{\mu\nu} \psi_b F^{\mu\nu},
\end{align}
where
\begin{equation}\label{magmomme}
  (\mu_{ab}) =
  U^\dag (M_{\lambda\lambda'}) U =
  \begin{pmatrix}
    \mu_{11} & \mu_{12} \\
    \mu_{21} & \mu_{22} \
  \end{pmatrix},
\end{equation}
is the magnetic moment matrix presented in the mass eigenstates
basis which, as we mentioned above, not necessarily to be
diagonal.

As in Secs.~\ref{VACUUM} and~\ref{MATTER} we will study the
situation of mass eigenstates neutrinos $\psi_a$ which are Dirac
particles. It means that the magnetic moments matrix $(\mu_{ab})$
can have both diagonal and nondiagonal elements. The diagonal
elements of this matrix correspond to usual magnetic moments and
the nondiagonal to the transition ones. The transition magnetic
moments are responsible for the transitions between left and right
polarized particles of different species.

Let us assume that the magnetic field is constant, uniform and
directed along the $z$-axis, $\mathbf{B}=(0, 0, B)$, and that the
electric field vanishes, $\mathbf{E}=0$. In this case we write
down the Pauli-Dirac equations for $\psi_a$, resulting from
Eq.~\eqref{LagrpsiB}, as follows:
\begin{align}\label{DireqpsiB}
  \mathrm{i} \dot{\psi}_a = & \mathcal{H}_a \psi_a +
  \mathcal{V} \psi_b,
  \quad
  a,b=1,2,
  \quad
  a \neq b,
  \notag
  \\
  \mathcal{H}_a = &
  (\bm{\alpha}\mathbf{p}) + \beta m_a - \mu_a \beta \Sigma_3 B,
  \quad
  \mathcal{V} = -\mu \beta \Sigma_3 B,
\end{align}
where $\mu_a=\mu_{aa}$, and $\mu=\mu_{ab}=\mu_{ba}$ are the
elements of the matrix $({\mu}_{ab})$~\eqref{magmomme}.

We should notice that, as in Sec.~\ref{MATTER}, the wave
equations~\eqref{DireqpsiB} are coupled due to the presence of the
interaction $\mathcal{V}$. Therefore we have to use a sort of the
perturbative approach to account for this term, whereas analogous
diagonal magnetic interaction $- \mu_a \beta \Sigma_3 B$ will be
taken into account exactly from the very beginning.

We will study the propagation of neutrinos in the transverse
magnetic field. Therefore it convenient to choose the initial
momentum along the $x$-axis $\mathbf{k} = (k,0,0)$ and the initial
spinor as $\nu_\beta^{(0)\mathrm{T}} = (1/2)(1,-1,-1,1)$. It is
easy to see that the wave function $\nu_\beta^{(0)}(\mathbf{r})$
describes an ultrarelativistic particle propagating along the
$x$-axis with its spin directed opposite to the $x$-axis, i.e. a
left polarized neutrino. The contribution of the longitudinal
magnetic field to the dynamics of neutrino oscillations is
suppressed by the factor $m_a/k \ll
1$~\cite{DvoStu02,AkhKhl88,EgoLobStu00}.

The general solution of Eq.~\eqref{DireqpsiB} can be presented as
follows:
\begin{align}\label{GenSolDirpsiB}
  \psi_{a}(\mathbf{r},t) = &
  \int \frac{\mathrm{d}^3\mathbf{p}}{(2\pi)^{3/2}}
  e^{\mathrm{i}\mathbf{p}\mathbf{r}}
  \notag
  \\
  & \times
  \sum_{\zeta=\pm 1}
  \big[
    a_a^{(\zeta)}(t)u_a^{(\zeta)}\exp{(-\mathrm{i}E_a^{(\zeta)} t)}
    \notag
    \\
    & +
    b_a^{(\zeta)}(t)v_a^{(\zeta)}\exp{(+\mathrm{i}E_a^{(\zeta)} t)}
  \big].
\end{align}
Our main goal is to determine the  coefficients $a_a^{(\zeta)}$
and $b_a^{(\zeta)}$ consistent with both the initial
condition~\eqref{inicondnulambdaexp} and~\eqref{inicondhel} and
the evolution equation~\eqref{DireqpsiB}. As in Sec.~\ref{MATTER}
these coefficients are in general functions of time.

We have already mentioned that the helicity of a neutral particle
with an anomalous magnetic moment is not conserved in an external
magnetic field. Therefore to classify the states of massive
neutrinos in Eq.~\eqref{GenSolDirpsiB} one has to use the
operator~\cite{DvoMaa07,TerBagKha65,Dvo10},
\begin{equation}\label{polarizoperator}
  \Pi_a = m_a \Sigma_3 +
  \mathrm{i}\gamma^0\gamma^5(\bm{\Sigma}\times\mathbf{p})_3 -
  \mu_a B,
\end{equation}
which commutes with the Hamiltonian $\mathcal{H}_a$ in
Eq.~\eqref{DireqpsiB} and thus characterizes the spin direction
with respect to the magnetic field. The quantum number $\zeta \pm
1$ is the sign of the eigenvalue of the
operator~\eqref{polarizoperator}.

The energy levels in Eq.~\eqref{GenSolDirpsiB} have the
form~\cite{DvoMaa07,Dvo10},
\begin{align}\label{energygenB}
  E_a^{(\zeta)} = & \sqrt{p_3^2 + \mathcal{E}_a^{(\zeta)2}},
  \quad
  \mathcal{E}_a^{(\zeta)} = \mathcal{K}_a - \zeta \mu_a B,
  \notag
  \\
  \mathcal{K}_a = & \sqrt{m_a^2 + p_1^2 + p_2^2}.
\end{align}
For our choice of the external magnetic field $\mathbf{B} = (0, 0,
B)$ and the initial momentum $\mathbf{k} = (k, 0, 0)$,
Eq.~\eqref{energygenB} reads
\begin{equation}\label{energyB}
  E_a^{(\zeta)} = \mathcal{K}_a - \zeta \mu_a B \approx
  k + \frac{m_a^2}{2k} - \zeta \mu_a B,
\end{equation}
for ultrarelativistic neutrinos with $k \gg m_a$. Here
$\mathcal{K}_a = \sqrt{k^2 + m_a^2}$ is the kinetic energy of
massive neutrinos.

The exact form for the basis spinors $u_a^{(\zeta)}$ and
$v_a^{(\zeta)}$ in Eq.~\eqref{GenSolDirpsiB} for arbitrary
neutrino momentum is presented in
Refs.~\cite{DvoMaa07,TerBagKha65}. We reproduce the basis spinors
for ultrarelativistic neutrinos,
\begin{align}\label{spinorsB}
  u^{-{}} = &
  \frac{1}{\sqrt{2}}
  \begin{pmatrix}
     0 \\
     1 \\
     1 \\
     0 \
  \end{pmatrix},
  \quad
  u^{+{}} =
  \frac{1}{\sqrt{2}}
  \begin{pmatrix}
     1 \\
     0 \\
     0 \\
     1 \
  \end{pmatrix},
  \notag
  \\
  v^{-{}} = &
  \frac{1}{\sqrt{2}}
  \begin{pmatrix}
     1 \\
     0 \\
     0 \\
     -1 \
  \end{pmatrix},
  \quad
  v^{+{}} =
  \frac{1}{\sqrt{2}}
  \begin{pmatrix}
     0 \\
     1 \\
     -1 \\
     0 \
  \end{pmatrix},
\end{align}
since we will be interested in the evolution of such particles. In
Eq.~\eqref{spinorsB} we omit the index ``$a$" since we examine the
case of $k \gg m_a$. The basis spinors $u^{-{}}$ and $v^{-{}}$
correspond to the negative eigenvalue of the
operator~\eqref{polarizoperator} (neutrino spin is directed
oppositely to the magnetic field), and $u^{+{}}$ and $v^{+{}}$ to
the positive one (neutrino spin is parallel to the magnetic
field).

Using the general solution~\eqref{GenSolDirpsiB} of the
Pauli-Dirac equation~\eqref{DireqpsiB} containing the undetermined
functions $a_a^{(\zeta)}$ and $b_a^{(\zeta)}$ and taking into
account the orthonormality of the basis spinors~\eqref{spinorsB}
we get the system of ordinary differential equations for these
functions $a_a^{(\zeta)}$ and $b_a^{(\zeta)}$:
\begin{align}\label{ODEDirB}
  \mathrm{i}\dot{a}_a^{(\zeta)} = &
  \exp{(+\mathrm{i}E_a^{(\zeta)}t)}u_a^{(\zeta)\dag}
  \mathcal{V}
  \notag
  \\
  & \times
  \displaystyle{\sum_{\zeta'=\pm 1}}
  \Big[
    a_b^{(\zeta')}u_b^{(\zeta')}\exp{(-\mathrm{i}E_b^{(\zeta')} t)}
    \notag
    \\
    & +
    b_b^{(\zeta')}v_b^{(\zeta')}\exp{(+\mathrm{i}E_b^{(\zeta')} t)}
  \Big],
  \notag
  \\
  \mathrm{i}\dot{b}_a^{(\zeta)} = &
  \exp{(-\mathrm{i}E_a^{(\zeta)} t)}v_a^{(\zeta)\dag}
  \mathcal{V}
  \notag
  \\
  & \times
  \displaystyle{\sum_{\zeta'=\pm 1}}
  \Big[
    a_b^{(\zeta')}u_b^{(\zeta')}\exp{(-\mathrm{i}E_b^{(\zeta')} t)}
    \notag
    \\
    & +
    b_b^{(\zeta')}v_b^{(\zeta')}\exp{(+\mathrm{i}E_b^{(\zeta')} t)}
  \Big],
\end{align}
which should be supplied with the initial
condition~\eqref{inicondamatt} but with different initial wave
function $\nu_\beta^{(0)\mathrm{T}} = (1/2)(1,-1,-1,1)$ (see
above).

With help of the obvious identities $\langle u_a^{\pm{}} |
\mathcal{V} | u_b^{\pm{}} \rangle=\mp \mu B$ and $\langle
u_a^{\pm{}} | \mathcal{V} | v_b^{\mp{}} \rangle=0$, which result
from Eq.~\eqref{spinorsB}, one can cast Eq.~\eqref{ODEDirB} into
the form
\begin{equation}\label{aeqB}
  \mathrm{i}\dot{a}_a^{\pm{}}=
  \mp a_b^{\pm{}}\mu B
  \exp{[\mathrm{i}(E_a^{\pm{}}-E_b^{\pm{}}) t]},
\end{equation}
which is analogous to Eq.~\eqref{aeqmatt} studied in
Sec.~\ref{MATTER}. Note that the ordinary differential equations
for the functions $a_a^{(\zeta)}$ and $b_a^{(\zeta)}$ again
decouple.

On the basis of the results of Appendix~\ref{DETAILSABSOL} we are
able to write down the solution of Eq.~\eqref{aeqB} as
\begin{align}\label{aeqsolB}
  a_1^{\pm{}}(t) = &
  F^{\pm{}}a_1^{\pm{}}(0)+G^{\pm{}}a_2^{\pm{}}(0),
  \notag
  \\
  a_2^{\pm{}}(t) = &
  F^{\pm{}*{}}a_2^{\pm{}}(0)-G^{\pm{}*{}}a_1^{\pm{}}(0),
\end{align}
where
\begin{align}\label{FGB}
  F^{\pm{}}= &
  \left[
    \cos\Omega_{\pm{}}t-
    \mathrm{i}\frac{\omega_{\pm{}}}{2\Omega_{\pm{}}}\sin\Omega_{\pm{}}t
  \right]\exp{(\mathrm{i}\omega_{\pm{}}t/2)},
  \notag
  \\
  G^{\pm{}}= &
  \pm\mathrm{i}\frac{\mu B}{\Omega_{\pm{}}}\sin\Omega_{\pm{}}t
  \exp{(\mathrm{i}\omega_{\pm{}}t/2)},
\end{align}
and
\begin{equation}\label{OmegaomegaB}
  \Omega_{\pm{}}=\sqrt{(\mu B)^2+(\omega_{\pm{}}/2)^2},
  \quad
  \omega_{\pm{}}=E_1^{\pm{}}-E_2^{\pm{}}.
\end{equation}
The details of the derivation of
Eqs.~\eqref{aeqsolB}-\eqref{OmegaomegaB} from Eqs.~\eqref{aeqB}
are also presented in Ref.~\cite{DvoMaa07}.

Using Eq.~\eqref{GenSolDirpsiB} and
Eqs.~\eqref{ODEDirB}-\eqref{OmegaomegaB} and the identity $\left(
v^{(\zeta)} \otimes v^{(\zeta)\dag} \right)\nu_\beta^{(0)}=0$ [see
Eq.~\eqref{spinorsB}] we obtain the wave functions  $\psi_a$,
$a=1,2$, as,
\begin{align}\label{psisolB}
  \psi_1(x,t)= &
  \exp{(-\mathrm{i}E_1^{+{}}t)}
  \left(
    u^{+{}}\otimes u^{+{}\dag}
  \right)
  \notag
  \\
  & \times
  [F^{+{}}\psi_1(x,0)+G^{+{}}\psi_2(x,0)]
  \notag
  \\
  & +
  \exp{(-\mathrm{i}E_1^{-{}}t)}
  \left(
    u^{-{}}\otimes u^{-{}\dag}
  \right)
  \notag
  \\
  & \times
  [F^{-{}}\psi_1(x,0)+G^{-{}}\psi_2(x,0)],
  \notag
  \\
  \psi_2(x,t)= &
  \exp{(-\mathrm{i}E_2^{+{}}t)}
  \left(
    u^{+{}}\otimes u^{+{}\dag}
  \right)
  \notag
  \\
  & \times
  [F^{+{}*{}}\psi_2(x,0)-G^{+{}*{}}\psi_1(x,0)]
  \notag
  \\
  & +
  \exp{(-\mathrm{i}E_2^{-{}}t)}
  \left(
    u^{-{}}\otimes u^{-{}\dag}
  \right)
  \notag
  \\
  & \times
  [F^{-{}*{}}\psi_2(x,0)-G^{-{}*{}}\psi_1(x,0)],
\end{align}
which satisfy the chosen initial condition since $G^{\pm{}}(0)=0$,
$F^{\pm{}}(0)=1$ [see Eqs.~\eqref{FGB}] and $[(u^{+{}}\otimes
u^{+{}\dag})+(u^{-{}}\otimes u^{-{}\dag})]\psi_a(x,0) =
\psi_a(x,0)$ [see Eq.~\eqref{spinorsB}].

To study the appearance of right polarized neutrinos of the type
``$\alpha$" we should act with the operator $P_{+{}} =
(1+\Sigma_1)/2$ defined in Eq.~\eqref{inicondhel} on the final
wave function $\nu_\alpha(x,t)$,
\begin{align}\label{nualphaRB}
  \nu_{\alpha}^\mathrm{R}(x,t)= &
  \frac{1}{2}(1+\Sigma_1)
  \notag
  \\
  & \times
  \left[
    \cos\theta\psi_a(x,t) - \sin\theta\psi_a(x,t)
  \right],
\end{align}
where $\psi_a(x,t)$ are shown in Eq.~\eqref{psisolB}.

With help of Eqs.~\eqref{matrtranslambda}, \eqref{Ulambdaa},
\eqref{psisolB}, and~\eqref{nualphaRB} we receive for the right
polarized component of $\nu_\alpha$ the expression
\begin{align}\label{nualphaRgenB}
  \nu_\alpha^{\mathrm{R}}(x,t)= &
  \frac{1}{2}
  \big\{
    \sin\theta\cos\theta
    \big[
      e^{-\mathrm{i}\mathcal{K}_1 t}
      (e^{\mathrm{i}\mu_1 B t}F^{+{}}-e^{-\mathrm{i}\mu_1 B t}F^{-{}})
      \notag
      \\
      & -
      e^{-\mathrm{i}\mathcal{K}_2 t}
      (e^{\mathrm{i}\mu_2 B t}F^{+{}*{}}-e^{-\mathrm{i}\mu_2 B t}F^{-{}*{}})
    \big]
    \notag
    \\
    & +
    \cos^2\theta
    e^{-\mathrm{i}\mathcal{K}_1 t}
    (e^{\mathrm{i}\mu_1 B t}G^{+{}}-e^{-\mathrm{i}\mu_1 B t}G^{-{}})
    \notag
    \\
    & +
    \sin^2\theta
    e^{-\mathrm{i}\mathcal{K}_2 t}
    (e^{\mathrm{i}\mu_2 B t}G^{+{}*{}}-e^{-\mathrm{i}\mu_2 B t}G^{-{}*{}})
  \big\}
  \notag
  \\
  & \times
  e^{\mathrm{i}kx} \nu_\alpha^{(0)\mathrm{R}},
\end{align}
where $\left( \nu_\alpha^{(0)\mathrm{R}} \right)^\mathrm{T} =
(1/2)(1,1,1,1)$ is the normalized spinor representing the right
polarized final neutrino state.

Finally, taking into account Eqs.~\eqref{FGB}
and~\eqref{OmegaomegaB} it is possible to express the wave
function in Eq.~\eqref{nualphaRgenB} in the form
\begin{align}\label{nualphaRexpB}
  \nu_\alpha^{\mathrm{R}}(x,t)= &
  \bigg\{
    \sin\theta\cos\theta
    \frac{1}{2\mathrm{i}}
    \bigg[
      \frac{\omega_{+{}}}{\Omega_{+{}}}\sin(\Omega_{+{}}t)
      \exp{(\mathrm{i}\bar{\mu}Bt)}
      \notag
      \\
      & -
      \frac{\omega_{-{}}}{\Omega_{-{}}}\sin(\Omega_{-{}}t)
      \exp{(-\mathrm{i}\bar{\mu}Bt)}
    \bigg]
    \notag
    \\
    & +
    \mathrm{i}\mu B
    \bigg[
      \frac{\sin(\Omega_{+{}}t)}{\Omega_{+{}}}\cos^2\theta
      \notag
      \\
      & -
      \frac{\sin(\Omega_{-{}}t)}{\Omega_{-{}}}\sin^2\theta
    \bigg]
    \cos(\bar{\mu}Bt)
  \bigg\}
  \notag
  \\
  & \times
  \exp{(-\mathrm{i} \sigma t+\mathrm{i}kx)}
  \nu_\alpha^{(0)\mathrm{R}},
\end{align}
where $\sigma = (\mathcal{K}_1 + \mathcal{K}_1)/2$ and $\bar{\mu}
= (\mu_1 + \mu_2)/2$. The magnetic moments $\mu$ and $\mu_a$ are
defined in Eq.~\eqref{magmomme}.

The transition probability for the process
$\nu_\beta^\mathrm{L}\to\nu_\alpha^\mathrm{R}$ can be directly
obtained as the squared modulus of $\nu_\alpha^{\mathrm{R}}(x,t)$
from Eq.~\eqref{nualphaRgenB} or Eq.~\eqref{nualphaRexpB}, that is
$P_{\nu_\beta^\mathrm{L}\to\nu_\alpha^\mathrm{R}}(t)=
|\nu_\alpha^{\mathrm{R}}(x,t)|^2$. Notice that the probability is
a function of time alone with no dependence on spatial
coordinates. This is of course obvious as we have taken the
initial wave function as a plane wave and the the magnetic field
spatially constant.

Let us now apply the general results Eq.~\eqref{nualphaRgenB} or
Eq.~\eqref{nualphaRexpB} to two special cases. We first consider
the situation where $\mu_{1,2}\gg\mu$, i.e. the case when the
transition magnetic moment is small compared with the diagonal
ones. Using Eqs.~\eqref{FGB} and~\eqref{OmegaomegaB} we find that,
in this case $F^{\pm{}}\approx 1$ and
$\Omega_{\pm{}}\approx\omega_{\pm{}}/2$, and
Eq.~\eqref{nualphaRgenB} takes the form
\begin{align}\label{nualphaRpertB}
  \nu_\alpha^{\mathrm{R}}(x,t)\approx &
  \mathrm{i}
  \bigg\{
    \sin\theta\cos\theta
    \big[
      e^{-\mathrm{i}\mathcal{K}_1 t}\sin\mu_1 B t
      \notag
      \\
      & -
      e^{-\mathrm{i}\mathcal{K}_2 t}\sin\mu_2 B t
    \big]
    \notag
    \\
    & +
    \cos 2\theta\frac{\mu B}{2}
    \bigg(
      e^{-\mathrm{i}\Sigma_{+{}} t}\frac{\sin\Delta_{+{}} t}{\Delta_{+{}}}
      \notag
      \\
      & +
      e^{-\mathrm{i}\Sigma_{-{}} t}\frac{\sin\Delta_{-{}} t}{\Delta_{-{}}}
    \bigg)
  \bigg\}
  e^{\mathrm{i}kx} \nu_\alpha^{(0)\mathrm{R}},
\end{align}
where
\begin{equation}
  \notag
  \Sigma_{\pm{}} = \sigma \pm \bar{\mu}B,
  \quad
  \Delta_{\pm{}} = \Phi \pm \delta \mu B,
  \quad
  \delta \mu = \frac{\mu_1 - \mu_2}{2},
\end{equation}
and the phase of vacuum oscillations $\Phi$ was defined in
Eq.~\eqref{Phisigma}. Eq.~\eqref{nualphaRpertB} was obtained in
Ref.~\cite{DvoMaa07} using the perturbative methods. Assuming that
$\mu \ll \mu_a$ the perturbation theory was developed in that
work. Now we rederive the same result as a particular case of the
more general result.

\begin{figure}
  \centering
  \includegraphics[scale=.45]{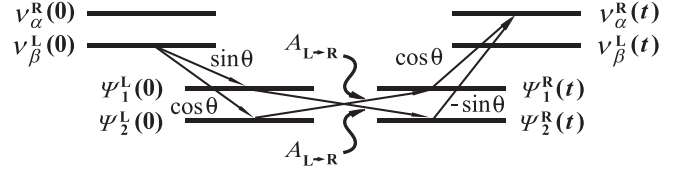}
  \caption{\label{diagram}
  The schematic illustration of the system evolution in the case
  $\mu\gg\mu_{1,2}$.
  The horizontal lines of the figure correspond to
  various neutrino eigenstates at different
  moments of time ($t=0$ and $t$).
  The expressions next to arrows correspond to the
  appropriate factors in the formula~\eqref{nualphaRbptB} of the
  wave function. The arrows from $\nu_\beta^\mathrm{L}(0)$ to
  $\psi_1^\mathrm{L}(0)$ and
  $\psi_2^\mathrm{L}(0)$, for example, indicate the vacuum mixing matrix
  transformation at $t=0$ and the
  arrow from $\psi_1^\mathrm{L}(0)$ to $\psi_2^\mathrm{R}(t)$
  the evolution of the mass eigenstates with the
  helicity change.
  The transitions $\psi_a^\mathrm{L}(0)\to\psi_b^\mathrm{R}(t)$
  can be described by the formula,
  $\psi_{1,2}^{\mathrm{R}}(t) =
  A_{\mathrm{L}\to\mathrm{R}}\psi_{2,1}^{\mathrm{L}}(0)$, where
  $A_{\mathrm{L}\to\mathrm{R}}=\mathrm{i}(\mu B/\Omega)\sin \Omega t$.
  This figure is taken from Ref.~\cite{DvoMaa07}.}
\end{figure}

As another application of our general result we will study the
situation, where the transition magnetic moments are much larger
than the diagonal ones, that is $\mu\gg\mu_{1,2}$.  In this case
Eqs.~\eqref{FGB} gives $F^{+{}}\approx F^{-{}}$ and
$G^{+{}}\approx - G^{-{}}$, and we receive from
Eq.~\eqref{nualphaRexpB} for the wave function
$\nu_\alpha^{\mathrm{R}}$ the expression
\begin{align}\label{nualphaRbptB}
  \nu_\alpha^{\mathrm{R}}(x,t) \approx &
  \mathrm{i}\exp{(-\mathrm{i} \sigma t+\mathrm{i}kx)}\cos (2\theta)
  \notag
  \\
  & \times
  \frac{\mu B}{\Omega_\mathrm{B}}\sin(\Omega_\mathrm{B} t)
  \nu_\alpha^{(0)\mathrm{R}},
\end{align}
where
\begin{equation}\label{OmegaMaj}
  \Omega_\mathrm{B} = \sqrt{(\mu B)^2+\Phi^2}.
\end{equation}
The transition probability for the process
$\nu_\beta^\mathrm{L}\to\nu_\alpha^\mathrm{R}$ is then given by
\begin{equation}\label{PtrbptB}
  P_{\nu_\beta^\mathrm{L}\to\nu_\alpha^\mathrm{R}}(t)=
  \cos^2(2\theta)
  \left(
    \frac{\mu B}{\Omega_\mathrm{B}}
  \right)^2
  \sin^2(\Omega_\mathrm{B} t).
\end{equation}
The behavior of the system in this case is schematically
illustrated in Fig~\ref{diagram}.
It should be noticed that the analog of Eq.~\eqref{PtrbptB} was
obtained in Ref.~\cite{LimMar88} where the authors studied the
resonant spin flavor precession of Dirac and Majorana neutrinos in
matter under the influence of an external magnetic field in frames
of the quantum mechanical approach.

Using Eq.~\eqref{nualphaRexpB} one can describe spin flavor
oscillations of Dirac neutrinos with arbitrary magnetic moments
matrix. It is the new result which was obtained using the
relativistic quantum mechanics approach. Nevertheless this result
is consistent with the conventional quantum mechanical description
of spin flavor oscillations. We will demonstrate the consistency
in Sec.~\ref{MATTERB} where more general case of neutrinos
propagating in matter and external magnetic field is studied.

Spin flavor oscillations of Dirac neutrinos with arbitrary initial
condition, not necessarily corresponding to ultrarelativistic
particles, were studied in Ref.~\cite{DvoMaa07} using the
perturbative approach. To effectively apply the perturbation
theory one has to study the situation of small transition magnetic
moment, $\mu \ll \mu_a$. One could derive
Eq.~\eqref{nualphaRpertB} using the results of
Ref.~\cite{DvoMaa07} in the limit $k \gg m_a$.

\section{Dirac neutrinos in matter under the influence of a
magnetic field\label{MATTERB}}

In this section, using the relativistic quantum mechanics method,
we study the general case of the mixed flavor neutrinos
propagating in background matter and interacting with an external
electromagnetic field~\cite{Dvo10}. We formulate the initial
condition problem for neutrino spin flavor oscillations. Then we
derive the effective Hamiltonian which governs spin flavor
oscillations and show the consistence of our approach to the usual
quantum mechanics method. The corrections to the standard
effective Hamiltonian are also obtained.

The Lagrangian for the system of two mixed flavor neutrinos $\nu =
(\nu_\alpha, \nu_\beta)$ interacting with background matter and
external electromagnetic field has the form,
\begin{align}\label{LagrnumattB}
  \mathcal{L} = &
  \sum_{\lambda=\alpha,\beta}
  \bar{\nu}_\lambda
  \mathrm{i}\gamma^\mu\partial_\mu
  \nu_{\lambda}
  -
  \sum_{\lambda\lambda'=\alpha,\beta}
  \bar{\nu}_\lambda
  \Big(
    m_{\lambda\lambda'} +
    \gamma_\mu^\mathrm{L} f^\mu_{\lambda\lambda'}
    \notag
    \\
    & +
    \frac{1}{2}
    M_{\lambda\lambda'}
    \sigma_{\mu\nu} F^{\mu\nu}
  \Big)
  \nu_{\lambda'},
\end{align}
where the mass matrix $(m_{\lambda\lambda'})$, matter interaction
matrix $(f^\mu_{\lambda\lambda'})$, and the magnetic moments
matrix $(M_{\lambda\lambda'})$ are defined in Secs.~\ref{VACUUM},
\ref{MATTER}, and~\ref{B} respectively.

In the following we will be interested in the standard model
neutrino interaction with matter which corresponds to the diagonal
matrix $f^\mu_{\lambda\lambda'} = \delta_{\lambda\lambda'}
f^\mu_{\lambda}$. Moreover we will study the situation of
nonmoving and unpolarized matter. In this case only zero-th
component of the four vector $f^\mu_{\lambda}$ is not equal to
zero. The explicit form of this component $f_{\lambda} \equiv
f^0_{\lambda}$ for the background matter composed of electrons,
protons, and neutrons is given in Eq.~\eqref{fdefinition}.

We choose the configuration of the electromagnetic field
$F_{\mu\nu} = (\mathbf{E},\mathbf{B})$ in Eq.~\eqref{LagrnumattB}
in the same form as in Sec.~\ref{B}. Namely, we suppose that the
electric field is absent $\mathbf{E} = 0$ and magnetic field is
constant and directed along the $z$-axis, $\mathbf{B} = (0, 0,
B)$.

To study the time evolution of flavor neutrinos we should supply
the Lagrangian~\eqref{LagrnumattB} with some initial condition. We
choose the initial wave functions in the same form as in
Sec.~\ref{B}, i.e. we suppose that $\nu_\alpha(\mathbf{r},0) = 0$
and $\nu_\beta(\mathbf{r},0) = e^{\mathrm{i} k x}
\nu_\beta^{(0)}$, where the spinor $\nu_\beta^{(0)}$ corresponds
to either left or right polarized neutrinos. The explicit form of
$\nu_\beta^{(0)}$ can be defined with help of the operators
$P_{\pm{}}$ in Eq.~\eqref{inicondhel}. It means that initially we
have a beam of neutrinos of the flavor ``$\beta$" with a specific
polarization propagating along the $x$-axis. If we study the
appearance of neutrinos of the flavor ``$\alpha$" of the opposite
polarization, it will correspond to a typical situation of
neutrino spin flavor oscillations in matter and transversal
magnetic field, $\nu_\beta^\mathrm{L,R} \leftrightarrow
\nu_\alpha^\mathrm{R,L}$.

Then we introduce the mass eigenstates $\psi_a$ using
Eqs.~\eqref{matrtranslambda} and~\eqref{Ulambdaa} to diagonalize
the mass matrix $(m_{\lambda\lambda'})$ in
Eq.~\eqref{LagrnumattB}. These mass eigenstates are again supposed
to be Dirac particles. Now the Lagrangian~\eqref{LagrnumattB}
expressed via the new mass eigenstates has the form,
\begin{align}\label{LagrpsimattB}
  \mathcal{L} = &
  \sum_{a=1,2}
  \bar{\psi}_a (\mathrm{i}\gamma^\mu \partial_\mu - m_a) \psi_a
  \notag
  \\
  & -
  \sum_{ab=1,2}
  \bar{\psi}_a
  \left(
    g_{ab}^{\mu} \gamma_\mu^\mathrm{L} +
    \frac{1}{2} \mu_{ab} \sigma_{\mu\nu} F^{\mu\nu}
  \right) \psi_b,
\end{align}
where $(g_{ab}^{\mu})$ and $(\mu_{ab})$ are the matrix of neutrino
interaction with matter and the neutrino magnetic moments matrix
expressed in the mass eigenstates basis, which are defined in
Eqs.~\eqref{gmatrmatt} and~\eqref{magmomme}. We remind that in
case of nonmoving and unpolarized matter the matrix
$(g_{ab}^{\mu})$ has only zero-th component~\eqref{gab0}.

On the basis of the mass eigenstates
Lagrangian~\eqref{LagrpsimattB}, we can derive the corresponding
wave equations which have the following form:
\begin{align}\label{DireqmattB}
  \mathrm{i}\dot{\psi}_a = & \mathcal{H}_a \psi_a + \mathcal{V} \psi_b,
  \quad
  a=1,2,
  \quad
  a \neq b,
  \notag
  \\
  \mathcal{H}_a = & (\bm{\alpha}\mathbf{p}) + \beta m_a -
  \mu_a\beta\Sigma_3 B + g_a(1-\gamma^5)/2,
  \notag
  \\
  \mathcal{V} = & -\mu\beta\Sigma_3 B + g(1-\gamma^5)/2.
\end{align}
Note that we cannot directly solve the wave
equations~\eqref{DireqmattB} because of the nondiagonal
interaction $\mathcal{V}$ which mixes different mass eigenstates
(see also Secs.~\ref{MATTER} and~\ref{B}). Nevertheless we can
point out an exact solution of the wave equation
$\mathrm{i}\dot{\psi}_a = \mathcal{H}_a\psi_a$, for a single mass
eigenstate $\psi_a$, that exactly accounts for the influence of
the external fields $g_a$ and $\mu_a B$. The contribution of the
mixing potential $\mathcal{V}$ can be then taken into account
using the perturbation theory, with all the terms in the expansion
series being accounted for exactly.

We look for the solution of Eq.~\eqref{DireqmattB} in the
following form~\cite{Dvo10}:
\begin{align}\label{GenSolDirpsimattB}
  \psi_{a}(\mathbf{r},t)= &
  e^{- \mathrm{i} g_a t/2}
  \int \frac{\mathrm{d}^3 \mathbf{p}}{(2\pi)^{3/2}}
  e^{\mathrm{i} \mathbf{p} \mathbf{r}}
  \notag
  \\
  & \times
  \sum_{\zeta=\pm 1}
  \Big[
    a_a^{(\zeta)}(t)u_a^{(\zeta)}\exp{(-\mathrm{i}E_a^{(\zeta)} t)}
    \notag
    \\
    & +
    b_a^{(\zeta)}(t)v_a^{(\zeta)}\exp{(\mathrm{i}E_a^{(\zeta)} t)}
  \Big],
\end{align}
where the energy levels, which were found in Ref.~\cite{Dvo08JPG},
have the form,
\begin{align}\label{EnergymattB}
  E^{(\zeta)}_a = & \sqrt{\mathcal{M}^2_a+m^2_a+p^2-2\zeta R^2_a},
  \notag
  \\
  \mathcal{M}_a = & \sqrt{(\mu_a B)^2 + g_a^2/4},
\end{align}
where $R^2_a=\sqrt{p^2 \mathcal{M}^2_a + (\mu_a B)^2 m^2_a}$.

The basis spinors in Eq.~\eqref{GenSolDirpsimattB} can be found in
the limit of the small neutrino mass~\cite{Dvo08JPG},
\begin{widetext}
\begin{align}\label{spinorsmattB}
  u^{(\zeta)}_a = &
  \frac{1}{2\sqrt{2\mathcal{M}_a(\mathcal{M}_a-\zeta g_a/2)}}
  \begin{pmatrix}
     \mu_a B+\zeta\mathcal{M}_a-g_a/2 \\
     \mu_a B-\zeta\mathcal{M}_a+g_a/2 \\
     \mu_a B-\zeta\mathcal{M}_a+g_a/2 \\
     \mu_a B+\zeta\mathcal{M}_a-g_a/2 \
  \end{pmatrix},
  \notag
  \\
  v^{(\zeta)}_a = &
  \frac{1}{2\sqrt{2\mathcal{M}_a(\mathcal{M}_a+\zeta g_a/2)}}
  \begin{pmatrix}
     \mathcal{M}_a-\zeta[\mu_a B-g_a/2] \\
     \mathcal{M}_a+\zeta[\mu_a B+g_a/2] \\
     -\mathcal{M}_a-\zeta[\mu_a B+g_a/2] \\
     -\mathcal{M}_a+\zeta[\mu_a B-g_a/2] \
  \end{pmatrix}.
\end{align}
\end{widetext}
It should be noted that the discrete quantum number $\zeta = \pm
1$ in Eqs.~\eqref{GenSolDirpsimattB}-\eqref{spinorsmattB} does not
correspond to the helicity quantum states.

Now our goal is to find the time dependent coefficients
$a_a^{(\zeta)}(t)$ and $b_a^{(\zeta)}(t)$. On the basis of the
general solution~\eqref{GenSolDirpsimattB} of the wave
equation~\eqref{DireqmattB} we obtain the ordinary differential
equations for these functions which formally coincide with
Eq.~\eqref{ODEDirmatt}. However the mixing potential $\mathcal{V}$
is now defined in Eq.~\eqref{DireqmattB}. To obtain the modified
Eq.~\eqref{ODEDirmatt} we again use the orthonormality of the
basis spinors~\eqref{spinorsmattB}. The initial condition for the
functions $a_a^{(\zeta)}(t)$ and $b_a^{(\zeta)}(t)$ also coincide
with Eq.~\eqref{inicondamatt}, with $\nu_\beta^{(0)}$
corresponding to a definite helicity spinor.

Taking into account the fact that $\langle u^{(\zeta)}_a |
\mathcal{V} | v^{(\zeta')}_b \rangle = 0$, we get that the
equations for $a_a^{(\zeta)}(t)$ and $b_a^{(\zeta)}(t)$ decouple,
i.e. the interaction $\mathcal{V}$ does not mix positive and
negative energy eigenstates. In the following we will consider the
evolution of only $a_a^{(\zeta)}(t)$ since the dynamics of
$b_a^{(\zeta)}(t)$ is studied analogously.

Let us rewrite the modified Eq.~\eqref{ODEDirmatt} in the more
conventional effective Hamiltonian form. For this purpose we
introduce the ``wave function"
$\Psi^{'\mathrm{T}}=(a_1^{-{}},a_2^{-{}},a_1^{+{}},a_2^{+{}})$.
Directly from the modified Eq.~\eqref{ODEDirmatt} for the
functions $a_a^{(\zeta)}(t)$ we derive the equation for $\Psi'$,
\begin{widetext}
\begin{equation}\label{Schr1QFT}
  \mathrm{i}\frac{\mathrm{d}\Psi'}{\mathrm{d}t} = H' \Psi',
  \quad
  H' =
  \begin{pmatrix}
    0 & h_{-{}} e^{\mathrm{i} \omega_{-{}} t} &
    0 & H_{-{}} e^{\mathrm{i} \Omega_{-{}} t} \\
    h_{-{}} e^{-\mathrm{i} \omega_{-{}} t} &
    0 & H_{+{}} e^{-\mathrm{i} \Omega_{+{}} t} & 0 \\
    0 & H_{+{}} e^{\mathrm{i} \Omega_{+{}} t} &
    0 & h_{+{}} e^{\mathrm{i} \omega_{+{}} t} \\
    H_{-{}} e^{-\mathrm{i} \Omega_{-{}} t} &
    0 & h_{+{}} e^{-\mathrm{i} \omega_{+{}} t} & 0 \
  \end{pmatrix},
\end{equation}
where
\begin{align}\label{hH}
  h_{\mp{}} = &
  \langle u^{\mp{}}_a | \mathcal{V} | u^{\mp{}}_b \rangle =
  \frac{1}{8\sqrt{\mathcal{M}_a\mathcal{M}_b
  (\mathcal{M}_a\pm g_a/2)(\mathcal{M}_b\pm g_b/2)}}
  \notag
  \\
  & \times
  [2\mu B (g_a \mu_b B + g_b \mu_a B) \pm
  4\mu B(\mu_a B\mathcal{M}_b+\mu_b B \mathcal{M}_a)
  \pm
  2g(g_a \mathcal{M}_b+g_b \mathcal{M}_a) +
  4 g \mathcal{M}_a\mathcal{M}_b+g g_a g_b],
  \quad
  a \neq b,
  \notag
  \\
  H_{\mp{}} = &
  \langle u^{\mp{}}_1 | \mathcal{V} | u^{\pm{}}_2 \rangle =
  \langle u^{\pm{}}_2 | \mathcal{V} | u^{\mp{}}_1 \rangle =
  \frac{1}{8\sqrt{\mathcal{M}_1\mathcal{M}_2
  (\mathcal{M}_1\pm g_1/2)(\mathcal{M}_2\mp g_2/2)}}
  \notag
  \\
  & \times
  [2\mu B (g_1 \mu_2 B + g_2 \mu_1 B) \mp
  4\mu B(\mu_1 B\mathcal{M}_2-\mu_2 B \mathcal{M}_1)
  \mp
  2g(g_1 \mathcal{M}_2-g_2 \mathcal{M}_1) -
  4 g \mathcal{M}_1\mathcal{M}_2+g g_1 g_2],
\end{align}
as well as $\omega_{\mp{}}=E_1^{\mp{}}-E_2^{\mp{}}+(g_1-g_2)/2$
and $\Omega_{\mp{}}=E_1^{\mp{}}-E_2^{\pm{}}+(g_1-g_2)/2$.

Instead of $\Psi'$ it is more convenient to use the transformed
``wave function" $\Psi$ defined by
\begin{equation}\label{PsiPsi'transmattB}
  \Psi' = \mathcal{U}\Psi,
  \quad
  \mathcal{U} = \mathrm{diag}
  \left\{
    e^{\mathrm{i}(\Omega+\omega_{-{}})t/2},
    e^{\mathrm{i}(\Omega-\omega_{-{}})t/2},
    e^{-\mathrm{i}(\Omega-\omega_{+{}})t/2},
    e^{-\mathrm{i}(\Omega+\omega_{+{}})t/2}
  \right\},
\end{equation}
where $\Omega=(\Omega_{-{}}-\Omega_{+{}})/2$, to exclude the
explicit time dependence of the effective Hamiltonian $H'$. Using
the property $\omega_{+{}} + \omega_{-{}} = \Omega_{+{}} +
\Omega_{-{}}$, we arrive to the new Schr\"{o}dinger equation for
the ``wave function" $\Psi$,
\begin{equation}\label{Schr2QFT}
  \mathrm{i}\frac{\mathrm{d}\Psi}{\mathrm{d}t} = H \Psi,
  \quad
  H =
  \mathcal{U}^\dag H' \mathcal{U} -
  \mathrm{i} \mathcal{U}^\dag \dot{\mathcal{U}} =
  \begin{pmatrix}
    (\Omega+\omega_{-{}})/2 & h_{-{}} & 0 & H_{-{}} \\
    h_{-{}} & (\Omega-\omega_{-{}})/2 & H_{+{}} & 0 \\
    0 & H_{+{}} & -(\Omega-\omega_{+{}})/2 & h_{+{}} \\
    H_{-{}} & 0 & h_{+{}} & -(\Omega+\omega_{+{}})/2 \
  \end{pmatrix}.
\end{equation}
Despite initially we used perturbation theory to account for the
influence of the potential $\mathcal{V}$ on the dynamics of the
system~\eqref{DireqmattB}, the contribution of this potential is
taken into account exactly in Eq.~\eqref{Schr2QFT}. It means that
our method allows one to sum up all terms in the perturbation
series.

As we mentioned above, the quantum number $\zeta$ does not
correspond to a definite helicity eigenstate. Thus the initial
condition, which we should add to Eq.~\eqref{Schr2QFT}, has to be
derived from Eqs.~\eqref{inicondamatt} and~\eqref{spinorsmattB}
and also depend on the neutrino oscillations channel. For example,
if we discuss $\nu_\beta^\mathrm{L}\to\nu_\alpha^\mathrm{R}$
neutrino oscillations, the proper initial condition for the ``wave
function" $\Psi(0)=\Psi_0$ is
\begin{equation}\label{inicond2QFT}
  \Psi_0^\mathrm{T}=
  \left(
    -\sin\theta \sqrt{\frac{\mathcal{M}_1+g_1/2}{2 \mathcal{M}_1}},
    -\cos\theta \sqrt{\frac{\mathcal{M}_2+g_2/2}{2 \mathcal{M}_2}},
    \sin\theta \sqrt{\frac{\mathcal{M}_1-g_1/2}{2 \mathcal{M}_1}},
    \cos\theta \sqrt{\frac{\mathcal{M}_2-g_2/2}{2 \mathcal{M}_2}}
  \right).
\end{equation}
Suppose that one has found the solution of the
system~\eqref{Schr2QFT} and~\eqref{inicond2QFT} as
$\Psi^\mathrm{T}(t)=(\psi_1,\psi_2,\psi_3,\psi_4)$. Then the
transition probability for
$\nu_\beta^\mathrm{L}\to\nu_\alpha^\mathrm{R}$ oscillations
channel can be found as
\begin{align}\label{PtrsfoQFT}
  P_{\nu_\beta^\mathrm{L}\to\nu_\alpha^\mathrm{R}}(t)= &
  \frac{1}{2}
  \Bigg\{
    \frac{\mu_1 B \cos\theta}{\sqrt{\mathcal{M}_1}}
    \left[
      \frac{\psi_1(t)}{\sqrt{\mathcal{M}_1+g_1/2}}+
      \frac{\psi_3(t)}{\sqrt{\mathcal{M}_1-g_1/2}}
    \right] 
    \notag
    \\
    & -
    \frac{\mu_2 B \sin\theta}{\sqrt{\mathcal{M}_2}}
    \left[
      \frac{\psi_2(t)}{\sqrt{\mathcal{M}_2+g_2/2}}+
      \frac{\psi_4(t)}{\sqrt{\mathcal{M}_2-g_2/2}}
    \right]
  \Bigg\}^2.
\end{align}
\end{widetext}
To obtain Eq.~\eqref{PtrsfoQFT} for simplicity we use the fact
that initially we have rather broad (in space) wave packet,
corresponding to the initial condition $\nu_\beta(\mathbf{r},0) =
e^{\mathrm{i} k x} \nu_\beta^{(0)\mathrm{L}}$ [see
Eq.~\eqref{inicondnulambdaexp}].

Now we demonstrate the consistency of the results of relativistic
quantum mechanics approach to the description of neutrino spin
flavor oscillations (see Eqs.~\eqref{Schr2QFT}-\eqref{PtrsfoQFT},
which look completely new) with the standard quantum mechanical
method developed in Ref.~\cite{LimMar88}. We remind that the
following effective Hamiltimian:
\begin{equation}\label{effHammatt}
  H_{QM}' =
  \begin{pmatrix}
    \Phi+g_1 & g & -\mu_1 B & -\mu B \\
    g & -\Phi+g_2 & -\mu B & -\mu_2 B \\
    -\mu_1 B & -\mu B & \Phi & 0 \\
    -\mu B & -\mu_2 B & 0 & -\Phi \
  \end{pmatrix},
\end{equation}
was proposed in Ref.~\cite{LimMar88} to describe the evolution of
neutrino mass eigenstates in matter under the influence of an
external magnetic field.

The effective Hamiltonian $H_{QM}'$ acts in the space with the
basis composed of helicity eigenstates of massive neutrinos. As we
mentioned above, the helicity operator
$(\bm{\Sigma}\mathbf{p})/|\mathbf{p}|$ does not commute with the
Hamiltonian $\mathcal{H}_a$ in Eq.~\eqref{DireqmattB}. Therefore
the choice of the helicity eigenstates as the basis functions is
justified only in the relatively weak external magnetic field case
(see the detailed discussion in Ref.~\cite{Dvo10}) or in case of
the small diagonal magnetic moments~\cite{DvoMaa09}. In our
approach we use the basis spinors~\eqref{spinorsmattB} which are
the eigenfunctions of the Hamiltonian $\mathcal{H}_a$ and exactly
take into account matter density and magnetic field strength. Thus
these spinors are more appropriate basis functions for the
description of spin flavor oscillations.

We have found that in frames of the relativistic quantum mechanics
approach the dynamics of the neutrino system can be described by
the Schr\"{o}dinger like equation with the effective
Hamiltonian~\eqref{Schr2QFT}. Let us decompose the energy
levels~\eqref{EnergymattB} supposing that neutrinos are
ultrarelativistic particles,
\begin{equation}\label{EnergymattBUR}
  E_a^{(\zeta)} = k + \frac{g_a}{2} - \zeta \mathcal{M}_a +
  \frac{m_a^2}{2k} + \zeta \frac{m_a^2 g_a^2}{8 k^2
  \mathcal{M}_a} + \dotsb.
\end{equation}
In Eq.~\eqref{EnergymattBUR} we keep the term $\sim m_a^2/k^2$ to
examine the corrections to the conventional quantum mechanical
approach.

Performing the similarity transformation of the effective
Hamiltonian $H$ in Eq.~\eqref{Schr2QFT} and using the orthogonal
matrix $\mathcal{R}$ ($\mathcal{R}^\mathrm{T}\mathcal{R}=I$) of
the following form:
\begin{widetext}
\begin{equation}\label{VQFTtoQM}
  \mathcal{R} = 
  \begin{pmatrix}
    \displaystyle
    -\frac{\sqrt{\mathcal{M}_1+g_1/2}}{\sqrt{2\mathcal{M}_1}} & 0 &
    \displaystyle
    \frac{\mu_1 B}{\sqrt{2\mathcal{M}_1(\mathcal{M}_1+g_1/2)}} & 0 \\
    0 &
    \displaystyle
    -\frac{\sqrt{\mathcal{M}_2+g_2/2}}{\sqrt{2\mathcal{M}_2}} &
    0 &
    \displaystyle
    \frac{\mu_2 B}{\sqrt{2\mathcal{M}_2(\mathcal{M}_2+g_2/2)}} \\
    \displaystyle
    \frac{\sqrt{\mathcal{M}_1-g_1/2}}{\sqrt{2\mathcal{M}_1}} & 0 &
    \displaystyle
    \frac{\mu_1 B}{\sqrt{2\mathcal{M}_1(\mathcal{M}_1-g_1/2)}} & 0 \\
    0 &
    \displaystyle
    \frac{\sqrt{\mathcal{M}_2-g_2/2}}{\sqrt{2\mathcal{M}_2}} &
    0 &
    \displaystyle
    \frac{\mu_2 B}{\sqrt{2\mathcal{M}_2(\mathcal{M}_2-g_2/2)}} \
  \end{pmatrix},
\end{equation}
we can see that the Hamiltonian $H$ transforms to
$\mathcal{R}^\mathrm{T} H \mathcal{R} \approx H_{QM} + \delta H$,
where
\begin{equation}\label{HQMfromQFT}
  H_{QM} = \\
  \begin{pmatrix}
    \Phi+3g_1/4-g_2/4 & g & -\mu_1 B & -\mu B \\
    g & -\Phi+3g_2/4-g_1/4 & -\mu B & -\mu_2 B \\
    -\mu_1 B & -\mu B & \Phi-(g_1+g_2)/4 & 0 \\
    -\mu B & -\mu_2 B & 0 & -\Phi-(g_1+g_2)/4 \
  \end{pmatrix},
\end{equation}
and
\begin{equation}\label{deltaH}
  \delta H =
  \frac{1}{16 k^2}
  \mathrm{diag}
  \left(
    -m_1^2\frac{g_1^3}{\mathcal{M}_1^2},
    -m_2^2\frac{g_2^3}{\mathcal{M}_2^2},
    m_1^2\frac{g_1^3}{\mathcal{M}_1^2},
    m_2^2\frac{g_2^3}{\mathcal{M}_2^2}
  \right),
\end{equation}
\end{widetext}
is the correction to the standard effective Hamiltonian. It should
be noted that the transformation matrix $\mathcal{R}$ in
Eq.~\eqref{VQFTtoQM} depends on the magnetic field strength and
the matter density.

The effective Hamiltonian $H_{QM}$ is equivalent to $H'_{QM}$ in
Eq.~\eqref{effHammatt} since $H_{QM} = H'_{QM} -
\mathrm{tr}(H'_{QM})/4 \cdot I$, where $I$ is the $4 \times 4$
unit matrix. It is known that the unit matrix does not change the
particles dynamics. Thus the relativistic quantum mechanics
approach is equivalent to the standard approach developed in
Ref.~\cite{LimMar88}.

Now let us discuss the correction $\delta H$ [see
Eq.~\eqref{deltaH}] to the quantum mechanical method. This
correction results from the fact that we use the correct energy
levels for a neutrino moving in dense matter and strong magnetic
field. Note that in Eq.~\eqref{deltaH} we keep only the diagonal
corrections $\sim m_a^2/k^2$ to the effective
Hamiltonian~\eqref{HQMfromQFT}. If we slightly change non-diagonal
elements of the Hamiltonian, it will result in the small changes
of the transition probability. However, if we add some small
quantity to diagonal elements, it can produce the resonance
enhancement of neutrino oscillations.

We should remind that the expressions for the basis
spinors~\eqref{spinorsmattB} were obtained in the approximation of
neutrinos with small masses, whereas in Eq.~\eqref{EnergymattBUR}
we expand the energy up to $\sim m_a^2/k^2$ terms. If we take into
account $\sim m_a^2/k^2$ corrections to the basis
spinors~\eqref{spinorsmattB}, we can expect that some non-diagonal
entries in the effective Hamiltonian $H$~\eqref{Schr2QFT} will
also obtain $\sim m_a^2/k^2$ corrections: $h_{\pm{}} \to h_{\pm{}}
+ \delta h_{\pm{}}$ and $H_{\pm{}} \to H_{\pm{}} + \delta
H_{\pm{}}$. However, using the explicit form of the effective
Hamiltonian $H$~\eqref{Schr2QFT} and the matrix
$\mathcal{R}$~\eqref{VQFTtoQM} we get that these additional
contributions are washed out in diagonal entries in
Eq.~\eqref{deltaH}. We analyze the validity of the approximations
made in the derivation of the correction~\eqref{deltaH} in
Appendix~\ref{ANALAPPR}.

\section{Spin flavor oscillations of Dirac neutrinos in the magnetized envelope
of a supernova\label{SPECMATT}}

In this section we study the application of the general formalism
for the description of neutrino spin flavor oscillations,
developed in Sec.~\ref{MATTERB}, to the situation of neutrinos
propagating in the expanding envelope formed after a supernova
explosion~\cite{DvoMaa09}. We find an exact solution of the
Schr\"{o}dinger equation with the effective
Hamiltonian~\eqref{HQMfromQFT} for the background matter profile
present in an expanding envelope and in the supernova magnetic
field. We also analyze the possibility of enhancement of neutrino
oscillations.

To describe the dynamics of neutrino spin flavor oscillations one
has to solve the evolution equation with the
Hamiltonian~\eqref{HQMfromQFT}. This problem, in its turn,
requires to solve a secular equation which is the fourth-order
algebraic equation in order to find the eigenvalues of the
effective Hamiltonian. Although one can express the solution to
such an equation in radicals, its actual form appears to be rather
cumbersome for arbitrary parameters.

If we, however, consider the case of a neutrino propagating in the
electrically neutral isoscalar matter, i.e. $n_e = n_p$ and $n_p =
n_n$, a reasonable solution is possible to find. We will
demonstrate later that it corresponds to a realistic physical
situation. As one can infer from Eq.~\eqref{fdefinition} for the
case of the $\nu_e^\mathrm{L}\to\nu_\mu^\mathrm{R}$ oscillations
channel, in a medium with this property one has the effective
potentials $f_\alpha \equiv f_\mu = V_\mu = -G_\mathrm{F}
n/\sqrt{2}$ and $f_\beta \equiv f_e = V_e = G_\mathrm{F}
n/\sqrt{2}$, where $n\equiv n_e = n_p = n_n$. Using
Eq.~\eqref{flambda} we obtain that $g_1 = -g_2 = g_0$, where $g_0
= -V\cos 2\theta$, $g=V\sin 2\theta$, and $V  =
G_\mathrm{F}n/\sqrt{2}$.

Let us point out that background matter with these properties may
well exist in some astrophysical environments. The matter profile
of presupernovae is poorly known, and a variety of presupernova
models with different profiles exist in the literature (see, e.g.,
Ref.~\cite{WooHegWea02}). Nevertheless, electrically neutral
isoscalar matter may well exist in the inner parts of
presupernovae consisting of elements heavier than hydrogen.
Indeed, for example, the model W02Z in Ref.~\cite{WooHegWea02}
predicts that in a 15$M_{\odot}$ presupernova one has
$Y_e=n_e/(n_p+n_n)=0.5$ in the O+Ne+Mg layer, between the Si+O and
He layers, in the radius range (0.007--0.2)$R_{\odot}$.

We also discuss the model of neutrino magnetic moments in which
the nondiagonal elements of the magnetic moments
matrix~\eqref{magmomme} are much bigger than the diagonal magnetic
moments. Such a magnetic moments matrix was previously discussed
in our works~\cite{DvoMaa07,Dvo08JPG,DvoMaa09} (see also
Sec.~\ref{B}). Note that in case of negligible diagonal magnetic
moments the helicity operator~\eqref{inicondhel} commutes with the
Hamiltonian $\mathcal{H}_a$ in Eq.~\eqref{DireqmattB} and hence
the effective Hamiltonian~\eqref{HQMfromQFT} acts in the helicity
eigenstates basis. In other words the effective
Hamiltonian~\eqref{effHammatt} proposed in the standard quantum
mechanical approach~\cite{LimMar88} is justified in our case.

For neutrinos having such magnetic moments and propagating in
isoscalar matter the effective Hamiltonian~\eqref{HQMfromQFT} is
replaced by
\begin{equation}\label{QMH4Dir}
  H_{QM} \to
  \begin{pmatrix}
    \Phi+g_0 & g & 0 & -\mu B \\
    g & -(\Phi+g_0) & -\mu B & 0 \\
    0 & -\mu B & \Phi & 0 \\
    -\mu B & 0 & 0 & -\Phi \
  \end{pmatrix}.
\end{equation}
We now look for the stationary solutions of the Schr\"odinger
equation with this Hamiltonian. After a straightforward
calculation one finds
\begin{align}\label{WFDir}
  \Psi(t)= &
  \sum_{\zeta = \pm 1}
  \Big[
    \left(
      U_\zeta \otimes U_\zeta^\dag
    \right)\exp{(-\mathrm{i}\mathcal{E}_\zeta t)}
    \notag
    \\
    & +
    \left(
      V_\zeta \otimes V_\zeta^\dag
    \right)\exp{(\mathrm{i}\mathcal{E}_\zeta t)}
  \Big]\Psi_0,
\end{align}
where we have denoted
\begin{align}\label{EnergyQMDir}
  \mathcal{E}_{\pm{}}= & \frac{1}{2}
  \sqrt{2V^2+4(\mu B)^2+4\Phi^2-4\Phi V\cos 2\theta \pm 2VR},
  \notag
  \\
  R = & \sqrt{(V-2\Phi\cos 2\theta)^2+4(\mu B)^2}.
\end{align}
The vectors $U_{\pm{}}$ and $V_{\pm{}}$ are the eigenvectors
corresponding to the energy eigenvalues $\mathcal{E}_{\pm{}}$ and
$-\mathcal{E}_{\pm{}}$, respectively. They are given by ($\zeta
=\pm{}$)
\begin{align}\label{UVQMspinorsDir}
  U_\zeta = &
  \frac{1}{N_\zeta}
  \begin{pmatrix}
    Z_\zeta \\
    \sin 2\theta (\mathcal{E}_\zeta-\Phi) \\
    -\mu B \sin 2\theta \\
    -\mu B Z_\zeta/(\mathcal{E}_\zeta + \Phi) \
  \end{pmatrix},
  \notag
  \\
  V_\zeta = &
  \frac{1}{N_\zeta}
  \begin{pmatrix}
    -\sin 2\theta (\mathcal{E}_\zeta-\Phi) \\
    Z_\zeta \\
    \mu B Z_\zeta/(\mathcal{E}_\zeta + \Phi) \\
    -\mu B \sin 2\theta \
  \end{pmatrix},
\end{align}
where
\begin{align}\label{ZNDir}
  Z_\zeta = & \frac{V + \zeta R}{2}-\mathcal{E}_\zeta \cos 2\theta,
  \notag
  \\
  N_\zeta^2 = & Z_\zeta^2
  \left[
    1+\frac{(\mu B)^2}{(\mathcal{E}_\zeta+\Phi)^2}
  \right]
  \notag
  \\
  & +
  \sin^2(2\theta)
  \left[
    (\mu B)^2+(\mathcal{E}_\zeta-\Phi)^2
  \right].
\end{align}
It should be noted that Eq.~\eqref{WFDir} is a general solution of
the evolution equation with the effective
Hamiltonian~\eqref{QMH4Dir} satisfying the initial condition
$\Psi(0)=\Psi_0$.

Note that we received the solution~\eqref{WFDir}-\eqref{ZNDir}
under some assumptions on the external fields such as isoscalar
matter with constant density and constant magnetic field. In
Sec.~\ref{MATTERB} we showed that our method is equivalent to the
quantum mechanical description of neutrino
oscillations~\cite{LimMar88} which can be used for a more general
case of coordinate dependent external fields. Nevertheless the
assumption of constant matter density and magnetic field is quite
realistic for certain astrophysical environments like a shock wave
propagating inside an expanding envelope after a supernova
explosion.

Consistently with Eqs.~\eqref{inicondnulambda}-\eqref{Ulambdaa},
and~\eqref{inicondnulambdaexp} with an initial spinor
$\nu_\beta^{(0)}$ corresponding to a left polarized
neutrino~\eqref{inicondhel}, we take the initial wave function
$\Psi(0)\equiv\Psi_0$ in Eq.~\eqref{WFDir} as
$\Psi_0^\mathrm{T}=(\psi_1^\mathrm{L}, \psi_2^\mathrm{L},
\psi_1^\mathrm{R},
\psi_2^\mathrm{R})=(\sin\theta,\cos\theta,0,0)$. Using
Eqs.~\eqref{WFDir}-\eqref{ZNDir} one finds the components of the
quantum mechanical wave function corresponding to the right
polarized neutrinos to be of the form
\begin{align}\label{psi1Rpsi2RDir}
  \psi_1^\mathrm{R}(t) = &
  \frac{\mu B}{N_{+{}}^2}
  \bigg\{
    \cos\theta
    \bigg[
      e^{\mathrm{i}\mathcal{E}_{+{}}t}
      \frac{Z_{+{}}^2}{\mathcal{E}_{+{}}+\Phi}
      \notag
      \\
      & -
      \sin^2(2\theta)(\mathcal{E}_{+{}}-\Phi)e^{-\mathrm{i}\mathcal{E}_{+{}}t}
    \bigg]
    \notag
    \\
    &
    -
    \sin \theta \sin 2\theta Z_{+{}}
    \bigg[
      e^{-\mathrm{i}\mathcal{E}_{+{}}t}
      \notag
      \\
      & +
      e^{\mathrm{i}\mathcal{E}_{+{}}t}
      \frac{\mathcal{E}_{+{}}-\Phi}{\mathcal{E}_{+{}}+\Phi}
    \bigg]
  \bigg\} +
  \{+{}\to{}-{}\},
  \notag
  \displaybreak[2]
  \\
  \psi_2^\mathrm{R}(t) = &
  \frac{\mu B}{N_{+{}}^2}
  \bigg\{
    \sin\theta
    \bigg[
      \sin^2(2\theta)(\mathcal{E}_{+{}}-\Phi)e^{\mathrm{i}\mathcal{E}_{+{}}t}
      \notag
      \\
      & -
      e^{-\mathrm{i}\mathcal{E}_{+{}}t}
      \frac{Z_{+{}}^2}{\mathcal{E}_{+{}}+\Phi}
    \bigg]
    \notag
    \displaybreak[2]
    \\
    &
    -
    \cos \theta \sin 2\theta Z_{+{}}
    \bigg[
      e^{\mathrm{i}\mathcal{E}_{+{}}t}
      \notag
      \\
      & +
      e^{-\mathrm{i}\mathcal{E}_{+{}}t}
      \frac{\mathcal{E}_{+{}}-\Phi}{\mathcal{E}_{+{}}+\Phi}
    \bigg]
  \bigg\} +
  \{+{}\to{}-{}\},
\end{align}
where the $\{+{}\to{}-{}\}$ stand for the terms similar to the
terms preceding each of them but with all quantities with a
subscript $+{}$ replaced with corresponding quantities with a
subscript $-{}$. The wave function of the right-handed neutrino of
the flavor ``$\alpha$", $\nu_\alpha^\mathrm{R}$, can be written
with help of Eqs.~\eqref{inicondnulambda}-\eqref{Ulambdaa},
and~\eqref{psi1Rpsi2RDir} as $\nu_\alpha^\mathrm{R}(t) =
\cos\theta\psi_1^\mathrm{R}(t) - \sin\theta\psi_2^\mathrm{R}(t)$.

The probability for the transition $\nu_\beta^\mathrm{L} \to
\nu_\alpha^\mathrm{R}$ is obtained as the square of the quantum
mechanical wave function $\nu_\alpha^\mathrm{R}$. One obtains
\begin{align}\label{PrtDir}
  P_{\nu_\beta^\mathrm{L} \to \nu_\alpha^\mathrm{R}}(t) = &
  \left|
    \nu_\alpha^\mathrm{R}
  \right|^2
  \notag
  \\
  & =
  [C_{+{}}\cos(\mathcal{E}_{+{}}t)+C_{-{}}\cos(\mathcal{E}_{-{}}t)]^2
  \notag
  \\
  & +
  [S_{+{}}\sin(\mathcal{E}_{+{}}t)+S_{-{}}\sin(\mathcal{E}_{-{}}t)]^2,
\end{align}
where ($\zeta =\pm$)
\begin{align}\label{CpmSpmDir}
  C_\zeta = & \frac{\mu B}{N_\zeta^2}
  \left\{
    \frac{Z_\zeta^2}{\mathcal{E}_\zeta+\Phi}-
    \sin^2(2\theta)(\mathcal{E}_\zeta-\Phi)
  \right\},
  \notag
  \\
  S_\zeta = & \frac{\mu B}{N_\zeta^2}
  \bigg\{
    \sin^2(2\theta)\frac{2\Phi Z_\zeta}{\mathcal{E}_\zeta+\Phi}
    \notag
    \\
    & +
    \cos 2\theta
    \left[
      \frac{Z_\zeta^2}{\mathcal{E}_\zeta+\Phi}+
      \sin^2(2\theta)(\mathcal{E}_\zeta-\Phi)
    \right]
  \bigg\}.
\end{align}
As a consistency check, one easily finds from
Eq.~\eqref{CpmSpmDir} that $C_{+{}}+C_{-{}}=0$ as required for
assuring $P(0)=0$.

In the following we will limit our considerations to the case
$\mathcal{E}_{+{}} \approx \mathcal{E}_{-{}}$, corresponding to
the situations where the  effect of the interactions of neutrinos
with matter ($V$) is small compared with that of the magnetic
interactions ($\mu B$) or the vacuum contribution ($\Phi$) or both
[see Eq.~\eqref{EnergyQMDir}]. Note that in this case one can
analyze the exact oscillation probability~\eqref{PrtDir}
analytically, which would be practically impossible in more
general situations.

In the case  $\mathcal{E}_{+{}} \approx \mathcal{E}_{-{}}$, one
can present the transition probability in Eq.~\eqref{PrtDir} in
the following form:
\begin{equation}\label{PrtEnvDir}
  P(t) = P_0(t) + P_c(t) \cos(2 \Omega t) + P_s(t) \sin(2 \Omega t),
\end{equation}
where
\begin{align}\label{P0PcPsDir}
  P_0(t) = & \frac{1}{2}
  \big[
    S_{+{}}^2+S_{-{}}^2+2 S_{+{}} S_{-{}} \cos(2 \delta \Omega t)
    \notag
    \\
    & -
    4 C_{+{}} C_{-{}} \sin^2(\delta \Omega t)
  \big],
  \notag
  \\
  P_c(t) = & -\frac{1}{2}
  \big[
    (S_{+{}}^2+S_{-{}}^2) \cos(2 \delta \Omega t) + 2 S_{+{}} S_{-{}}
    \notag
    \\
    & -
    4 C_{+{}} C_{-{}} \sin^2(\delta \Omega t)
  \big],
  \notag
  \\
  P_s(t) = & \frac{1}{2}
  \left(
    S_{+{}}^2-S_{-{}}^2
  \right)
  \sin(2 \delta \Omega t),
\end{align}
and
\begin{equation}\label{FrDir}
  \Omega = \frac{\mathcal{E}_{+{}}+\mathcal{E}_{-{}}}{2},
  \quad
  \delta \Omega = \frac{\mathcal{E}_{+{}}-\mathcal{E}_{-{}}}{2}.
\end{equation}
As one can infer from these expressions, the transition
probability $P(t)$ is a rapidly oscillating function,  with the
frequency $\Omega$, enveloped from up and down by the slowly
varying functions $P_{u,d} = P_0 \pm \sqrt{P_c^2+P_s^2}$,
respectively.

\begin{figure*}
  \centering
  \includegraphics[scale=.93]{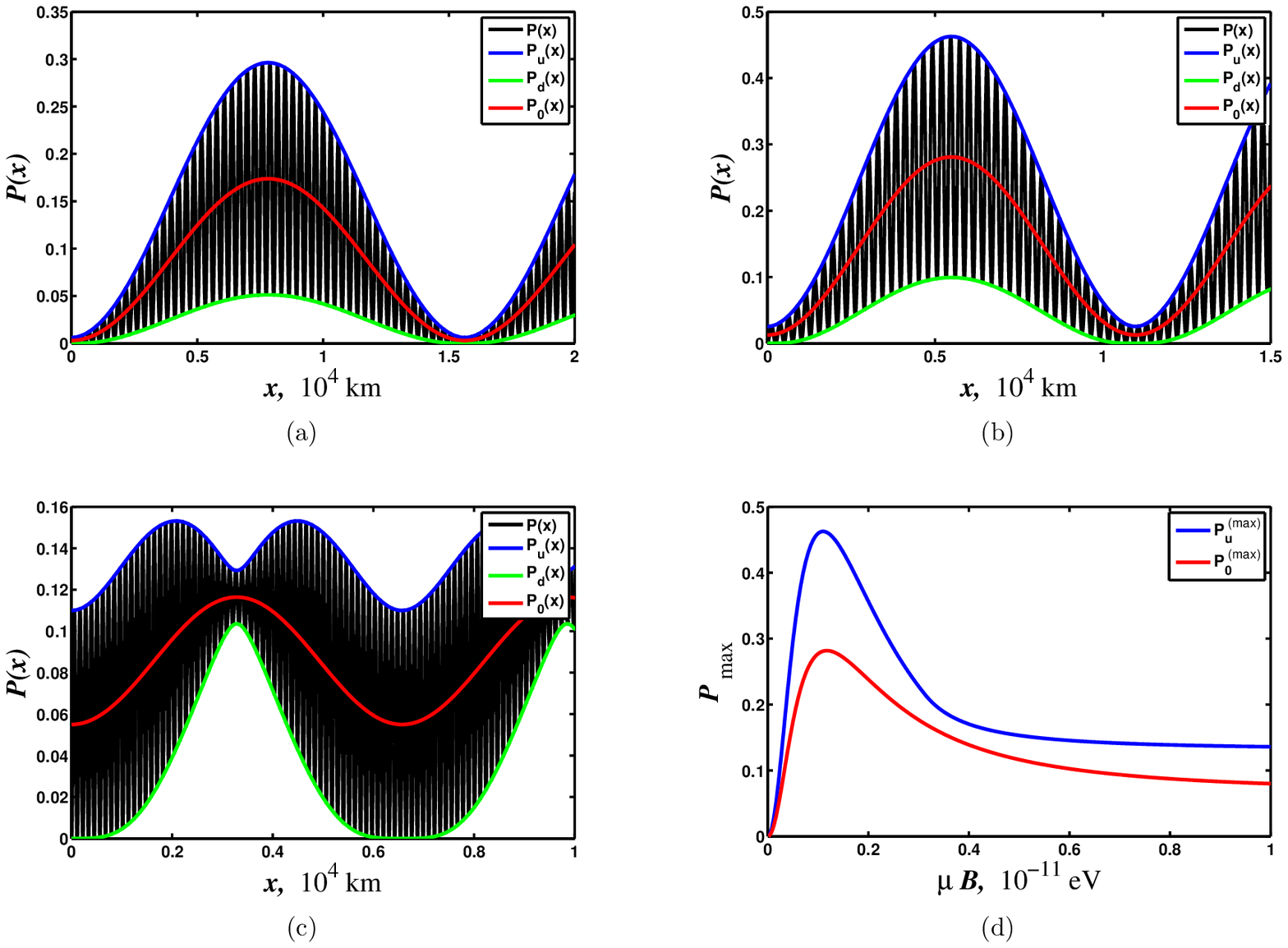}
  \caption{\label{Dirnulowdens} (color online)
  (a)-(c) The transition probability versus the distance passed by a neutrino
  beam in matter with the density $\rho = 10\thinspace\text{g/cc}$;
  (a) $\mu B = 5 \times 10^{-13}\thinspace\text{eV}$,
  (b) $\mu B = 1.1 \times 10^{-12}\thinspace\text{eV}$,
  (c) $\mu B = 5 \times 10^{-12}\thinspace\text{eV}$.
  We take that $E_\nu = 10\thinspace\text{MeV}$,
  $\delta m^2 = 8 \times 10^{-5}\thinspace\text{eV}^2$ and
  $\theta = 0.6$, which is quite close to the solar neutrinos'
  oscillations parameters.
  The black line is the function $P(x)$,
  the blue and green lines are the envelope functions $P_{u,d}(x)$, and
  the red line is the averaged transition probability $P_0(x)$.
  (d) The dependence of the maximal values of the functions $P(x)$ and
  $P_0(x)$, blue and red lines, respectively, on the magnetic energy $\mu B$
  for the given density.
  This figure is taken from Ref.~\cite{DvoMaa09}.}
\end{figure*}

\begin{figure*}
  \centering
  \includegraphics[scale=.93]{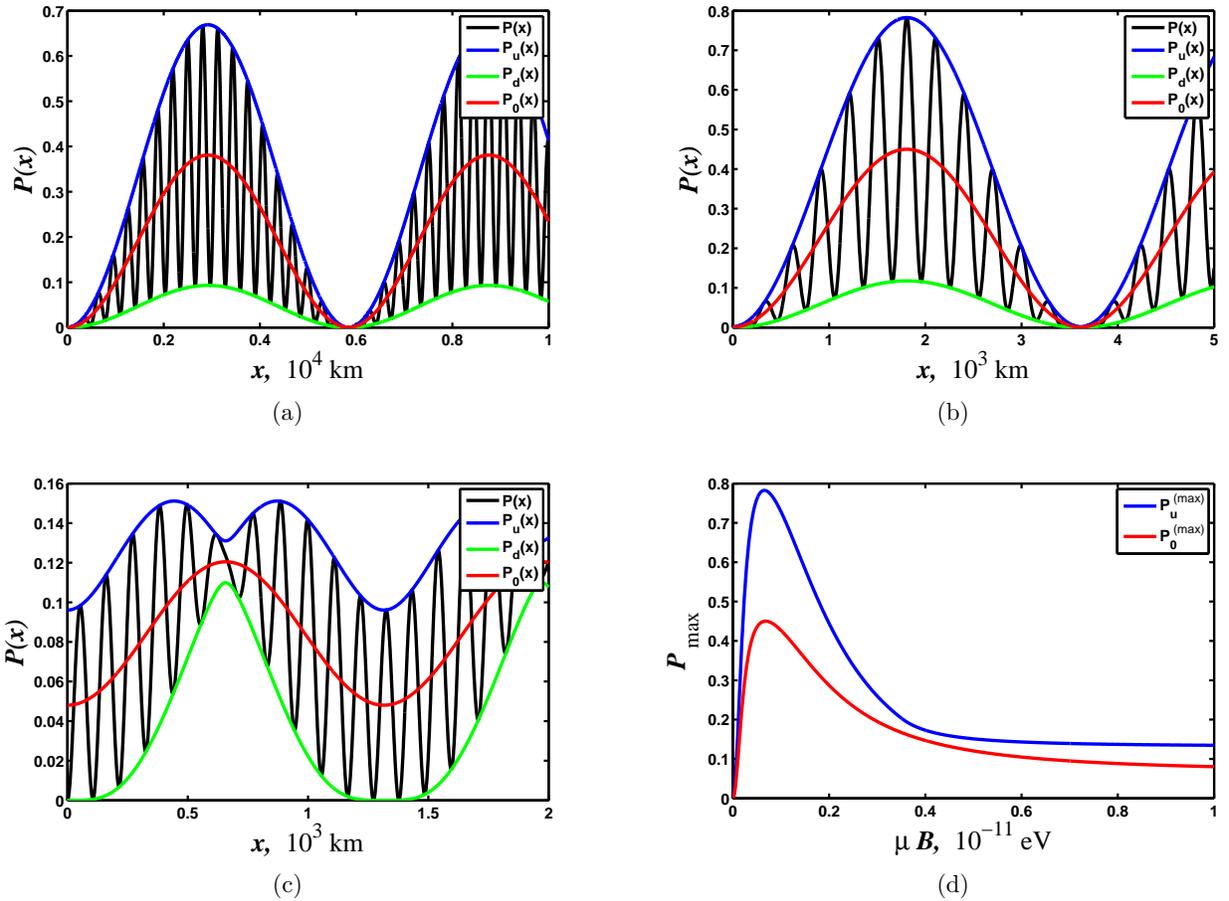}
  \caption{\label{Dirnumeddens} (color online)
  The same as in Fig.~\ref{Dirnulowdens} for the density
  $\rho = 50\thinspace\text{g/cc}$;
  (a) $\mu B = 3.5 \times 10^{-13}\thinspace\text{eV}$,
  (b) $\mu B = 6.6 \times 10^{-13}\thinspace\text{eV}$,
  (c) $\mu B = 5 \times 10^{-12}\thinspace\text{eV}$.
  After Ref.~\cite{DvoMaa09}.}
\end{figure*}

\begin{figure*}
  \centering
  \includegraphics[scale=.93]{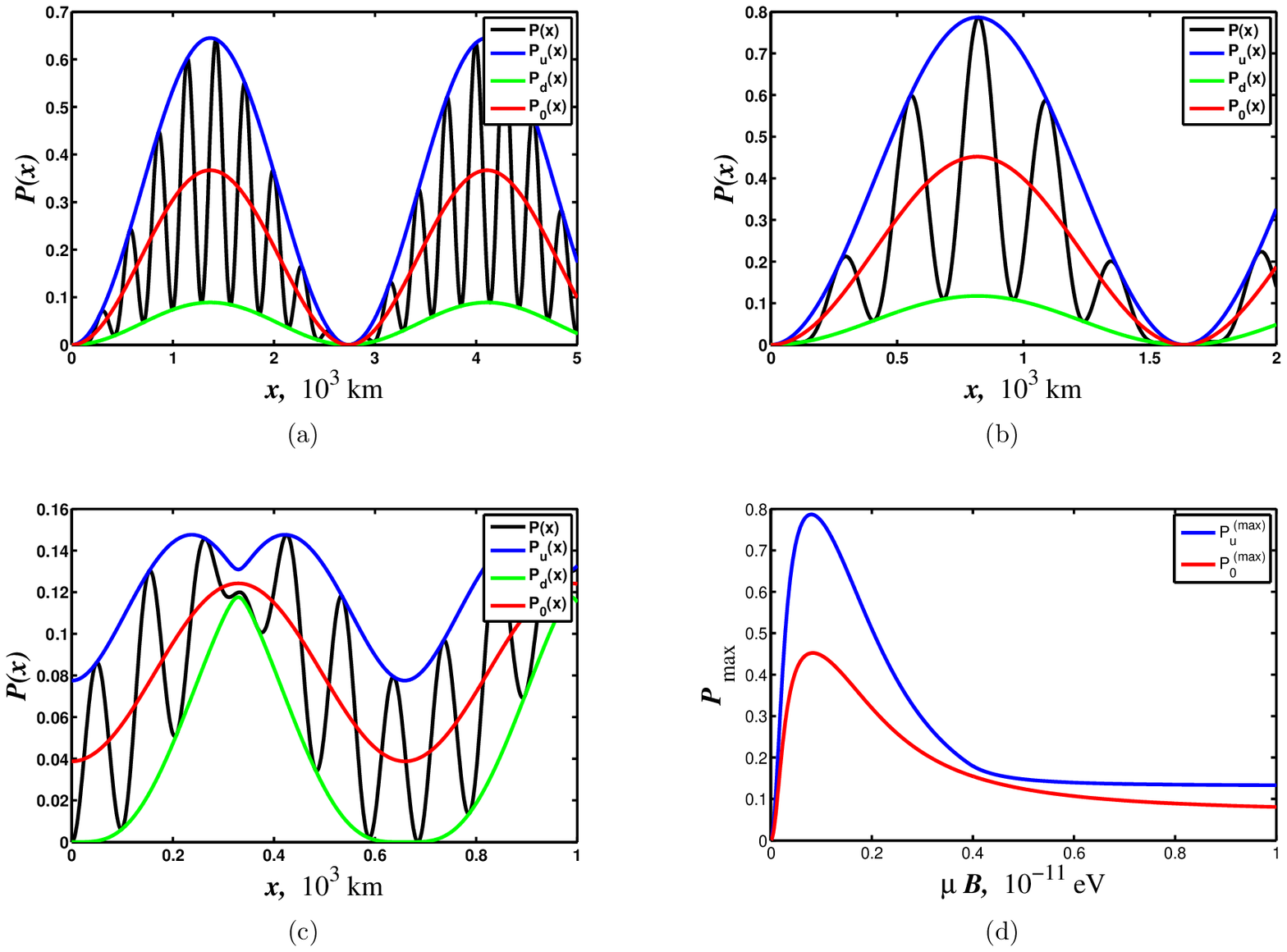}
  \caption{\label{Dirnuhighdens} (color online)
  The same as in Fig.~\ref{Dirnulowdens} for the density
  $\rho = 100\thinspace\text{g/cc}$;
  (a) $\mu B = 4 \times 10^{-13}\thinspace\text{eV}$,
  (b) $\mu B = 8 \times 10^{-13}\thinspace\text{eV}$,
  (c) $\mu B = 5 \times 10^{-12}\thinspace\text{eV}$.
  After Ref.~\cite{DvoMaa09}.}
\end{figure*}

The behavior of the transition probability for various matter
densities $\rho$ and the values of $\mu B$ and for a fixed
neutrino energy of $E=10\thinspace\text{MeV}$ and squared mass
difference of $\delta m^2 = 8 \times 10^{-5}\thinspace\text{eV}^2$
is illustrated in Figs.~\ref{Dirnulowdens}-\ref{Dirnuhighdens}.

As these plots show, at low matter densities the envelope
functions give, at each propagation distance, the range of the
possible values of the oscillation probability. At greater matter
densities, where the probability oscillates less intensively, the
envelope functions are not that useful in analyzing the physical
situation.

One can find the maximum value of the upper envelope function,
which is also the upper bound for the transition probability,
given as
\begin{widetext}
\begin{equation}\label{PumaxDir}
  P_u^\mathrm{(max)} =
  \begin{cases}
    (S_{+{}}-S_{-{}})^2, & \text{if $B < B'$}, \\
    \displaystyle
    \frac{C_{+{}}C_{-{}}(S_{+{}}^2-S_{-{}}^2)^2}
    {C_{+{}}C_{-{}}(S_{+{}}^2+S_{-{}}^2) +
    (C_{+{}}C_{-{}})^2+
    (S_{+{}}S_{-{}})^2}, &
    \text{if $B > B'$},
  \end{cases}
\end{equation}
\end{widetext}
where the value $B'$ is the solution of the transcendent algebraic
equation, $C_{+{}}C_{-{}}=S_{+{}}S_{-{}}$. The corresponding
maximum values of the averaged transition probability $P_0(x)$ are
given by
\begin{equation}\label{P0maxDir}
  P_0^\mathrm{(max)} =
  \frac{1}{2}[(S_{+{}}S_{-{}})^2-4C_{+{}}C_{-{}}],
\end{equation}
for arbitrary values of $B$. The values of these maxima depend  on
the size of the quantity $\mu B$. These dependencies are plotted
in Figs.~\ref{Dirnulowdens}(d)-\ref{Dirnuhighdens}(d). In the case
of rapid oscillations the physically relevant quantities, rather
than the maxima, are the averaged values of the transition
probability, which are also plotted in these figures.

As Figs.~\ref{Dirnulowdens}(d)-\ref{Dirnuhighdens}(d) show, the
interplay of the matter effect and the magnetic interaction can
lead, for a given magnetic moment $\mu$, to an enhanced spin
flavor transition if the magnetic field $B$ has a suitable
strength relative to the density of matter $\rho$. In our
numerical examples this occurs at $\mu B_\mathrm{max} = 1.1 \times
10^{-12}\thinspace\text{eV}$ for $\rho = 10\thinspace\text{g/cc}$,
at $\mu B_\mathrm{max} = 6.6 \times 10^{-13}\thinspace\text{eV}$
for $\rho = 50\thinspace\text{g/cc}$, and at $\mu B_\mathrm{max} =
8 \times 10^{-13}\thinspace\text{eV}$ for $\rho =
100\thinspace\text{g/cc}$. For these values of $\mu B$ both the
maxima and the average of the transition probability become
considerably larger than for any other values of $\mu B$.
Figs.~\ref{Dirnulowdens}(b)-\ref{Dirnuhighdens}(b) correspond to
the situation of maximal enhancement, whereas
Figs.~\ref{Dirnulowdens}(a)-\ref{Dirnuhighdens}(a) and
Figs.~\ref{Dirnulowdens}(c)-\ref{Dirnuhighdens}(c) illustrate the
situation above and below the optimal strength $B_\mathrm{max}$ of
the magnetic field.

It is noteworthy that the enhanced transition probability is
achieved towards the lower end of the  $\mu B$ region where
substantial transitions all occur, that is, at relatively moderate
magnetic fields. At larger values of $\mu B$ the maximum of the
transition probability approaches towards $\cos^2(2\theta)$.
Indeed, if $\mu B \gg \max(\Phi,V)$, the transition probability
can be written in the form $P(t)=\cos^2(2\theta)\sin^2(\mu B t)$
[see Eq.~\eqref{PtrbptB}]. It was found in Ref.~\cite{LikStu95}
that neutrino spin flavor oscillations can be enhanced in a very
strong magnetic field, with the transition probability being
practically equal to unity. This phenomenon can be realized only
for Dirac neutrinos with small nondiagonal magnetic moments and
small mixing angle. As we can see from
Figs.~\ref{Dirnulowdens}(d)-\ref{Dirnuhighdens}(d) the situation
is completely different for big nondiagonal magnetic moments.

One should notice that for long propagation distances consisting
of several oscillation periods of the envelope functions, the
enhancement effect would diminish considerably because of
averaging. In the numerical examples presented in
Figs.~\ref{Dirnulowdens}-\ref{Dirnuhighdens} the period of the
envelope function is of the order of
$10^3-10^4\thinspace\text{km}$, which is a typical size of a shock
wave with the matter densities we have used in the plots (see,
e.g., Ref.~\cite{Kaw06}). Thus the enhanced spin flavor transition
could take place when neutrinos traverse a shock wave.

Let us recall that the above analysis was made by assuming
neutrinos to be Dirac particles. We will see below (see
Sec.~\ref{MAJSPECMATT}) that the corresponding results are quite
different in the case of Majorana neutrinos.

\section{Spin flavor oscillations between electron and
sterile astrophysical neutrinos\label{STERILE}}

In this section we continue with the studies of oscillations of
supernova neutrinos on the basis of the effective Hamiltonian
derived in Sec.~\ref{MATTERB}. In particular, we discuss the
possibility of spin flavor oscillations between right polarized
electron neutrino and additional sterile neutrino in an expanding
envelope formed after a supernova explosion under the influence of
a strong magnetic field. It is shown that the resonance
enhancement of neutrino oscillations is possible if we take into
account the correction to the effective
Hamiltonian~\eqref{deltaH}.

As we mentioned in Sec.~\ref{SPECMATT}, in general case the
solution of the Schr\"{o}dinger equation based on the effective
Hamiltonians~\eqref{HQMfromQFT} and~\eqref{deltaH} is quite
cumbersome and should be analyzed numerically. To study spin
flavor oscillations analytically we consider, just for simplicity,
the situation when the vacuum mixing angle between different
neutrino eigenstates is small, $\theta \ll 1$. If $\theta \ll 1$,
the matter interaction term and the the magnetic moments coincide
in both flavor and mass eigenstates bases.

A resonance of neutrino oscillations can appear if the difference
between two diagonal elements in the effective Hamiltonian is
small~\cite{AkhLanSci97}. In table~\ref{oscchannels}, where we
take into account the contributions of both
Eqs.~\eqref{HQMfromQFT} and~\eqref{deltaH},
\begin{table*}
  \begin{center}
    \caption{\label{oscchannels}
    The resonance conditions for the
    $\nu_\alpha^\mathrm{L,R} \leftrightarrow \nu_\beta^\mathrm{L,R}$
    oscillations channels
    which take into account the Hamiltonians~\eqref{HQMfromQFT} and~\eqref{deltaH}.}
    \begin{ruledtabular}
      \begin{tabular}{l l l l}
        No. & Oscillations channel & Resonance condition \\
        \hline
        1 & $\nu_\beta^\mathrm{L} \leftrightarrow \nu_\alpha^\mathrm{R}$ &
        $\displaystyle \Phi =
        \frac{f_\beta}{2} - \frac{1}{32 k^2}
        \left(
          \frac{m_1^2 f_\alpha^3}{\mathcal{M}_1^2}+\frac{m_2^2 f_\beta^3}{\mathcal{M}_2^2}
        \right)$ \\
        2 & $\nu_\beta^\mathrm{R} \leftrightarrow \nu_\alpha^\mathrm{L}$ &
        $\displaystyle \Phi =
        -\frac{f_\alpha}{2} + \frac{1}{32 k^2}
        \left(
          \frac{m_1^2 f_\alpha^3}{\mathcal{M}_1^2}+\frac{m_2^2 f_\beta^3}{\mathcal{M}_2^2}
        \right)$ \\
        3 & $\nu_\beta^\mathrm{L} \leftrightarrow \nu_\alpha^\mathrm{L}$ &
        $\displaystyle \Phi =
        \frac{f_\beta-f_\alpha}{2} + \frac{1}{32 k^2}
        \left(
          \frac{m_1^2 f_\alpha^3}{\mathcal{M}_1^2}-\frac{m_2^2 f_\beta^3}{\mathcal{M}_2^2}
        \right)$ \\
        4 & $\nu_\beta^\mathrm{R} \leftrightarrow \nu_\alpha^\mathrm{R}$ &
        $\displaystyle \Phi =
        -\frac{1}{32 k^2}
        \left(
          \frac{m_1^2 f_\alpha^3}{\mathcal{M}_1^2}-\frac{m_2^2 f_\beta^3}{\mathcal{M}_2^2}
        \right)$ \\
      \end{tabular}
    \end{ruledtabular}
  \end{center}
\end{table*}
we list the resonance conditions for various oscillations
channels. It can be seen from this table that the
corrections~\eqref{deltaH} to the effective Hamiltonian become
important when we discuss $\nu_{e,\mu,\tau}^\mathrm{R}
\leftrightarrow \nu_s^\mathrm{L,R}$ oscillations channel, where
$\nu_s$ is the additional sterile neutrino.

As an example, we can study the case of $\nu_e^\mathrm{R}
\leftrightarrow \nu_s^\mathrm{L}$ oscillations. Putting $\alpha
\equiv s$ and $\beta \equiv e$ in table~\ref{oscchannels} and
using Eq.~\eqref{fdefinition}, we obtain that $f_s = 0$, $f_e =
\sqrt{2} G_\mathrm{F}(n_e-n_n/2)$ and $\mathcal{M}_2 \equiv
\mathcal{M}_e = \sqrt{(\mu_{\nu_e} B)^2 + f_e^2}$.
From table~\ref{oscchannels} we obtain the resonance condition for
this oscillations channel as,
\begin{equation}\label{rescond1}
  \delta m^2 = \frac{1}{4\sqrt{2}E_\nu} G_\mathrm{F}
  \left(
    n_e-\frac{n_n}{2}
  \right)
  m_{\nu_e}^2,
\end{equation}
where we discuss the case of small diagonal magnetic moment of an
electron neutrino: $\mu_{\nu_e} B \ll f_e$ and $E_\nu = k$ is the
neutrino energy. In Appendix~\ref{ANALAPPR} we discuss other small
factors which can also contribute to the resonance
condition~\eqref{rescond1}.

As an example we consider $\nu_e^\mathrm{R} \leftrightarrow
\nu_s^\mathrm{L}$ oscillations in an expanding envelope of a
supernova explosion. It is known (see, e.g.,
Refs.~\cite{Not88,BarMoh88} for the most detailed analysis) that
right polarized electron neutrinos can be created during the
explosion of a core-collapse supernova. Indeed, if a neutrino is a
Dirac particle, the spin flip can occur in the following reaction:
$\nu_\mathrm{L}+(e^{-{}},p,N) \to \nu_\mathrm{R}+(e^{-{}},p,N)$,
with electrons $e^{-{}}$, protons $p$, and nuclei $N$ in the dense
matter of a forming neutron star. This spin flip is due to the
interaction of the neutrino diagonal magnetic moment with charged
particles. Hence left polarized neutrinos are converted into right
polarized ones with energy in the range
$(100-200)\thinspace\text{MeV}$~\cite{AyaDOlTor00}.

Besides the fact that the creation of right polarized neutrinos
can provide additional supernova cooling~\cite{KuzMikOkr09}, these
particles can be observed in a terrestrial detector. When right
polarized neutrinos propagate from the supernova explosion site
towards the Earth, interacting with the galactic magnetic field,
their helicity can change back and they become left polarized
neutrinos. Although the flux of this kind of neutrinos is smaller
than that of generic left polarized
neutrinos~\cite{BarMoh88,LycBli10}, potentially we can detect
these particles. The analysis of the neutrino spin precession in
the galactic magnetic field is given in Ref.~\cite{LycBli10}.

Suppose that the flux of right polarized electron neutrinos is
crossing an expanding envelope of a supernova. A shock wave can be
formed in the envelope~\cite{Tom05}. Approximately
$1\thinspace\text{s}$ after the core collapse, the matter density
in the shock wave region $L \sim 10^8\thinspace\text{cm}$ can be
up to $10^6\thinspace\text{g/cm}^3$. We can also suppose that the
matter density is approximately constant inside the shock wave
(see also Sec.~\ref{SPECMATT}).

For the electroneutral matter Eq.~\eqref{rescond1} reads
\begin{align}\label{rescond2}
  \delta m^2 \approx &
  5.0 \times 10^{-17}\thinspace\text{eV}^2
  \times
  (3Y_e-1)
  \notag
  \\
  & \times
  \left(
    \frac{\rho}{10^6\thinspace\text{g/cm}^3}
  \right)
  \left(
    \frac{E_\nu}{100\thinspace\text{MeV}}
  \right)^{-1}
  \left(
    \frac{m_{\nu_e}}{1\thinspace\text{eV}}
  \right)^2,
\end{align}
where $Y_e=n_e/(n_e+n_n)$ is the electrons fraction. From
Eq.~\eqref{rescond2} we can see that for the matter with $Y_e >
1/3$ and $\rho \sim 10^6\thinspace\text{g/cm}^3$ and an electron
neutrino with $E_\nu \sim 100\thinspace\text{MeV}$ and $m_{\nu_e}
\sim  1\thinspace\text{eV}$ the mass squared difference should be
$\delta m^2 \sim 10^{-17}\thinspace\text{eV}^2$.

Note that the possibility of existence of sterile neutrinos
closely degenerate in mass with active neutrinos was recently
discussed. In Ref.~\cite{Ker03} it was examined how the fluxes of
supernova neutrinos can be altered is presence of the almost
degenerate neutrinos. The effect of spin flavor oscillations
between active and sterile neutrinos with small $\delta m^2$ on
the solar neutrino fluxes was discussed in Ref.~\cite{snusolar}.
The implications of the CP phases of the mixing between active and
sterile neutrinos in the scenario with small $\delta m^2$ to the
effective mass of an electron neutrinos were considered in
Ref.~\cite{MaaRii10}. In Ref.~\cite{Esm10} it was examined the
possibility to experimentally confirm, e.g., in the IceCube
detector, the existence of almost degenerate in masses sterile
neutrinos emitted with high energies from extragalactic sources.
The range of $\delta m^2$ studied in Ref.~\cite{Ker03} is
$10^{-16}\thinspace\text{eV}^2 < \delta m^2 <
10^{-12}\thinspace\text{eV}^2$ which is quite close to the
estimates obtained from Eq.~\eqref{rescond2}. In Ref.~\cite{Esm10}
the sterile neutrinos with even smaller $\delta m^2$ were studied:
$10^{-19}\thinspace\text{eV}^2 < \delta m^2 <
10^{-12}\thinspace\text{eV}^2$.

Besides the fulfillment of the resonance
condition~\eqref{rescond1}, to have the significant
$\nu_e^\mathrm{R} \leftrightarrow \nu_s^\mathrm{L}$ transitions
rate the strength of the magnetic field and distance traveled by
the neutrinos beam, which we take to be equal to the shock wave
size $L$, should satisfy the condition $\mu B L \approx \pi/2$. We
can rewrite this condition as,
\begin{equation}\label{rescond3}
  B \approx 5.3 \times 10^7\thinspace\text{G}
  \left(
    \frac{\mu}{10^{-12}\thinspace\mu_\mathrm{B}}
  \right)^{-1}
  \left(
    \frac{L}{10^{3}\thinspace\text{km}}
  \right)^{-1}.
\end{equation}
%
For the transition magnetic moment $\mu = 3 \times
10^{-12}\thinspace\mu_\mathrm{B}$~\cite{Raf90} and $L \sim
10^{8}\thinspace\text{cm}$ (see above), we get that $B \sim
10^7\thinspace\text{G}$. Supposing that the magnetic field of a
neutron star depends on the distance as $B(r) = B_0 (R/r)^3$,
where $R = 10\thinspace\text{km}$ is the typical protoneutron star
radius and $B_0 = 10^{13}\thinspace\text{G}$ is the magnetic field
on the surface of a protoneutron star, we get that at $r =
10^8\thinspace\text{cm}$ the magnetic field reaches
$10^7\thinspace\text{G}$, which is consistent with the estimates
of Eq.~\eqref{rescond3}. Note that magnetic fields in a supernova
explosion can be even higher than $10^{13}\thinspace\text{G}$ and
can reach the values of $\sim
10^{16}\thinspace\text{G}$~\cite{Aki03}.

\section{Majorana neutrinos in vacuum\label{MAJVACUUM}}

In this section we apply the formalism, developed in
Secs.~\ref{VACUUM}-\ref{MATTERB} for the description of the Dirac
neutrinos evolution in various external fields, for the studies of
oscillations of Majorana neutrinos. Here we study the evolution of
mixed Majorana neutrinos in vacuum. We solve the initial condition
problem for the two component neutrino wave functions and find the
transition probability. However, besides flavor oscillations, for
Majorana particles one can consider transitions between neutrinos
and antineutrinos. We study this process also on the basis of
relativistic quantum mechanics.

In general case the system of flavor neutrinos can be described
with help of the appropriate number of left and right handed
spinors~\cite{Kob80}. We can however suggest that only left handed
chirality components $\nu_\lambda^\mathrm{L} =
(1/2)(1-\gamma^5)\nu_\lambda$ are present in our system. It is the
case, for example, for neutrinos in frames of the standard model
where the right-handed neutrinos are sterile. The general
situation when one has both left and right handed particles will
be studied in Sec.~\ref{INHOMOGEN}.

The general mass matrix, involving left-handed flavor neutrinos
$\nu_\lambda^\mathrm{L}$, can be diagonalized with help of the
matrix transformation~\cite{Kob80},
\begin{equation}\label{matrtransMaj}
  \nu_\lambda^\mathrm{L} = \sum_a U_{\lambda a} \eta_a,
\end{equation}
where $\lambda=\alpha,\beta$ is the flavor index and $\eta_a$,
$a=1,2$, corresponds to a Majorana particle with a definite mass
$m_a$. In the simplest case the mixing of the flavor states arises
purely from Majorana mass terms between the left-handed neutrinos,
and then the mixing matrix $U_{\lambda a}$ is a $2 \times 2$ and
unitary matrix, i.e., $a=1,2$ and, assuming no CP violation, it
can be parameterized in the same way as in Eq.~\eqref{Ulambdaa}.

We study the evolution of this system with the following initial
condition [see also Eq.~\eqref{inicondnulambdaexp}]:
\begin{equation}\label{IniCondMaj}
  \nu_\alpha^\mathrm{L}(\mathbf{r},0) = 0,
  \quad
  \nu_\beta^\mathrm{L}(\mathbf{r},0) =
  \nu_\beta^{(0)}e^{\mathrm{i}\mathbf{k}\mathbf{r}},
\end{equation}
where $\mathbf{k}=(0,0,k)$ is the initial momentum and
$\nu_\beta^{(0)\mathrm{T}}=(0,1)$. The initial state is thus a
left-handed neutrino of flavor ``$\beta$" propagating along the
$z$-axis to the positive direction.

As both the left-handed state $\nu_\lambda^\mathrm{L}$ and
Majorana state $\eta_a$ have two degrees of freedom, we will
describe them in the following by using two-component Weyl
spinors. The Weyl spinor of a free Majorana particle obeys the
wave equation (see, e.g.,~\cite{FukYan03p292}),
\begin{equation}\label{WEMajVac}
  \mathrm{i}\dot{\eta}_a+(\bm{\sigma}\mathbf{p}) \eta_a+
  \mathrm{i} m_a\sigma_2\eta_a^{*{}}=0.
\end{equation}
Note that Eq.~\eqref{WEMajVac} can be formally derived from
Eq.~\eqref{Direqpsivac} if impose the Majorana condition $\psi_a =
(\psi_a)^c$, where the index ``$c$" means the charge conjugation,
on the four component spinor $\psi_a$. The Majorana condition
means that this spinor is represented as $\psi_a^\mathrm{T} =
(\mathrm{i} \sigma_2 \eta_a^{*{}}, \eta_a)$.

The general solution of Eq.~\eqref{WEMajVac} can be presented
as~\cite{Cas57}
\begin{align}\label{GenSolMajVac}
  \eta_a(\mathbf{r},t)= &
  \int \frac{\mathrm{d}^3\mathbf{p}}{(2\pi)^{3/2}}
  e^{\mathrm{i}\mathbf{p}\mathbf{r}}
  \notag
  \\
  & \times
  \sum_{\zeta = \pm 1}
  \big[
    a_a^{(\zeta)}(\mathbf{p})u_a^{(\zeta)}(\mathbf{p}) e^{-\mathrm{i}E_a t}
    \notag
    \\
    & +
    a_a^{(\zeta)*{}}(-\mathbf{p})v_a^{(\zeta)}(-\mathbf{p}) e^{\mathrm{i}E_a t}
  \big],
\end{align}
where $E_a=\sqrt{m_a^2+|\mathbf{p}|^2}$. As in Sec.~\ref{VACUUM}
the coefficient $a_a^{(\zeta)}(\mathbf{p})$ is time independent
and its value is determined by the initial
condition~\eqref{IniCondMaj}.

The basis spinors $u_a^{(\zeta)}$ and $v_a^{(\zeta)}$ in
Eq.~\eqref{GenSolMajVac} have the form
\begin{align}\label{SpinorsMajVac}
  u_a^{-{}}(\mathbf{p}) = & \lambda_a w_{-{}},
  \quad
  u_a^{+{}}(\mathbf{p}) = - \lambda_a \frac{m_a}{E_a+|\mathbf{p}|}w_{+{}},
  \notag
  \\
  v_a^{+{}}(\mathbf{p}) = & \lambda_a w_{-{}},
  \quad
  v_a^{-{}}(\mathbf{p}) = \lambda_a \frac{m_a}{E_a+|\mathbf{p}|}w_{+{}},
\end{align}
where $w_{\pm{}}$ are  helicity amplitudes given
by~\cite{BerLifPit89}
\begin{align}\label{HelAmplMaj}
  w_{+{}} = &
  \begin{pmatrix}
    e^{-i\phi/2}\cos(\vartheta/2) \\
    e^{i\phi/2}\sin(\vartheta/2) \
  \end{pmatrix},
  \notag
  \\
  w_{-{}} = &
  \begin{pmatrix}
    -e^{-i\phi/2}\sin(\vartheta/2) \\
    e^{i\phi/2}\cos(\vartheta/2) \
  \end{pmatrix},
\end{align}
the angles $\phi$ and $\vartheta$ giving the direction of the
momentum of the particle, $\mathbf{p} = |\mathbf{p}| \times
(\sin\vartheta\cos\phi,\sin\vartheta\sin\phi,\cos\vartheta)$. The
normalization factor $\lambda_a$ in Eq.~\eqref{SpinorsMajVac} can
be chosen as
\begin{equation}\label{NormFactMaj}
  \lambda_a^{-2}=
  1-\frac{m_a^2}{(E_a+|\mathbf{p}|)^2}.
\end{equation}
Let us mention the following  properties of the helicity
amplitudes $w_{\pm{}}$:
\begin{widetext}
\begin{gather}
  (\bm{\sigma}\mathbf{p})w_{\pm{}} = \pm|\mathbf{p}|w_{\pm{}},
  \quad
  \mathrm{i}\sigma_2 w_{\pm{}}^{*{}} = \mp w_{\mp{}},
  \quad
  w_{\pm{}}(-\mathbf{p}) = \mathrm{i} w_{\mp{}}(\mathbf{p}),
  \notag
  \\
  \left( w_{+{}} \otimes w_{-{}}^\mathrm{T} \right) -
  \left( w_{-{}} \otimes w_{+{}}^\mathrm{T} \right) =
  \mathrm{i}\sigma_2,
  \quad
  \label{HelAmplProp}
  \left( w_{+{}} \otimes w_{+{}}^\dag \right) +
  \left( w_{-{}} \otimes w_{-{}}^\dag \right) = 1,
\end{gather}
\end{widetext}
which can be immediately obtained from Eq.~\eqref{HelAmplMaj} and
which are useful in deriving the results given below.

The time-independent coefficients $a_a^{\pm{}}(\mathbf{p})$ in
Eq.~\eqref{GenSolMajVac} have the following form~\cite{Cas57}:
\begin{align}\label{apmMajVac}
  a_a^{+{}}(\mathbf{p}) = & \frac{1}{(2\pi)^{3/2}}
  \bigg[
    \eta_a^{(0)\dag}(-\mathbf{p}) v_a^{+{}}(\mathbf{p})
    \notag
    \\
    & +
    \frac{\mathrm{i}m_a}{E_a+|\mathbf{p}|}
    v_a^{+{}\dag}(-\mathbf{p})\eta_a^{(0)}(\mathbf{p})
  \bigg],
  \notag
  \\
  a_a^{-{}}(\mathbf{p}) = & \frac{1}{(2\pi)^{3/2}}
  \bigg[
    u_a^{-{}\dag}(\mathbf{p}) \eta_a^{(0)}(\mathbf{p})
    \notag
    \\
    & -
    \frac{\mathrm{i}m_a}{E_a+|\mathbf{p}|}
    \eta_a^{(0)\dag}(-\mathbf{p}) u_a^{-{}}(-\mathbf{p})
  \bigg],
\end{align}
where $\eta^{(0)}_a(\mathbf{p})$ is the Fourier transform of the
initial wave function $\eta_a$,
\begin{equation*}
  \eta^{(0)}_a(\mathbf{p}) =
  \int \mathrm{d}^3\mathbf{p}
  e^{-\mathrm{i}\mathbf{p}\mathbf{r}}
  \eta^{(0)}_a(\mathbf{r}).
\end{equation*}
Using Eqs.~\eqref{GenSolMajVac}-\eqref{apmMajVac} we then obtain
the following expression for the wave function for the neutrino
mass eigenstate:
\begin{align}\label{etaaMajVac}
  \eta_a(\mathbf{r},t) = &
  \int \frac{\mathrm{d}^3\mathbf{p}}{(2\pi)^3}
  e^{\mathrm{i}\mathbf{p}\mathbf{r}}
  \lambda_a^2
  \notag
  \\
  & \times
  \Bigg[
    \Bigg\{
      \left(
        e^{-\mathrm{i}E_a t}-
        \left[
          \frac{m_a}{E_a+|\mathbf{p}|}
        \right]^2
        e^{\mathrm{i}E_a t}
      \right)
      \notag
      \\
      & \times
      \left( w_{-{}} \otimes w_{-{}}^\dag \right)
      \notag
      \\
      & +
      \left(
        e^{\mathrm{i}E_a t}-
        \left[
          \frac{m_a}{E_a+|\mathbf{p}|}
        \right]^2
        e^{-\mathrm{i}E_a t}
      \right)
      \notag
      \\
      & \times
      \left( w_{+{}} \otimes w_{+{}}^\dag \right)
    \Bigg\}
    \eta^{(0)}_a(\mathbf{p})
    \notag
    \\ & -
    2\frac{m_a}{E_a+|\mathbf{p}|}
    \sin(E_a t) \sigma_2 \eta^{(0)*{}}_a(-\mathbf{p})
  \Bigg].
\end{align}

From Eqs.~\eqref{HelAmplProp} and~\eqref{etaaMajVac} it follows
that  a mass eigenstate particle initially in the left polarized
state $\eta_a^{(0)}(\mathbf{r}) \sim
w_{-{}}(\mathbf{k})e^{\mathrm{i}\mathbf{k}\mathbf{r}}$ is
described at later times by
\begin{align}\label{etaaMajVacSC}
  \eta_a(\mathbf{r},t) \sim &
  \lambda_a^2
  \bigg\{
    \left(
      e^{-\mathrm{i}E_a t}-
      \left[
        \frac{m_a}{E_a+|\mathbf{k}|}
      \right]^2
      e^{\mathrm{i}E_a t}
    \right)
    e^{\mathrm{i}\mathbf{k}\mathbf{r}}w_{-{}}(\mathbf{k})
    \notag
    \\ & -
    2\mathrm{i}\frac{m_a}{E_a+|\mathbf{k}|}
    \sin(E_a t) e^{-\mathrm{i}\mathbf{k}\mathbf{r}}w_{+{}}(\mathbf{k})
  \bigg\}.
\end{align}
Let us notice that the second term in Eq.~\eqref{etaaMajVacSC}
describes an antineutrino state. Indeed the spinor
$w_{+{}}(\mathbf{k})$ satisfies the relation,
$(\bm{\sigma}\mathbf{k})w_{+{}}(\mathbf{k}) =
|\mathbf{k}|w_{+{}}(\mathbf{k})$, see Eq.~\eqref{HelAmplProp}.
Therefore it corresponds to an antiparticle (see
Ref.~\cite{BerLifPit89p137}). This term is responsible for the
neutrino-to-antineutrino flavor state transition
$\nu_\beta^\mathrm{L} \leftrightarrow (\nu_\alpha^\mathrm{L})^c$.

According to Eqs.~\eqref{Ulambdaa} and~\eqref{matrtransMaj}, the
wave function of the left-handed neutrino of the flavor
``$\alpha$" is $\nu_\alpha^\mathrm{L} = \cos \theta
\eta_1^\mathrm{L} - \sin \theta \eta_2^\mathrm{L}$. From
Eqs.~\eqref{matrtransMaj} and~\eqref{etaaMajVacSC} it then follows
that the probability of the transition $\nu_\beta^\mathrm{L}\to
\nu_\alpha^\mathrm{L}$ in vacuum is given by
\begin{align}\label{PtrMajVac}
  P_{\nu_\beta^\mathrm{L}\to \nu_\alpha^\mathrm{L}}(t) = &
  |\nu_\alpha^\mathrm{L}|^2 = \sin^2(2\theta)
  \bigg\{
    \sin^2(\Phi t)
    \notag
    \\
    & +
    \frac{1}{2|\mathbf{k}|^2}
    \cos(\sigma t)\sin(\Phi t)
    \notag
    \\
    & \times
    [m_1^2 \sin(E_1 t) - m_2^2 \sin(E_2 t)]
  \bigg\}
  \notag
  \\
  & +
  \mathcal{O}
  \left(
    \frac{m_a^4}{|\mathbf{k}|^4}
  \right),
\end{align}
where $\sigma$ and $\Phi$ are introduced in Eq.~\eqref{Phisigma}.

We can compare Eq.~\eqref{PtrMajVac} with analogous expression for
Dirac particles~\eqref{PtrPsurvac}. The leading terms in both
equations coincide, whereas the next-to-leading terms have
different signs. Therefore one can reveal the nature of neutrinos,
i.e. say if neutrino mass eigenstates are Dirac or Majorana
particles, studying the corrections to the transition probability
formulas.

Analogously we can calculate the transition probability for the
process $\nu_\beta^\mathrm{L}\to (\nu_\alpha^\mathrm{L})^c$ using
the second term in Eq.~\eqref{etaaMajVacSC},
\begin{align}\label{PtrnuanuMajVac}
  P_{\nu_\beta^\mathrm{L}\to (\nu_\alpha^\mathrm{L})^c}(t) = &
  |(\nu_\alpha^\mathrm{L})^c|^2
  \notag
  \\
  & =
  \frac{\sin^2(2\theta)}{4|\mathbf{k}|^2}
  [m_1 \sin(E_1 t) - m_2 \sin(E_2 t)]^2
  \notag
  \\
  & +
  \mathcal{O}
  \left(
    \frac{m_a^4}{|\mathbf{k}|^4}
  \right).
\end{align}
Note that the next-to-leading term in Eq.~\eqref{PtrMajVac} and
leading term in Eq.~\eqref{PtrnuanuMajVac} have the same order of
magnitude $\sim m_a^2/|\mathbf{k}|^2$.

Before moving to consider Majorana neutrinos in magnetic fields in
Sec.~\ref{MAJNUMATTB}, we make a general comment concerning the
validity of our approach based on relativistic field theory
involving (classical) first quantized Majorana fields. It has been
stated~\cite{SchVal81} that the dynamics of massive Majorana
fields cannot be described within the classical field theory
approach due to the fact that the mass term of the Lagrangian,
$\eta^\mathrm{T} \mathrm{i}\sigma_2 \eta$, vanishes when $\eta$ is
represented as a $c$-number function (see, e.g.,
Eq.~\eqref{spinmassLargM} below). Note that Eq.~\eqref{WEMajVac}
is a direct consequence of the Dirac equation if we suggest that
the four-component wave function satisfies the Majorana condition.
Therefore a solution of Eq.~\eqref{WEMajVac}, i.e., wave functions
and energy levels, in principle does not depend on the existence
of a Lagrangian resulting in this equation. The wave equations
describing elementary particles should follow from the quantum
field theory principles. However quite often these quantum
equations allow classical solutions (see Ref.~\cite{Giu96} for
many interesting examples). We demonstrate in
Secs.~\ref{VACUUM}-\ref{MATTERB} (see also
Refs.~\cite{Dvo05,Dvo06,DvoMaa07,Dvo08JPG,Dvo10,Dvo08JPCS,DvoMaa09})
that oscillations of Dirac neutrinos in vacuum and various
external fields can be described in the framework of the classical
field theory. The main result of this section was to show that the
quantum Eq.~\eqref{WEMajVac} for massive Majorana particles can be
solved [see Eq.~\eqref{etaaMajVacSC}] in the framework of the
classical field theory as well.

\section{Majorana neutrinos in matter and transversal magnetic
field\label{MAJNUMATTB}}

In this sections we apply the formalism developed in
Sec.~\ref{MAJVACUUM} to the more general case of Majorana
neutrinos propagating in background matter and interacting with an
external magnetic field. We show that the initial condition
problem of the mass eigenstates can be expressed in terms of the
Schr\"{o}dinger equation with an effective Hamiltonian which
coincides with previously proposed one~\cite{LimMar88}.

For describing the evolution of two Majorana mass eigenstates in
matter under the influence of an external magnetic field, the wave
equation~\eqref{WEMajVac} is to be modified to the following form:
\begin{align}\label{WEMajmattB}
  \mathrm{i}\dot{\eta}_a + &
  \left(
    \bm{\sigma}\mathbf{p}-\frac{g_a}{2}
  \right)\eta_a +
  \mathrm{i} m_a\sigma_2\eta_a^{*{}}
  \notag
  \\
  & -
  \frac{g}{2}\eta_b -
  \mathrm{i} \mu (\bm{\sigma}\mathbf{B}) \sigma_2 \epsilon_{ab} \eta_b^{*{}}= 0,
  \quad
  a \neq b,
\end{align}
where $\epsilon_{ab}=\mathrm{i}(\sigma_2)_{ab}$, and $g_a$ and $g$
were defined in Eq.~\eqref{gab0}. Note that Eq.~\eqref{WEMajmattB}
can be formally derived from Eq.~\eqref{DireqmattB} if one
neglects vector current interactions, i.e., replace
$(1-\gamma^5)/2$ with $-\gamma^5/2$, and takes into account the
fact that the magnetic moment matrix of Majorana neutrinos is
antisymmetric (see, e.g., Ref.~\cite{FukYan03p477}). We will apply
the same initial condition~\eqref{IniCondMaj} as in the vacuum
case. It should be mentioned that the evolution of Majorana
neutrinos in matter and in a magnetic field has been previously
discussed in Ref.~\cite{Pas96}.

The general solution of Eq.~\eqref{WEMajmattB} can be expressed in
the following form:
\begin{widetext}
\begin{equation}\label{GenSolMajmattB}
  \eta_a(\mathbf{r},t)= 
  \int \frac{\mathrm{d}^3\mathbf{p}}{(2\pi)^{3/2}}
  e^{\mathrm{i}\mathbf{p}\mathbf{r}}
  \sum_{\zeta = \pm 1}
  \big[
    a_a^{(\zeta)}(\mathbf{p},t)u_a^{(\zeta)}(\mathbf{p})
    \exp(-\mathrm{i}E_a^{(\zeta)} t) +
    a_a^{(\zeta)*{}}(-\mathbf{p},t)v_a^{(\zeta)}(-\mathbf{p})
    \exp(\mathrm{i}E_a^{(\zeta)} t)
  \big],
\end{equation}
where the energy levels are given in Eq.~\eqref{energymatt} (see
Ref.~\cite{StuTer05}). The basis spinors in
Eq.~\eqref{GenSolMajmattB} can be chosen as
\begin{equation}
  u_a^{-{}}(\mathbf{p}) = \lambda_a^{-{}} w_{-{}},
  \
  u_a^{+{}}(\mathbf{p}) = - \lambda_a^{+{}}
  \frac{m_a}{E_a^{+{}}+(|\mathbf{p}|-g_a/2)}w_{+{}},
  \
  v_a^{+{}}(\mathbf{p}) = \lambda_a^{+{}} w_{-{}},
  \
  v_a^{-{}}(\mathbf{p}) = \lambda_a^{-{}}
  \frac{m_a}{E_a^{-{}}+(|\mathbf{p}|+g_a/2)}w_{+{}},
\end{equation}
where the normalization factors $\lambda_a^{(\zeta)}$,
$\zeta=\pm{}$,  are given by
\begin{equation}
  (\lambda_a^{(\zeta)})^{-2}=
  1-\frac{m_a^2}{[E_a+(|\mathbf{p}|-\zeta g_a/2)]^2}.
\end{equation}

Let us consider the propagation of Majorana neutrinos in the
transversal magnetic field. Using a similar technique as in the
Dirac case in Sec.~\ref{MATTERB} and assuming $k \gg m_a$, we end
up with the following ordinary differential equations for the
coefficients  $a_a^{(\zeta)}$,
\begin{equation}\label{Ham1Maj}
  \mathrm{i}\frac{\mathrm{d}\Psi'}{\mathrm{d}t} = H' \Psi',
  \quad
  H' =
  \begin{pmatrix}
    0 & g e^{\mathrm{i} \omega_{-{}} t}/2 & 0 &
    \mu B e^{\mathrm{i} \Omega_{-{}} t} \\
    g e^{-\mathrm{i} \omega_{-{}} t}/2  & 0 &
    -\mu B e^{-\mathrm{i} \Omega_{+{}} t} & 0 \\
    0 & -\mu B e^{\mathrm{i} \Omega_{+{}} t} &
    0 & -g e^{\mathrm{i} \omega_{+{}} t}/2 \\
    \mu B e^{-\mathrm{i} \Omega_{-{}} t} & 0 &
    -g e^{-\mathrm{i} \omega_{+{}} t}/2 & 0 \
  \end{pmatrix},
\end{equation}
where
$\Psi^{'\mathrm{T}}=(a_1^{-{}},a_2^{-{}},a_1^{+{}},a_2^{+{}})$ and
\begin{equation}
  \omega_{\pm{}} = E_1^{\pm{}}-E_2^{\pm{}} \approx 2\Phi \mp \frac{g_1-g_2}{2},
  \quad
  \Omega_{\mp{}} = E_1^{\mp{}}-E_2^{\pm{}} \approx 2\Phi \pm
  \frac{g_1+g_2}{2},
\end{equation}
which should be compared with Eq.~\eqref{Schr1QFT}.

By making the matrix transformation
%
\begin{equation}\label{matrtransfPsi}
  \Psi' = \mathcal{U}\Psi,
  \quad
  \mathcal{U} = \mathrm{diag}
  \left\{
    e^{\mathrm{i}(\Phi+g_1/2)t},
    e^{-\mathrm{i}(\Phi-g_2/2)t},
    e^{\mathrm{i}(\Phi-g_1/2)t},
    e^{-\mathrm{i}(\Phi+g_2/2)t}
  \right\},
\end{equation}
we can recast Eq.~\eqref{Ham1Maj} into the form
\begin{equation}\label{Ham2Maj}
  \mathrm{i}\frac{\mathrm{d}\Psi}{\mathrm{d}t} = H \Psi,
  \quad
  H =
  \mathcal{U}^\dag H' \mathcal{U} -
  \mathrm{i} \mathcal{U}^\dag \dot{\mathcal{U}}
  =
  \begin{pmatrix}
    \Phi + g_1/2 & g/2 & 0 & \mu B \\
    g/2 & -\Phi + g_2/2 & -\mu B & 0 \\
    0 & -\mu B & \Phi - g_1/2 & -g/2 \\
    \mu B & 0 & -g/2 & -\Phi - g_2/2 \
  \end{pmatrix}.
\end{equation}
\end{widetext}
Let us note that the analogous effective Hamiltonian has been used
in describing the spin flavor oscillations of Majorana neutrinos
within the quantum mechanical approach (see, e.g.,
Ref.~\cite{LimMar88}) if we use the basis $\Psi_{QM}^\mathrm{T} =
(\psi_1^\mathrm{L}, \psi_2^\mathrm{L}, [\psi_1^\mathrm{L}]^c,
[\psi_2^\mathrm{L}]^c)$.

Note that the consistent derivation of the master
Eq.~\eqref{WEMajmattB} should be done in the framework of the
quantum field theory (see, e.g., Ref.~\cite{SchVal81}), supposing
that the spinors $\eta_a$ are expressed via anticommuting
operators. This quantum field theory treatment is important to
explain the asymmetry of the magnetic moment matrix. However, it
is possible to see that the main Eq.~\eqref{WEMajmattB} can also
be reduced to the standard Schr{\"o}dinger evolution
Eq.~\eqref{Ham2Maj} for neutrino spin flavor oscillations if we
suppose that the wave functions $\eta_a$ are $c$-number objects.
That is why one can again conclude that classical and quantum
field theory methods for studying Majorana neutrinos' propagation
in external fields are equivalent.

\section{Spin flavor oscillations of Majorana neutrinos in the expanding
envelope of a supernova\label{MAJSPECMATT}}

In this section we discuss the application of the formalism
developed in Sec.~\ref{MAJNUMATTB} to the description of spin
flavor oscillations of Majorana neutrinos in an expanding envelope
of a supernova and compare the results with the case of Dirac
neutrinos studied in Sec.~\ref{SPECMATT}.

The dynamics of the system of two Majorana neutrinos in matter
under the influence of an external magnetic field is governed by
the Schr\"{o}dinger equation~\eqref{Ham2Maj}. However, as in
Sec.~\ref{SPECMATT}, the explicit analytical solution of this
equation is quite cumbersome. That is why we again consider the
situation when $n_e = n_p = n_n = n$, which results in $g_1 =
-g_2$ (see Eqs.~\eqref{fdefinition} and~\eqref{flambda} as well as
Sec.~\ref{SPECMATT}). In this case the eigenvalues of the
Hamiltonian~\eqref{Ham2Maj} $\lambda = \pm \mathcal{E}_{\pm{}}$
are given by
\begin{align}
  \mathcal{E}_{\pm{}} = & \frac{1}{2}
  \sqrt{V^2+4(\mu B)^2+4\Phi^2 \pm 4VR},
  \quad
  \notag
  \\
  R = & \sqrt{(\Phi\cos 2\theta)^2+(\mu B)^2},
\end{align}
where $V = G_\mathrm{F} n/\sqrt{2}$ as in Sec~\ref{SPECMATT}. The
time evolution of the wave function is described by the formula,
\begin{align}\label{PsimattBMaj}
  \Psi(t) = & \sum_{\zeta = \pm 1}
  \Big[
    \left(
      U_\zeta \otimes U_\zeta^\dag
    \right)\exp{(-\mathrm{i}\mathcal{E}_\zeta t)}
    \notag
    \\
    &
    +
    \left(
      V_\zeta \otimes V_\zeta^\dag
    \right)\exp{(\mathrm{i}\mathcal{E}_\zeta t)}
  \Big]\Psi_0,
\end{align}
where  $U_\zeta$ and $V_\zeta$ are the eigenvectors of the
Hamiltonian~\eqref{Ham2Maj}, given as
\begin{equation}\label{BasisSpinorsMaj}
  U_\zeta =
  \frac{1}{N_\zeta}
  \begin{pmatrix}
    -x_\zeta \\
    -y_\zeta \\
    1 \\
    -z_\zeta \
  \end{pmatrix},
  \quad
  V_\zeta =
  \frac{1}{N_\zeta}
  \begin{pmatrix}
    -y_\zeta \\
    x_\zeta \\
    z_\zeta \\
    1 \
  \end{pmatrix},
\end{equation}
where
\begin{align}\label{xyzSigma}
  x_\zeta = & \frac{\mu B (\mathcal{E}_\zeta+\Phi)}{\Sigma_\zeta} V \sin 2\theta,
  \notag
  \\
  y_\zeta = & \frac{\mu B}{\mathcal{E}_\zeta+\Phi-V \cos 2\theta /2}
  \left[
    1+\frac{(\mathcal{E}_\zeta+\Phi)}{2 \Sigma_\zeta} V^2 \sin^2(2\theta)
  \right],
  \notag
  \\
  z_\zeta = & \frac{V\sin 2\theta}{2(\mathcal{E}_\zeta+\Phi+V \cos 2\theta /2)}
  \left[
    1+\frac{2(\mu B)^2(\mathcal{E}_\zeta+\Phi)}{\Sigma_\zeta}
  \right],
  \notag
  \\
  \Sigma_\zeta = & \frac{V}{2}
  [2\mathcal{E}_\zeta(\mathcal{E}_\zeta+\Phi)-V^2/2 + \Phi V \cos 2\theta]
  \cos 2\theta
  \notag
  \\
  & +
  \zeta R V
  (\mathcal{E}_\zeta+\Phi-V \cos 2\theta /2).
\end{align}
The normalization coefficient $N_\zeta$ in
Eq.~\eqref{BasisSpinorsMaj} is given by
$N_\zeta=\sqrt{1+x_\zeta^2+y_\zeta^2+z_\zeta^2}$.

Proceeding along the same lines as in Sec.~\ref{SPECMATT}, we
obtain from Eqs.~\eqref{matrtransMaj}
and~\eqref{PsimattBMaj}-\eqref{xyzSigma}  the probability of the
process $\nu_\beta^\mathrm{L} \to \nu_\alpha^\mathrm{R}$ as,
\begin{align}\label{PtrmattBMaj}
  P_{\nu_\beta^\mathrm{L} \to \nu_\alpha^\mathrm{R}}(t) = &
  [C_{+{}}\cos(\mathcal{E}_{+{}}t)+C_{-{}}\cos(\mathcal{E}_{-{}}t)]^2
  \notag
  \\
  & +
  [S_{+{}}\sin(\mathcal{E}_{+{}}t)+S_{-{}}\sin(\mathcal{E}_{-{}}t)]^2,
\end{align}
where
\begin{align}\label{CpmSpmMaj}
  C_\zeta = & -\frac{1}{N_\zeta^2}[\sin 2\theta (x_\zeta + y_\zeta z_\zeta) +
  \cos 2\theta (y_\zeta - x_\zeta z_\zeta)],
  \notag
  \\
  S_\zeta = & \frac{1}{N_\zeta^2}(y_\zeta + x_\zeta z_\zeta).
\end{align}
Consistently with Eq.~\eqref{IniCondMaj}, we have taken the
initial  wave function as
\begin{equation}\label{IniCondQMMaj}
  \Psi_0^\mathrm{T}=(\sin\theta,\cos\theta,0,0).
\end{equation}
With help of Eqs.~\eqref{xyzSigma} and~\eqref{CpmSpmMaj} it is
easy to check that $C_{+{}}+C_{-{}}=0$  guaranteeing $P(0)=0$.

Note that formally Eq.~\eqref{PtrmattBMaj} corresponds to the
transitions $\nu_\beta^\mathrm{L} \to \nu_\alpha^\mathrm{R}$.
However, virtually it describes oscillations between active
neutrinos $\nu_\beta^\mathrm{L} \leftrightarrow
(\nu_\alpha^\mathrm{L})^c$ since
$\nu_\alpha^\mathrm{R}=(\nu_\alpha^\mathrm{L})^c$ for Majorana
particles.

\begin{figure*}
  \centering
  \includegraphics[scale=.93]{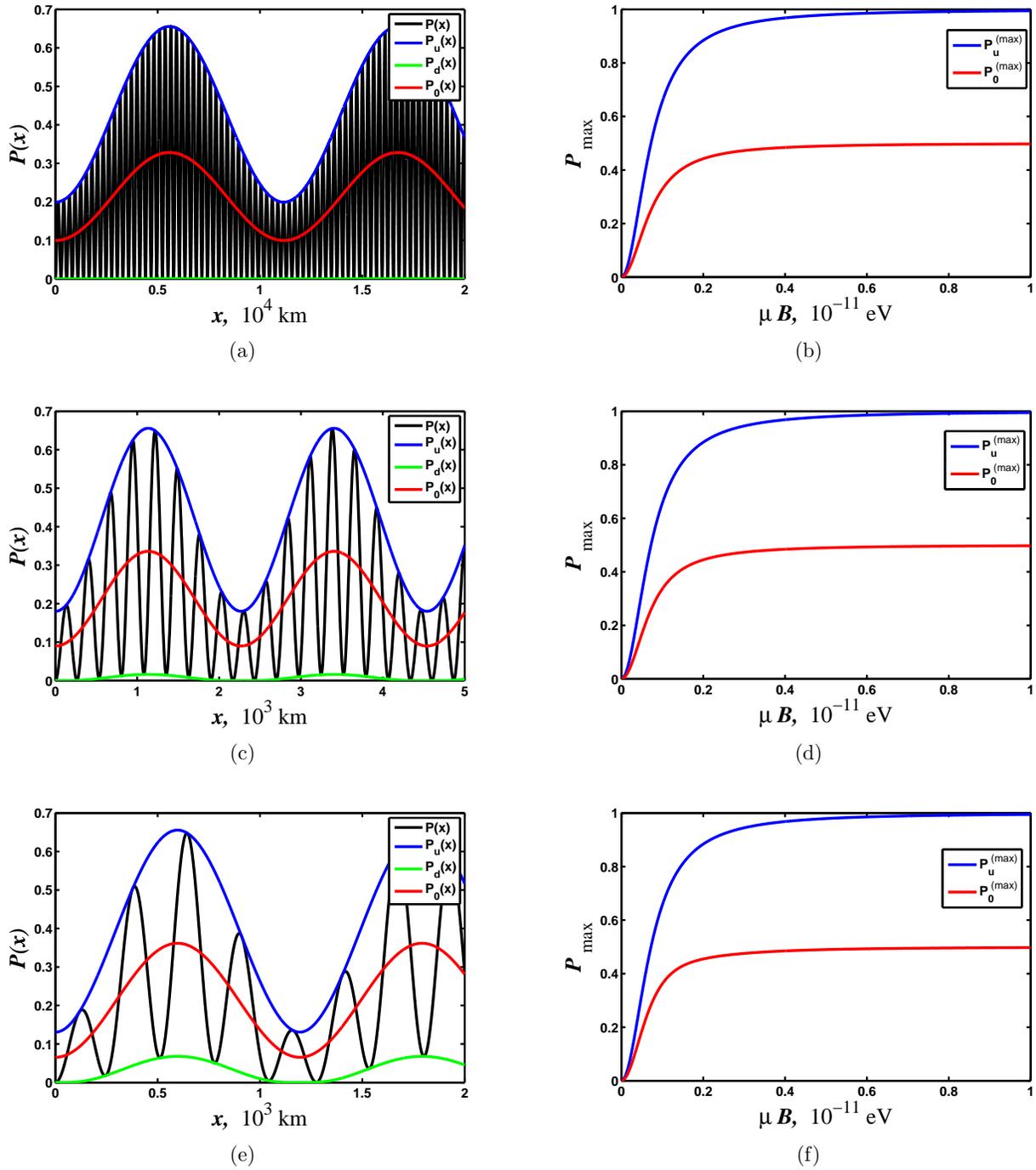}
  \caption{\label{Majnuenv} (color online)
  (a), (c) and (e) The transition probability versus the
  distance passed by a neutrino beam. The neutrino parameters
  have the same values as in Figs.~\ref{Dirnulowdens}-\ref{Dirnuhighdens},
  $E_\nu = 10\thinspace\text{MeV}$,
  $\delta m^2 = 8 \times 10^{-5}\thinspace\text{eV}^2$ and
  $\theta = 0.6$. The magnetic energy is equal to
  $\mu B = 10^{-12}\thinspace\text{eV}$.
  The black line is the function $P(x)$,
  the blue and green lines are the envelope functions $P_{u,d}(x)$ and
  the red line is the averaged transition probability $P_0(x)$.
  (b), (d) and (f)
  The dependence of the maximal values of the functions $P(x)$ and
  $P_0(x)$, blue and red lines respectively, on the magnetic energy
  for the given density.
  Panels (a) and (b) correspond to the matter density
  $\rho = 10\thinspace\text{g/cc}$,
  (c) and (d) to $\rho = 50\thinspace\text{g/cc}$ and
  (e) and (f) to $\rho = 100\thinspace\text{g/cc}$.
  This figure is taken from Ref.~\cite{DvoMaa09}.}
\end{figure*}

As in the case of Eq.~\eqref{PrtDir}, Eq.~\eqref{PtrmattBMaj} can
be treated  analytically for relatively small values of the
effective potential $V$. The ensuing envelope functions $P_{u,d} =
P_0 \pm \sqrt{P_c^2+P_s^2}$ depend on the coefficients $C_\zeta$
and $S_\zeta$ in the same way as in Eq.~\eqref{P0PcPsDir}. The
transition probabilities at various  values of the matter density
and the magnetic field are presented in Fig.~\ref{Majnuenv}.

Despite the formal similarity between Dirac and Majorana
transition probabilities [see Eqs.~\eqref{PrtDir}
and~\eqref{PtrmattBMaj}] the actual dynamics is quite different in
these two cases, as one can see by comparing
Figs.~\ref{Dirnulowdens}(d)-\ref{Dirnuhighdens}(d) and
Fig.~\ref{Majnuenv}, panels (b), (d) and (f). In particular, in
the Majorana case $P_u^\mathrm{(max)}=4|C_{+{}}C_{-{}}|$ for
arbitrary $B$, to be compared with Eq.~\eqref{PumaxDir}, while the
function $P_0^\mathrm{(max)}$ has the same form as in the Dirac
case given in Eq.~\eqref{P0maxDir}. In contrast with the Dirac
case, the averaged transition probability does not achieve its
maximal value at some moderate magnetic field $B_\mathrm{max}$
value, but both $P_u^\mathrm{max}$ and $P_0^\mathrm{max} $ are
monotonically increasing functions of the strength of the magnetic
field with 1 and 1/2 as their asymptotic values, respectively. We
can understand this behavior when we recall that, at $\mu B \gg
\max(\Phi,V)$, the effective Hamiltonian  Eq.~\eqref{Ham2Maj}
becomes
\begin{equation}\label{HinfMaj}
  H_\infty =
  \mathrm{i} \mu B \gamma^2,
  \quad
  \mathrm{i}\gamma^2 =
  \begin{pmatrix}
    0 & 0 & 0 & 1 \\
    0 & 0 & -1 & 0 \\
    0 & -1 & 0 & 0 \\
    1 & 0 & 0 & 0 \
  \end{pmatrix}.
\end{equation}
The Schr\"odinger equation with the effective
Hamiltonian~\eqref{HinfMaj} has the formal solution
\begin{equation}\label{FormSolMaj}
  \Psi(t)= \exp(-\mathrm{i} H_\infty t)\Psi(0)
  =
  [\cos(\mu B t)+\gamma^2\sin(\mu B t)]\Psi(0).
\end{equation}
Using Eqs.~\eqref{matrtransMaj}, \eqref{IniCondQMMaj}
and~\eqref{FormSolMaj} we then immediately arrive to the following
expression for the transition probability,
$P(t)=|\nu_\alpha^\mathrm{R}|^2=\sin^2(\mu B t)$, which explains
the behavior of the function $P_u^\mathrm{(max)}$ at strong
magnetic fields. Note that the analogous result was also obtained
in Ref.~\cite{LikStu95}.

Finally, it is worth of noticing that in contrast to the Dirac
case, the behavior of the transition probability in the Majorana
case is qualitatively similar for different matter densities and
different magnetic fields [see Fig.~\ref{Majnuenv}, panels (a),
(c), and (e)].

The problem of Majorana neutrinos' spin flavor oscillations was
studied in Refs.~\cite{AndSatPRD03,AndSatJCAP03} with help of
numerical codes. For example, in Ref.~\cite{AndSatPRD03} the
evolution equation for three neutrino flavors propagating inside a
presupernova star with zero metallicity, e.g., corresponding to
W02Z model~\cite{WooHegWea02}, was solved for the realistic matter
and magnetic field profiles. Although our analytical transition
probability formula~\eqref{PtrmattBMaj} is valid only for the
constant matter density and magnetic field strength, it is
interesting to compare our results with the numerical simulations
of Ref.~\cite{AndSatPRD03}. In those calculations the authors used
magnetic fields $B \sim 10^{10}\thinspace\text{G}$ and magnetic
moments $\sim 10^{-12}\thinspace\mu_\mathrm{B}$ that give us the
magnetic energy $\mu B \sim 10^{-11}\thinspace\text{eV}$. This
value is the maximal magnetic energy used in our work.

It was found in Ref.~\cite{AndSatPRD03} that spin flavor
conversion is practically adiabatic for low-energy neutrinos
corresponding to $E_\nu \sim 5\thinspace\text{MeV}$ inside the
region where $Y_e \approx 0.5$ and the averaged transition
probability for the channel $\nu_\mu \to \bar{\nu}_e$ is close
$0.5$. This big transition probability is due to the RSF-H and
RSF-L resonances at the distance $\approx 0.01 R_\odot$. Even so
that we study two neutrino oscillations scheme, we obtained the
analogous behavior of $P_0^\mathrm{(max)}$ (see
Fig.~\ref{Majnuenv}). However, in our case this big transition
probability is due to the presence of the strong magnetic field
(see Refs.~\cite{GiuStu09,LikStu95}). We cannot compare our
transition probability formula~\eqref{PtrmattBMaj} with the
results of Ref.~\cite{AndSatPRD03} for higher energies, $E_\nu
> 25\thinspace\text{MeV}$, since spin flavor oscillations become
strongly nonadiabatic for these kinds of energies and one has to
take into account the coordinate dependence of the matter density
which should decrease with radius as $1/r^3$~\cite{Kac02}.

\section{Evolution of neutrinos emitted by classical sources\label{INHOMOGEN}}

In this section we present an alternative approach to the
description of neutrino flavor oscillations which involves the
studies of the evolution of mixed massive neutrinos in vacuum
under the influence of external classical fields~\cite{Dvo09}.
These fields are supposed to be localized in space and treated as
``sources" which emit flavor neutrinos.

As in Secs.~\ref{VACUUM}-\ref{MATTERB} we consider a number of
flavor neutrinos $\nu_\lambda$, which, in principle, can be
arbitrary: $\lambda = e, \mu, \tau, \dotsc$. Note that we will
formulate the dynamics of the system in terms of the left
$\nu_\lambda^\mathrm{L}$ and right $\nu_\lambda^\mathrm{R}$ handed
chiral components of the neutrino spinors.

The external sources are denoted as $J_\lambda^\mu$. It should be
noted that, if one describes the emission of mixed massive
neutrinos in any process within the quantum field theory, there is
a nonzero probability of emission of neutrinos of different
flavors~\cite{Shr80}. Although a source of one particular flavor
can prevail, one should always consider the situation when the
sources of all flavors are present.

The Lorentz invariant Lagrangian describing the evolution of this
system has the form,
\begin{align}\label{LagrnuvacNHE}
  \mathcal{L} = &
  \sum_{\lambda}
  \left(
    \overline{\nu_\lambda^\mathrm{L}}
    \mathrm{i}\gamma^\mu\partial_\mu
    \nu_\lambda^\mathrm{L}+
    \overline{\nu_\lambda^\mathrm{R}}
    \mathrm{i}\gamma^\mu\partial_\mu
    \nu_\lambda^\mathrm{R}
  \right)
  \notag
  \\
  & -
  \sum_{\lambda\lambda'}
  \Big(
    m_{\lambda\lambda'}^\mathrm{D}
    \overline{\nu_\lambda^\mathrm{L}}\nu_{\lambda'}^\mathrm{R}+
    m_{\lambda\lambda'}^\mathrm{L}
    \left(
      \nu_\lambda^\mathrm{L}
    \right)^\mathrm{T}
    C \nu_{\lambda'}^\mathrm{L}
    \notag
    \\
    & +
    m_{\lambda\lambda'}^\mathrm{R}
    \left(
      \nu_\lambda^\mathrm{R}
    \right)^\mathrm{T}
    C \nu_{\lambda'}^\mathrm{R}
    +\text{h.c.}
  \Big)
  \notag
  \\
  & +
  \sum_{\lambda}
  \left(
    \overline{\nu_\lambda^\mathrm{L}} \gamma_\mu \ell_\lambda^\mathrm{L}
    J_\lambda^\mu + \text{h.c.}
  \right),
\end{align}
where $\left( m_{\lambda\lambda'}^\mathrm{D} \right)$, $\left(
m_{\lambda\lambda'}^\mathrm{L} \right)$, $\left(
m_{\lambda\lambda'}^\mathrm{R} \right)$ are the Dirac as well as
(left and right) Majorana mass matrices and
$C=\mathrm{i}\gamma^2\gamma^0$ is the charge conjugation matrix.
These mass matrices should satisfy certain requirements which are
discussed in Ref.~\cite{Kob80}. The
Lagrangian~\eqref{LagrnuvacNHE} is CPT-invariant. In this section
we adopt the notations for Dirac matrices as in
Ref.~\cite{ItzZub84p385}. The analogous mass terms are generated
in theoretical models based on the type-II seesaw mechanism (see,
e.g., Ref.~\cite{SchVal80}).

The external fields $J_\lambda^\mu = J_\lambda^\mu(\mathbf{r},t)$
can be arbitrary functions. If we suppose that neutrinos interact
with external sources in frames of the electroweak model, the
spinor $\ell_\lambda$ in Eq.~\eqref{LagrnuvacNHE} is the charged
$\mathrm{SU}(2)$ isodoublet partner of $\nu_\lambda$. For example,
we can study the neutrino emission in a process like the inverse
$\beta$-decay: $p+e^{-}\to\nu_e+n$. In this case the external
fields in Eq.~\eqref{LagrnuvacNHE} are (see, e.g.,
Ref.~\cite{Oku90p36})
\begin{align}\label{fmuexample}
  J_{\nu_e}^\mu = & -\sqrt{2}G_\mathrm{F}
  \bar{\varPsi}_n \gamma^\mu (1-\alpha\gamma^5) \varPsi_p,
  \notag
  \\
  J_{\nu_\mu}^\mu = & 0,
  \quad
  J_{\nu_\tau}^\mu = 0,
\end{align}
where $\varPsi_p$ and $\varPsi_n$ are the wave functions of a
proton and a neutron, $\alpha\approx 1.25$. In
Eq.~\eqref{fmuexample} we assume that a source consists of
electrons, protons and neutrons.

To proceed in the analysis of the dynamics of the
system~\eqref{LagrnuvacNHE} we should diagonalize the mass term of
the Lagrangian. The diagonalized Lagrangian is expressed in terms
of the Majorana fields with different masses~\cite{Kob80}. In
general case the number of these Majorana fields is double than
the number of flavors in the initial Lagrangian. In the following
we discuss the case when only Dirac mass term is presented. Then
we study the general situation.

\subsection{Dirac mass term\label{DMT}}

In this section we suppose that only Dirac mass matrix is present
in Eq.~\eqref{LagrnuvacNHE}. Analogously to the discussion in
Sec.~\ref{VACUUM} we introduce the new set of spinor fields
$\psi_a$. In our situation the mixing matrix is a square unitary
matrix $(U_{\lambda a})$,
\begin{equation}\label{spinnupsirel}
  \nu_\lambda=
  \sum_a
  U_{\lambda a} \psi_a.
\end{equation}
By definition the mass eigenstates $\psi_a$ are Dirac particles.

When we transform the Lagrangian~\eqref{LagrnuvacNHE} using
Eq.~\eqref{spinnupsirel}, it is rewritten in the following way:
\begin{equation}\label{spinmassLarg}
  \mathcal{L} =
  \sum_a
  \bar{\psi}_a (\mathrm{i}\gamma^\mu\partial_\mu - m_a) \psi_a +
  \sum_a
  (\bar{\psi}_a \xi_a + \text{h.c.}),
\end{equation}
where $\xi_a$ is the external source for the fermion $\psi_a$,
\begin{equation}\label{spinmasssour}
  \xi_a=
  \sum_\lambda
  U_{a \lambda}^\dag
  \gamma_\mu^\mathrm{L} \ell_\lambda f_\lambda^\mu.
\end{equation}
Using Eqs.~\eqref{spinmassLarg} and~\eqref{spinmasssour} we
receive the inhomogeneous Dirac equation for the fermion $\psi_a$,
\begin{equation}\label{inhomDir}
  (\mathrm{i}\gamma^\mu\partial_\mu-m_a)\psi_a=-\xi_a.
\end{equation}
As in Sec.~\ref{VACUUM} the masses $m_a$ are the eigenvalues of
the matrix $(m_{\lambda\lambda'}^\mathrm{D})$.

The solution to Eq.~\eqref{spinmasssour} for the arbitrary spinor
$\xi_a$ is expressed with help of the retarded Green function for
a spinor field (see Ref.~\cite{BogShi80p140}),
\begin{equation}\label{spinmasseigensol1}
  \psi_a(\mathbf{r},t)=\int\mathrm{d}^3\mathbf{r}'\mathrm{d}t'
  S^\mathrm{ret}_a(\mathbf{r}-\mathbf{r}',t-t')\xi_a(\mathbf{r}',t').
\end{equation}
The explicit form of $S^\mathrm{ret}_a(\mathbf{r},t)$ can be also
found in Ref.~\cite{BogShi80p140},
\begin{equation}\label{spinretGfun}
  S^\mathrm{ret}_a(\mathbf{r},t)=
  (\mathrm{i}\gamma^\mu\partial_\mu+m_a)D^\mathrm{ret}_a(\mathbf{r},t).
\end{equation}
In Eq.~\eqref{spinretGfun} $D^\mathrm{ret}_a(\mathbf{r},t)$ is the
retarded Green function for a scalar field~\cite{BogShi80p140},
\begin{align}\label{retGfunScal}
  D^\mathrm{ret}_a(\mathbf{r},t) = &
  \int\frac{\mathrm{d}^4 p}{(2\pi)^4}
  \frac{e^{\mathrm{i}px}}{m_a^2-p^2+\mathrm{i}\epsilon p^0}
  \notag
  \\
  = &
  \frac{1}{2\pi}\theta(t)
  \left\{
    \delta(s^2)-\theta(s^2)\frac{m_a}{2s}J_1(m_a s)
  \right\},
\end{align}
where $\theta(t)$ is the Heaviside step function, $J_1(z)$ is the
first order Bessel function, $s=\sqrt{t^2-r^2}$ and
$r=|\mathbf{r}|$.

To proceed in further calculations it is necessary to define the
behavior of the external sources. We assume that the sources are
localized in space and emit harmonic radiation,
\begin{equation}\label{spinflavsour}
  \ell_\lambda(\mathbf{r},t) J_\lambda^\mu(\mathbf{r},t) =
  \theta(t) l_\lambda J_\lambda^{(0)\mu}
  e^{-\mathrm{i}E t}\delta^3(\mathbf{r}),
\end{equation}
where $l_\lambda=\ell^{(0)}_\lambda$ is the time independent
component of the spinor $\ell_\lambda$, $J_\lambda^{(0)\mu}$ is
the amplitude of the function $J_\lambda^\mu(\mathbf{r},t)$, and
$E$ is the frequency of the source. Using
Eqs.~\eqref{spinmasssour} and~\eqref{spinflavsour} we obtain for
$\xi_a$,
\begin{align}\label{spinamplchi}
  \xi_a(\mathbf{r},t) = & \theta(t)\xi_a^{(0)}
  e^{-\mathrm{i}E t}\delta^3(\mathbf{r}),
  \notag
  \\
  \xi_a^{(0)} = &
  \sum_\lambda U_{a \lambda}^\dag \gamma_\mu^\mathrm{L}
  l_\lambda J_\lambda^{(0)\mu}.
\end{align}
With help of Eqs.~\eqref{spinretGfun}-\eqref{spinamplchi} we can
rewrite Eq.~\eqref{spinmasseigensol1} in the following way:
\begin{align}\label{psiRet}
  \psi_a(\mathbf{r},t) = &
  \left(
    \mathrm{i}\gamma^\mu\frac{\partial}{\partial x^\mu}+m_a
  \right)
  \notag
  \\
  & \times
  \left\{
    e^{-\mathrm{i}E t}
    \int_0^t\mathrm{d}\tau
    D^\mathrm{ret}_a(\mathbf{r},\tau)e^{\mathrm{i}E\tau}
  \right\}
  \xi_a^{(0)}.
\end{align}

One can see that two major terms appear in Eq.~\eqref{psiRet} [see
also Eq.~\eqref{retGfunScal}] while integrating over $\tau$
\begin{equation}
  \int_0^t\mathrm{d}\tau
  D^\mathrm{ret}_a(\mathbf{r},\tau)e^{\mathrm{i}E\tau}=
  \mathcal{I}_1+\mathcal{I}_2.
\end{equation}
They are
\begin{align}
  \label{int1}
  \mathcal{I}_1 & =\frac{1}{2\pi}\int_0^t\mathrm{d}\tau
  \delta(s^2)e^{\mathrm{i}E\tau} =
  \frac{\theta(t-r)}{4\pi r}e^{\mathrm{i}E r},
  \\
  \label{int2}
  \mathcal{I}_2 & = -\frac{m_a}{4\pi}\int_0^t\mathrm{d}\tau
  \theta(s^2)\frac{J_1(m_a s)}{s}e^{\mathrm{i}E\tau}
  \notag
  \\
  & =
  -\frac{m_a}{4\pi}\theta(t-r)\int_0^{x_m}\mathrm{d}x
  \frac{J_1(m_a x)}{\sqrt{r^2+x^2}}e^{\mathrm{i}E\sqrt{r^2+x^2}},
\end{align}
where $x_m=\sqrt{t^2-r^2}$ and $s=\sqrt{\tau^2-r^2}$. It is
interesting to notice that both $\mathcal{I}_1$ and
$\mathcal{I}_2$ in Eqs.~\eqref{int1} and~\eqref{int2} are equal to
zero if $t<r$. It means that the initial perturbation from a
source placed at the point $\mathbf{r}=0$ reaches a detector
placed at the point $\mathbf{r}$ only after the time $t>r$.

Despite the integral $\mathcal{I}_1$ in Eq.~\eqref{int1} is
expressed in terms of the elementary functions, the integral
$\mathcal{I}_2$ in Eq.~\eqref{int2} cannot be computed
analytically for arbitrary $r$ and $t$. However some reasonable
assumptions can be made to simplify the considered expression.
Suppose that an observer is at the fixed distance from a source.
As we have already mentioned one detects a signal starting from
$t>r$. It is obvious that a nonstationary rapidly oscillating
signal is detected when the wave front just arrives to a detector,
i.e. when $t \gtrsim r$. The situation is analogous to waves
propagating on the water surface. Therefore, if we suppose that
one starts observing particles when the non-stationary signal
attenuates, i.e. $t\gg r$ or $x_m\to\infty$, we can avoid
relaxation phenomena. In this case the integral $\mathcal{I}_2$
can be computed analytically,
\begin{equation}\label{int2fin}
  \mathcal{I}_2= - \frac{1}{4\pi r}
  (e^{\mathrm{i}E r}-e^{\mathrm{i}p_a r}),
\end{equation}
where $p_a=\sqrt{E^2-m_a^2}$ is the analog of the particle
momentum.

To obtain Eq.~\eqref{int2fin} we use the known values of the
integrals,
\begin{align}
  \int_0^\infty \mathrm{d}x &
  \frac{J_\nu(m x)}{\sqrt{r^2+x^2}}\sin(E\sqrt{r^2+x^2})
  \notag
  \\
  = &
  \frac{\pi}{2}
  J_{\nu/2}
  \left[
    \frac{r}{2}
    \left(
      E-\sqrt{E^2-m^2}
    \right)
  \right]
  \notag
  \\
  & \times
  J_{-\nu/2}
  \left[
    \frac{r}{2}
    \left(
      E+\sqrt{E^2-m^2}
    \right)
  \right],
  \notag
  \\
  \int_0^\infty \mathrm{d}x &
  \frac{J_\nu(m x)}{\sqrt{r^2+x^2}}\cos(E\sqrt{r^2+x^2})
  \notag
  \\
  = &
  -\frac{\pi}{2}
  J_{\nu/2}
  \left[
    \frac{r}{2}
    \left(
      E-\sqrt{E^2-m^2}
    \right)
  \right]
  \notag
  \\
  & \times
  N_{-\nu/2}
  \left[
    \frac{r}{2}
    \left(
      E+\sqrt{E^2-m^2}
    \right)
  \right],
\end{align}
and the fact that the Bessel and Neumann functions of $\pm 1/2$
order
\begin{align}\label{J12}
  J_{1/2}(z)= & N_{-1/2}(z)= \sqrt{\frac{2}{\pi z}}\sin z,
  \notag
  \\
  J_{-1/2}(z)= & \sqrt{\frac{2}{\pi z}}\cos z,
\end{align}
are expressed in terms of the elementary functions. The
approximations made in derivation of Eq.~\eqref{int2fin} are
analysed in Appendix~\ref{ERROR}.

Using Eqs.~\eqref{psiRet}-\eqref{J12} we obtain the field
distribution of the fermion $\psi_a$ in the following form:
\begin{equation}\label{spinmasseigensol2}
  \psi_a(\mathbf{r},t)=
  e^{-\mathrm{i}E t+\mathrm{i}p_a r} O_a
  \frac{\xi_a^{(0)}}{4\pi r},
\end{equation}
where $O_a=\gamma^0E-(\bm{\gamma}\mathbf{n})p_a+m_a$ and
$\mathbf{n}$ is the unit vector towards a detector. It should be
noted that in deriving Eq.~\eqref{spinmasseigensol2} we
differentiate only exponential rather than the factor $1/r$
because the derivative of $1/r$ is proportional to $1/r^2$. Such a
term is negligible at large distances from a source. We also
remind that Eq.~\eqref{spinmasseigensol2} is valid for $t\gg r$.

Let us turn to the description of the evolution of the fields
$\nu_\lambda$. Using Eqs.~\eqref{spinnupsirel}
and~\eqref{spinmasseigensol2}, we obtain the corresponding wave
function,
\begin{equation}\label{spinflavfin}
  \nu_\lambda(\mathbf{r},t)=
  e^{-\mathrm{i}E t}
  \sum_{a\lambda'}
  U_{\lambda a} U_{a\lambda'}^\dag
  e^{\mathrm{i}p_a r}
  O_a\gamma_\mu^\mathrm{L} l_{\lambda'}
  \frac{J_{\lambda'}^{(0)\mu}}{4\pi r}.
\end{equation}

As in Secs.~\ref{VACUUM}-\ref{MATTERB} we study the evolution of
only two fermions. The mixing matrix is given in
Eq.~\eqref{Ulambdaa}. We also choose the amplitudes of the sources
in the following way: $J_\alpha^{(0)\mu}=0$,
$J_\beta^{(0)\mu}\equiv J^\mu \neq 0$ and $l_\beta\equiv l$. This
choice of the sources corresponds to the emission of neutrinos of
the flavor ``$\beta$" and detection of the neutrinos belonging to
the flavor ``$\alpha$" at the distance $r$ from a source. Using
Eqs.~\eqref{Ulambdaa} and~\eqref{spinflavfin} we obtain the wave
functions of each of the particles $\nu_{\alpha,\beta}$,
\begin{align}\label{spinnu12wf}
  \nu_\alpha(\mathbf{r},t)= &
  \sin\theta\cos\theta
  e^{-\mathrm{i}E t}
  \frac{J^\mu}{4\pi r}
  \notag
  \\
  & \times
  (e^{\mathrm{i}p_1 r}O_1-e^{\mathrm{i}p_2 r}O_2)
  \gamma_\mu^\mathrm{L} l,
  \notag
  \\
  \nu_\beta(\mathbf{r},t)= &
  e^{-\mathrm{i}E t}
  \frac{J^\mu}{4\pi r}
  \notag
  \\
  & \times
  (\sin^2\theta e^{\mathrm{i}p_1 r}O_1+\cos^2\theta e^{\mathrm{i}p_2 r}O_2)
  \gamma_\mu^\mathrm{L} l.
\end{align}
In the following we discuss the high frequency approximation,
$E\gg m_{1,2}$, which corresponds to the emission of
ultrarelativistic neutrinos.

The probability to detect a neutrino of the flavor ``$\lambda$"
can be calculated as $P_\lambda(\mathbf{r},t) =
|\nu_\lambda(\mathbf{r},t)|^2$. Finally using
Eq.~\eqref{spinnu12wf} we get for the probabilities,
\begin{align}\label{spinint1}
  P_\alpha(r)= & -2E^2
  \frac{J^{\dag\mu} J^\nu}{(4\pi r)^2}
  \langle T_{\mu\nu} \rangle
  \notag
  \\
  & \times
  \left\{
    \sin^2(2\theta)\sin^2
    \left(
      \frac{\delta m^2}{4E}r
    \right)+
    \mathcal{O}
    \left(
      \frac{m_a^2}{E^2}
    \right)
  \right\},
  \notag
  \\
  P_\beta(r)= & -2E^2
  \frac{J^{\dag\mu} J^\nu}{(4\pi r)^2}
  \langle T_{\mu\nu} \rangle
  \notag
  \\
  & \times
  \bigg\{
    1-\sin^2(2\theta)\sin^2
    \left(
      \frac{\delta m^2}{4E}r
    \right)
    \notag
    \\
    & +
    \mathcal{O}
    \left(
      \frac{m_a^2}{E^2}
    \right)
  \bigg\},
\end{align}
where $\langle T_{\mu\nu} \rangle=
l^\dag\gamma_\mu^\dag(\bm{\alpha}\mathbf{n})\gamma_\nu l$. It is
possible to calculate the components of the tensor $\langle
T_{\mu\nu} \rangle$. They depend on the properties of the fermion
$\ell_\beta\equiv\ell$,
\begin{gather}
  \langle T_{00} \rangle=-(\mathbf{v}\mathbf{n}),
  \quad
  \langle T_{0i} \rangle=\langle T_{i0} \rangle^{*}=n_i-
  \mathrm{i}[\mathbf{n}\times\bm{\zeta}]_i,
  \notag
  \\
  \label{Tmunucomp}
  \langle T_{ij} \rangle =
  \delta_{ij}(\mathbf{v}\mathbf{n})-
  (v_i n_j + v_j n_i) +
  \mathrm{i}\varepsilon_{ijk}n_k,
\end{gather}
where $\mathbf{v}=\langle \bm{\alpha} \rangle$ is the velocity of
fermion $\ell$ and $\bm{\zeta}=\langle \bm{\Sigma} \rangle$ is its
spin.

Let us discuss the simplified case when the spatial components of
the four-vector $J^\mu$ are equal to zero: $\mathbf{J}=0$. It
corresponds to the neutrino emission by a nonmoving and
unpolarized source. Using Eqs.~\eqref{spinint1}
and~\eqref{Tmunucomp} we can find the transition and survival
probabilities in the following way:
\begin{align}
  \label{tranprspin}
  P_{\beta \to \alpha}(r) \sim &
  \sin^2(2\theta)\sin^2
  \left(
    \frac{\delta m^2}{4E}r
  \right),
  \\
  \label{survprspin}
  P_{\beta \to \beta}(r) \sim &
  1-\sin^2(2\theta)\sin^2
  \left(
    \frac{\delta m^2}{4E}r
  \right),
\end{align}
where we drop the common factor $ 2 E^2 (\mathbf{v}\mathbf{n})
|J^0|^2 / (4\pi r)^2$ to have the probabilities normalized to
unity. It should be noted that Eqs.~\eqref{tranprspin}
and~\eqref{survprspin} are the same as the common formulae for the
description of neutrino flavor oscillations in
vacuum~\cite{Kob80,GriPon69}.

\subsection{General mass term\label{GMT}}

When both Dirac and Majorana mass matrices are present in
Eq.~\eqref{LagrnuvacNHE}, left and right handed chiral components
should be transformed independently (see also Ref.~\cite{Kob80}),
\begin{equation}\label{spinnupsirelM}
  \nu_\lambda^\mathrm{L} =
  \sum_a
  U_{\lambda a} \Psi_a^\mathrm{L},
  \quad
  \nu_\lambda^\mathrm{R} =
  \sum_a
  V_{\lambda a} \Psi_a^\mathrm{R},
\end{equation}
with help of the matrices $(U_{\lambda a})$ and $(V_{\lambda a})$
to diagonalize the general mass term. Note that these matrices are
rectangular and nonhermitian. The modern parameterization for
these matrices is given in Ref.~\cite{Xin08}.

However one cannot say whether particles $\Psi_a^\mathrm{L,R}$ are
Majorana or Dirac although they correspond to definite mass
eigenstates. That is why we introduce a new field $\psi_a =
\Psi_a^\mathrm{L} + (\Psi_a^\mathrm{L})^c$ which is Majorana by
definition.

The Lagrangian for the particles $\psi_a$ takes the form,
\begin{align}\label{spinmassLargM}
  \mathcal{L} = &
  \sum_a
  \Big(
    \overline{\psi_a^\mathrm{L}}
    \mathrm{i}\gamma^\mu\partial_\mu \psi_a^\mathrm{L} -
    m_a \overline{\psi_a^\mathrm{L}} \psi_a^\mathrm{R}
    \notag
    \\
    & +
    \overline{\psi_a^\mathrm{L}} \xi_a^\mathrm{R} + \text{h.c.}
  \Big),
\end{align}
where the external sources $\xi_a \equiv \xi_a^\mathrm{R}$ have
the same form as in Eq.~\eqref{spinmasssour}.

As we mentioned in Sec.~\ref{MAJVACUUM}, Majorana spinors are
equivalent to two component Weil spinors~\cite{Cas57}. Hence we
can rewrite the spinors $\psi_a^\mathrm{L,R}$ and $\xi_a$ as
\begin{equation}
  \psi_a^\mathrm{L} =
  \begin{pmatrix}
    0 \\
    \eta_a
  \end{pmatrix},
  \quad
  \psi_a^\mathrm{R} =
  \begin{pmatrix}
    \mathrm{i}\sigma_2\eta_a^{*{}} \\
    0
  \end{pmatrix},
  \quad
  \xi_a =
  \begin{pmatrix}
    \phi_a \\
    0
  \end{pmatrix},
\end{equation}
where
\begin{equation}\label{phisour}
  \phi_a=
  \sum_\lambda
  U_{a \lambda}^\dag
  \chi_\lambda J_\lambda^0.
\end{equation}
To obtain Eq.~\eqref{phisour} we suppose that, as in
Sec.~\ref{DMT}, the vectors $J_\lambda^\mu$ have only time
component $J_\lambda^0$. We also assume that $\left(
\ell^\mathrm{L}_\lambda \right)^\mathrm{T}=(0,\chi_\lambda)$. To
derive Eq.~\eqref{phisour} it is crucial that only left handed
currents interactions are presented in Eq.~\eqref{LagrnuvacNHE}.

It is useful to rewrite the Lagrangian~\eqref{spinmassLargM} in
terms of the two component spinors $\eta_a$ and
$\phi_a$~\cite{FukYan03p292},
\begin{align}\label{Largetaeta*}
  \mathcal{L} = &
  \sum_a
  \mathrm{i} \eta_a^\dag (\partial_t - \bm{\sigma}\bm{\nabla}) \eta_a
  \notag
  \\
  & +
  \sum_a
  \left(
    \frac{\mathrm{i}}{2} m_a \eta_a^\dag \sigma_2 \eta_a^{*{}} +
    \eta_a^\dag \phi_a + \text{h.c.}
  \right).
\end{align}
Using Eq.~\eqref{Largetaeta*} we can receive the wave equation for
two-component spinors,
\begin{equation}\label{inhomDirWeil}
  \left(
    \frac{\partial}{\partial t}-\bm{\sigma}\bm{\nabla}
  \right)
  \eta_a+m_a \sigma_2 \eta_a^{*{}} =
  \mathrm{i} \phi_a.
\end{equation}
The solutions of Eq.~\eqref{inhomDirWeil} have the form (see,
e.g., Ref.~\cite{FukYan03p292}),
\begin{align}
  \label{solWeileta}
  \eta_a(\mathbf{r},t)=\int\mathrm{d}^3\mathbf{r}'\mathrm{d}t'
  \mathcal{S}^\mathrm{ret}_a(\mathbf{r}-\mathbf{r}',t-t')
  \phi_a(\mathbf{r}',t'),
  \\
  \label{solWeileta*}
  \eta_a^{*{}}(\mathbf{r},t)=\int\mathrm{d}^3\mathbf{r}'\mathrm{d}t'
  \mathcal{R}^\mathrm{ret}_a(\mathbf{r}-\mathbf{r}',t-t')
  \phi_a(\mathbf{r}',t'),
\end{align}
where the retarded Green functions are expressed as (see also
Ref.~\cite{FukYan03p292}),
\begin{align}\label{SRretM}
  \mathcal{S}_a^\mathrm{ret}(\mathbf{r},t) = &
  \mathrm{i} \tilde{\sigma}^\mu \partial_\mu D_a^\mathrm{ret}(\mathbf{r},t),
  \notag
  \\
  \mathcal{R}_a^\mathrm{ret}(\mathbf{r},t) = &
  \mathrm{i} m_a \sigma_2 D_a^\mathrm{ret}(\mathbf{r},t).
\end{align}
Here $\partial_\mu = (\partial_t,\bm{\nabla})$,
$\tilde{\sigma}^\mu = (\sigma^0,\bm{\sigma})$, $\sigma^0$ is the
$2 \times 2$ unit matrix, $\bm{\sigma}$ are Pauli matrices, and
$D_a^\mathrm{ret}(\mathbf{r},t)$ is given in
Eq.~\eqref{retGfunScal}. One can check that Eq.~\eqref{solWeileta}
and~\eqref{solWeileta*}, along with the definition of the retarded
Green functions~\eqref{SRretM}, represent the solutions to
Eq.~\eqref{inhomDirWeil} by means of direct substituting.

Let us assume that the sources $\phi_a$ depend on time and spatial
coordinates as in Sec.~\ref{DMT},
\begin{align}\label{Weilsour}
  \phi_a(\mathbf{r},t) = &
  \theta(t) \phi_a^{(0)} e^{-\mathrm{i} E t} \delta^3(\mathbf{r}),
  \notag
  \\
  \phi_a^{(0)} = & \sum_\lambda
  U^\dag_{a \lambda} J_\lambda \chi^{(0)}_\lambda,
\end{align}
where $\chi^{(0)}_\lambda$ is the time independent component of
$\chi_\lambda$. In deriving of Eq.~\eqref{Weilsour} from
Eq.~\eqref{phisour} we introduce the new quantities $J_\lambda
\equiv J_\lambda^{(0)0}$, where $J_\lambda^{(0)0}$ is the time
independent part of $J_\lambda^{0}$, to simplify the notations.

On the basis of Eqs.~\eqref{solWeileta}-\eqref{Weilsour} and using
the technique developed in Sec.~\ref{DMT} we get the particles
wave functions,
\begin{align}
  \label{solWeiletafin}
  \eta_a(\mathbf{r},t)= & \mathrm{i} \tilde{\sigma}^\mu \partial_\mu
  [e^{- \mathrm{i} E t + \mathrm{i} p_a r}]
  \frac{\phi_a^{(0)}}{4 \pi r},
  \\
  \label{solWeileta*fin}
  \eta_a^{*{}}(\mathbf{r},t)= & \mathrm{i} m_a \sigma_2
  e^{- \mathrm{i} E t + \mathrm{i} p_a r}
  \frac{\phi_a^{(0)}}{4 \pi r}.
\end{align}
To derive Eqs.~\eqref{solWeiletafin} and~\eqref{solWeileta*fin} we
suppose that $t \gg r$. The derivatives in
Eq.~\eqref{solWeiletafin} are applied on the exponent only because
of the same reasons as in Sec.~\ref{DMT}.

Using Eqs.~\eqref{spinnupsirelM}, \eqref{solWeiletafin}
and~\eqref{solWeileta*fin} as well as the following identity:
\begin{align}\label{ident}
  \left(
    \nu_\lambda^\mathrm{L}
  \right)^c = &
  \left(
    \nu_\lambda^c
  \right)^\mathrm{R}=
  \sum_a U_{\lambda a}^{*{}}\psi_a^\mathrm{R},
  \notag
  \\
  \left(
    \nu_\lambda^\mathrm{L}
  \right)^c = & C
  \left(
    \overline{\nu_\lambda^\mathrm{L}}
  \right)^\mathrm{T},
\end{align}
we get the wave functions of $\nu_\lambda^\mathrm{L}$ and $\left(
\nu_\lambda^\mathrm{L} \right)^c$ as
\begin{align}
  \label{nulambda}
  \nu_\lambda^\mathrm{L}(\mathbf{r},t) = &
  \frac{2E}{4 \pi r}
  e^{-\mathrm{i}E t}
  \sum_{a \lambda'}
  e^{\mathrm{i} p_a r}
  U_{\lambda a} U_{a \lambda'}^\dag
  \notag
  \\
  & \times
  J_{\lambda'}
  \begin{pmatrix}
    0 \\
    \tilde{\chi}^{(0)}_{\lambda'}
  \end{pmatrix},
  \\
  \label{nulambdac}
  \left(
    \nu_\lambda^\mathrm{L}
  \right)^c(\mathbf{r},t) = &
  -\frac{2E}{4 \pi r}
  e^{-\mathrm{i}E t}
  \sum_{a \lambda'}
  \frac{m_a}{2E} e^{\mathrm{i} p_a r}
  U_{\lambda a}^{*{}} U_{a \lambda'}^\dag
  \notag
  \\
  & \times
  J_{\lambda'}
  \begin{pmatrix}
    \chi^{(0)}_{\lambda'} \\
    0
  \end{pmatrix},
\end{align}
where $\tilde{\chi}^{(0)}_{\lambda'} =
(1/2)[1-(\bm{\sigma}\mathbf{n})] \chi^{(0)}_{\lambda'}$. In
deriving of Eq.~\eqref{nulambda} we suppose that
\begin{equation}
  [E-p_a(\bm{\sigma}\mathbf{n})] \approx E [1-(\bm{\sigma}\mathbf{n})],
\end{equation}
that is valid for relativistic neutrinos.

It is possible to construct two four component Majorana spinors
from two-component Weil spinors,
\begin{equation}\label{psi1psi2}
  \psi_a^{(1)} =
  \begin{pmatrix}
    \mathrm{i} \sigma_2 (\eta_a)^{*{}} \\
    \eta_a
  \end{pmatrix},
  \quad
  \psi_a^{(2)} =
  \begin{pmatrix}
    \mathrm{i} \sigma_2 \eta_a^{*{}} \\
    (\eta_a^{*{}})^{*{}}
  \end{pmatrix},
\end{equation}
where $\eta_a$ and $\eta_a^{*{}}$ are defined in
Eqs.~\eqref{solWeiletafin} and~\eqref{solWeileta*fin}. One can see
that these spinors satisfy the Majorana condition $\left[
\psi_a^{(1,2)} \right]^c = \psi_a^{(1,2)}$. We use the spinor
$\psi_a^{(1)}$ to receive Eq.~\eqref{nulambda} and $\psi_a^{(2)}$
for Eq.~\eqref{nulambdac}. In this case we get that
$\nu_\lambda^\mathrm{L}$ and $\left( \nu_\lambda^\mathrm{L}
\right)^c$ are obtained as a result of the evolution of particles
emitted from the same source.

It should be mentioned that both $\nu_\lambda^\mathrm{L}$ and
$\left( \nu_\lambda^\mathrm{L} \right)^c$ in Eqs.~\eqref{nulambda}
and~\eqref{nulambdac} propagate forward in time. To explain this
fact let us discuss the complex conjugated
equation~\eqref{solWeileta}. Performing the same computations one
arrives to the analog of Eq.~\eqref{nulambdac} which would depend
on time as $e^{\mathrm{i}E t}$. However this wave function
describes a particle emitted by the source different from that
discussed here. Indeed, if we studied the complex conjugated
Eq.~\eqref{solWeileta}, the integrand there would be
$\mathcal{S}^\mathrm{ret*{}}_a \phi_a^{*{}}$. It would mean that
the source in Eq.~\eqref{Weilsour} would be proportional to
$\phi_a^{(0)*{}}e^{\mathrm{i} E t}$, that, in its turn, would
signify that $\left( \nu_\lambda^\mathrm{L} \right)^c$ would be
emitted in a process involving a lepton which is a charge
conjugated counterpart to that discussed here.

Let us illustrate this problem on a more physical example. Suppose
that we study a \emph{neutrino} emission in a process like the
inverse $\beta$-decay,
\begin{equation}\label{exnuem}
  p + \ell_\lambda^{-{}} \to n + \nu_\lambda \Rightarrow
  n + \sum_a U_{\lambda a}\psi_a^\mathrm{L},
\end{equation}
where $p$, $n$ and $\ell_\lambda^{-{}}$ stand for a proton, a
neutron and for a negatively charged lepton. The complex
conjugated Eq.~\eqref{solWeileta} would correspond to a process,
\begin{align}\label{exantinuem}
  n + \ell_\lambda^{+{}} \to p + \tilde{\nu}_\lambda
  \Rightarrow &
  p + \sum_a U_{\lambda a}^{*{}}
  \left(
    \psi_a^\mathrm{L}
  \right)^c
  \notag
  \\
  & \Rightarrow
  p + \sum_a U_{\lambda a}^{*{}}
  \psi_a^\mathrm{R},
\end{align}
where $\tilde{\nu}_\lambda$ and $\ell_\lambda^{+{}}$ denote an
\emph{antineutrino} and a positively charged lepton, which is the
charge conjugated counterpart to $\ell_\lambda^{-{}}$. In
Eq.~\eqref{exantinuem} we use the facts that only left handed
interactions exist in nature and the fields $\psi_a$ describe
Majorana particles. As one can see, Eqs.~\eqref{exnuem}
and~\eqref{exantinuem} represent two different processes. Hence,
if we used $(\eta_a)^{*{}}$ to obtain $\left(
\nu_\lambda^\mathrm{L} \right)^c(\mathbf{r},t)$, i.e. after a beam
of neutrinos passes some distance $r$, it would correspond to the
initial reaction~\eqref{exantinuem} rather than~\eqref{exnuem}.
Note that the same result also follows from Eq.~\eqref{psi1psi2}
if we replace $\psi_a^{(1)} \leftrightarrow \psi_a^{(2)}$ there.

Let us suppose for simplicity that the momentum of the fermion
$\ell_\lambda$ is parallel to the neutrino momentum. It takes
place if a relativistic incoming lepton is studied. We also assume
that this fermion is in a state with the definite helicity,
\begin{equation}\label{helstatefer}
  \frac{1}{2}[1-(\bm{\sigma}\mathbf{n})]\chi^{(0)}_\lambda =
  \chi^{(0)}_\lambda.
\end{equation}
This expression is again natural for the relativistic fermion
$\ell_\lambda$. One can notice that the
Lagrangian~\eqref{LagrnuvacNHE} is written in terms of the left
handed chiral projections of $\ell_\lambda$. Therefore, if we
study a relativistic lepton, it will have its spin directed
oppositely to the particle momentum as one can see from
Eq.~\eqref{helstatefer}.

Using Eqs.~\eqref{nulambda}, \eqref{nulambdac}
and~\eqref{helstatefer} as well as the orthonormality of the
two-component spinors $\chi^{(0)}_\lambda$, $\left(
\chi^{(0)\dag}_\lambda \chi^{(0)}_{\lambda'} \right) =
\delta_{\lambda\lambda'}$, we get the probabilities to detect
$\nu_\lambda^\mathrm{L}$ and $\left( \nu_\lambda^\mathrm{L}
\right)^c$ as
\begin{align}
  \label{Pnulambda}
  P_{\nu_\lambda^\mathrm{L}}(r) \sim &
  \sum_{ab, \lambda'}
  e^{\mathrm{i} (p_a-p_b) r}
  \notag
  \\
  & \times
  U_{\lambda a} U_{a \lambda'}^\dag U_{b \lambda}^\dag U_{\lambda' b}
  |J_{\lambda'}|^2,
  \\
  \label{Pnulambdac}
  P_{\left( \nu_\lambda^\mathrm{L} \right)^c}(r) \sim &
  \sum_{ab, \lambda'}
  \frac{m_a m_b}{(2E)^2}
  e^{\mathrm{i} (p_a-p_b) r}
  \notag
  \\
  & \times
  U_{\lambda a}^{*{}} U_{a \lambda'}^\dag
  U_{b \lambda}^\mathrm{T} U_{\lambda' b}
  |J_{\lambda'}|^2.
\end{align}
In Eqs.~\eqref{Pnulambda} and~\eqref{Pnulambdac} we drop the
factor $(2E)^2/(4 \pi r)^2$.

We can see that Eqs.~\eqref{Pnulambda} and~\eqref{Pnulambdac}
contain the oscillating exponent. This our result reproduces the
usual formulae for neutrino oscillations in vacuum~\cite{Kob80}.
It should be also noticed that the expressions for the
probabilities depend on $r$ rather than on $t$ in contrast to our
previous
works~\cite{Dvo05,Dvo06,Dvo08JPCS,DvoMaa07,Dvo10,Dvo08JPG,DvoMaa09}.
Note that the problem whether neutrino oscillations happen in
space or in time was also discussed in Ref.~\cite{AkhSmi09}. The
similar coordinate dependence of the probabilities was obtained in
Refs.~\cite{Kob82,GiuKimLeeLee93} where the problem of neutrino
oscillations in vacuum was studied.

It was mentioned in Ref.~\cite{Kob82} that oscillations between
active and sterile neutrinos are possible in case of the
nonunitary matrix $(U_{\lambda a})$, i.e. the presence of only
Majorana mass terms is not sufficient for the existence of this
kind of transitions. The situation is analogous to that considered
in the pioneering work~\cite{PontecorvoOsc} where oscillations
between neutrinos and antineutrinos were studied. In
Eq.~\eqref{Pnulambdac} we obtain that the probability to detect
$\left( \nu_\lambda^\mathrm{L} \right)^c$ is suppressed by the
factor $m_a m_b/E^2$. It is in agreement with the results of
Refs.~\cite{Kob82,SchVal81v23} as well as with
Eq.~\eqref{PtrnuanuMajVac}.

\section{Quantum field theory description of neutrino oscillations
in background matter\label{NU2ANTINU}}

In the present section we study neutrino oscillations in
background matter (see Sec.~\ref{MATTER}) in frames of the quantum
field theory approach. In particular we will be interested to
examine the influence of background matter on the transitions
between neutrinos and antineutrinos discussed in
Secs.~\ref{MAJVACUUM} and~\ref{GMT}.

Despite we could receive satisfactory results for the description
of neutrino-to-antineutrino transitions in vacuum in frames of the
relativistic quantum mechanics approach [see
Eqs.~\eqref{PtrnuanuMajVac} and~\eqref{Pnulambdac}] the accurate
treatment of this process has to be done within the quantum field
theory because of the following reason. For the case of Majorana
neutrinos, particles are identical to their antiparticles and only
in frames of the quantum field theory one has the most accurate
description of antiparticle states. Moreover, as we have seen in
Sec.~\ref{MAJVACUUM}, ``antineutrino" states appear along with the
small factor $m_a/E_\nu$. Typically various approaches to the
description of neutrino oscillations give contradictory results at
this order of accuracy~\cite{Beu02}. That is why we will use the
quantum field theory treatment to capture such a tiny effect.

As in Sec.~\ref{INHOMOGEN}, we will study the system of flavor
neutrinos $\nu_\lambda$, $\lambda = e,\mu,\tau,\dotsc$,
propagating in dense background matter between two spatial points,
$\mathbf{x}_1$ and $\mathbf{x}_2$. We suggest that emission and
absorption of neutrinos is due to the interaction with leptons
$l_\lambda^{\pm{}}$ and heavy nucleons $N$. i.e these interactions
are localized in two spatial regions: a ``source" and a
``detector". On the contrary, the neutrino interaction with
background matter is uniformly distributed along the total
neutrino propagation distance $\mathbf{L} = \mathbf{x}_2 -
\mathbf{x}_1$.

In general case the dynamics of the system of mixed massive
neutrinos should be formulated in frames of the left
$\nu_\lambda^\mathrm{L}$ and right $\nu_\lambda^\mathrm{R}$ handed
projections of flavor neutrinos [see also
Eq.~\eqref{LagrnuvacNHE}],
\begin{align}\label{Lagrnu2antinu}
  \mathcal{L} = &
  \sum_\lambda
  \left(
    \overline{\nu_\lambda^\mathrm{L}}
    \mathrm{i} \gamma^\mu \partial_\mu
    \nu_\lambda^\mathrm{L} +
    \overline{\nu_\lambda^\mathrm{R}}
    \mathrm{i} \gamma^\mu \partial_\mu
    \nu_\lambda^\mathrm{R}
  \right)
  \notag
  \\
  & -
  \sum_{\lambda\lambda'}
  \Big(
    m_{\lambda\lambda'}^\mathrm{D}
    \overline{\nu_\lambda^\mathrm{L}}
    \nu_{\lambda'}^\mathrm{R} +
    m_{\lambda\lambda'}^\mathrm{L}
    \left(
      \nu_\lambda^\mathrm{L}
    \right)^\mathrm{T}
    C \nu_{\lambda'}^\mathrm{L}
    \notag
    \\
    & +
    m_{\lambda\lambda'}^\mathrm{R}
    \left(
      \nu_\lambda^\mathrm{R}
    \right)^\mathrm{T}
    C \nu_{\lambda'}^\mathrm{R}
    + \text{h.c.}
  \Big)
  \notag
  \\
  & -
  \sum_\lambda
  \overline{\nu_\lambda^\mathrm{L}}
  \gamma_\mu
  \nu_{\lambda'}^\mathrm{L} f_{\lambda\lambda'}^\mu -
  \sqrt{2}G_\mathrm{F}
  \left(
    j^\mu J_\mu + \text{h.c.}
  \right),
\end{align}
where $m_{\lambda\lambda'}^\mathrm{D}$ and
$m_{\lambda\lambda'}^\mathrm{L,R}$ are Dirac and Majorana mass
matrices defined in Sec.~\ref{INHOMOGEN}, and
\begin{equation}
  j^\mu = \sum_\lambda
  \overline{\nu_\lambda^\mathrm{L}} \gamma^\mu
  l_\lambda^\mathrm{L},
\end{equation}
is the neutrino-lepton current. The
effective potential of the neutrino interaction with background
matter $f_{\lambda\lambda'}^\mu$ [see Eq.~\eqref{Lagrnumatt}] is
supposed to be coordinate independent, whereas the nuclear current
$J_\mu$ is localized in space. We will take into account the
interaction with background matter exactly. As in
Sec.~\ref{MATTER} here we study the general case and consider the
matrix $(f_{\lambda\lambda'}^\mu)$ to be nondiagonal. Note that in
Eq.~\eqref{Lagrnu2antinu} right handed neutrinos do not
participate in interactions with other particles, i.e. they are
sterile.

\begin{figure}
  \centering
  \includegraphics[scale=.8]{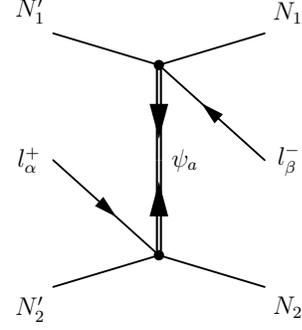}
  \caption{\label{nu2antinuFD}
  The Feynmann diagram corresponding to the lepton-nuclei
  scattering, $(N_1,N_2) + l_\beta^{-{}} \to
  (\text{neutrinos}) \to (N_1',N_2') + l_\alpha^{+{}}$, with the $S$-matrix
  element given in Eq.~\eqref{Smatrixnu}.}
\end{figure}

As we will see below, the transitions between neutrinos and
antineutrinos manifest in the process like $(N_1,N_2) +
l_\beta^{-{}} \to (\text{neutrinos}) \to (N_1',N_2') +
l_\alpha^{+{}}$. It should be noted that in this process massive
neutrinos appear as virtual particles rather than show up
explicitly~\cite{Kob82,GiuKimLeeLee93}. The $S$-matrix element for
this kind of reaction has the form~\cite{Kob82},
\begin{align}\label{Smatrixnu}
  S = & - \frac{1}{2!}
  \left(
    \sqrt{2}G_\mathrm{F}
  \right)^2
  \int \mathrm{d}^4 x \mathrm{d}^4 y
  \\
  \notag
  & \times
  \langle
    l_\alpha^{+{}}; N_1' N_2'
    \left|
      T \{ j^\mu(x) J_\mu(x) j^\nu(y) J_\nu(y) \}
    \right|
    l_\beta^{-{}}; N_1 N_2
  \rangle,
\end{align}
%
where $N_{1,2}$ and $N_{1,2}'$ are states of the initial and final
nucleons. This process is schematically illustrated in
Fig~\ref{nu2antinuFD}.
Note that the propagation of internal
neutrinos is shown by a broad line which means that it accounts
for the interaction with background matter exactly.

Now we have to diagonalize the mass matrices
$m_{\lambda\lambda'}^\mathrm{D}$ and
$m_{\lambda\lambda'}^\mathrm{L,R}$ in Eq.~\eqref{Lagrnu2antinu} by
means of the matrix transformations~\eqref{spinnupsirelM}, with
the rectangular matrices $(U_{\lambda a})$ and $(V_{\lambda a})$
being nonunitary and independent.
%
%
The amplitude of the process shown in Fig.~\ref{nu2antinuFD} is
nonzero if virtual neutrinos correspond to Majorana fields. That
is why we introduce a new Majorana field $\psi_a =
\Psi_a^\mathrm{L} + (\Psi_a^\mathrm{L})^c$ as in Sec.~\ref{GMT}.

Let us rewrite the Lagrangian~\eqref{Lagrnu2antinu} in terms of
the Majorana fields $\psi_a$,
\begin{align}\label{Lagrpsi2antipsi}
  \mathcal{L} = &
  \sum_a
  \bar{\psi}_a
  (\mathrm{i} \gamma^\mu \partial_\mu - m_a)
  \psi_a -
  \notag
  \\
  & -
  \sum_{ab}
  g_{ab}^\mu \overline{\psi_a^\mathrm{L}} \gamma_\mu \psi_b^\mathrm{L} -
  (\text{sources}),
\end{align}
where $(g_{ab}^\mu)$ is the matrix of the effective potentials of
the neutrino interaction with matter in the mass eigenstates basis
which is defined in Eq.~\eqref{gmatrmatt}.
In Eq.~\eqref{Lagrpsi2antipsi} we do not present the expression
for the sources in the explicit form since they are supposed to be
localized in space.

With help of Eq.~\eqref{spinnupsirelM} the $T$-product of the
neutrino-lepton currents in Eq.~\eqref{Smatrixnu} takes the form,
\begin{widetext}
\begin{equation}\label{Tjj}
  \langle
    l_\alpha^{+{}}
    \left|
      T \{ j_\mu(x) j_\nu(y) \}
    \right|
    l_\beta^{-{}}
  \rangle = 
  \frac{e^{\mathrm{i} p_\alpha x - \mathrm{i} p_\beta y}}
  {2\mathcal{V}\sqrt{E_\alpha E_\beta}}
  \sum_{ab}
  U_{\alpha a}^{*{}} U_{\beta b}^{*{}}
  u^\mathrm{T}(-p_\alpha)
  \left(
    \gamma_\mu^\mathrm{L}
  \right)^\mathrm{T}
  \langle
    0 | T \{
    \left[
      \bar{\psi}_a(x)
    \right]^\mathrm{T} \bar{\psi}_b(y) \} | 0
  \rangle
  \gamma_\nu^\mathrm{L}
  u(p_\beta),
\end{equation}
\end{widetext}
where $u(p_{\alpha,\beta})$ are the outgoing and incoming spinors
with four momenta $p_{\alpha,\beta} =
(E_{\alpha,\beta},\mathbf{p}_{\alpha,\beta})$ and $\mathcal{V}$ is
the normalization volume. In Eq.~\eqref{Tjj} we use the
representation of lepton states in the form of a plane wave from
Ref.~\cite{BerLifPit89}.

Since the mass of the nucleons is much bigger that the energies of
leptons, we can suggest that they are at rest, i.e. we replace the
nucleon currents
\begin{equation}
  J_\mu(x) = \delta_{0\mu} \delta(\mathbf{x}-\mathbf{x}_2),
  \quad
  J_\mu(x) = \delta_{0\mu} \delta(\mathbf{y}-\mathbf{x}_1),
\end{equation}
in Eq.~\eqref{Smatrixnu}.

The $T$-product of Majorana neutrino fields in Eq.~\eqref{Tjj} can
be rewritten as (see, e.g., Ref.~\cite{FukYan03p292})
\begin{equation}\label{Majnuprop}
  \langle
    0 | T \{
    \left[
      \bar{\psi}_a(x)
    \right]^\mathrm{T} \bar{\psi}_b(y) \} | 0
  \rangle =
  - C S_{ab}(x-y),
\end{equation}
where $C$ is the charge conjugation matrix and $S_{ab}(x-y) =
\langle 0 | T \{ \psi_a(x) \bar{\psi}_b(y) \} | 0 \rangle$ is the
usual Feynmann propagator corresponding to a Dirac particle.

To simplify the calculations we suggest that the matrix
$(g_{ab}^\mu)$ in Eq.~\eqref{Lagrpsi2antipsi} is close to
diagonal. Under this assumption the propagator in
Eq.~\eqref{Majnuprop} has the diagonal form: $S_{ab}(x) =
\delta_{ab} S_a(x)$. The explicit form of the function $S_a(x)$ is
given in Appendix~\ref{NU2ANTINUINT}. Moreover, as in
Secs.~\ref{MATTER} and~\ref{MATTERB}, we will study the situation
of nonmoving and unpolarized matter, with matrix $(g_{ab}^\mu)$
having only the zeroth component.

Using the results of Appendix~\ref{NU2ANTINUINT} we can represent
the $S$-matrix element in Eq.~\eqref{Smatrixnu} as
\begin{align}\label{Smatrixfinal}
  S = & 2 \pi \delta(E_\alpha - E_\beta)
  \frac{G_\mathrm{F}^2}{4\mathcal{V}\sqrt{E_\alpha E_\beta}}
  M_{\beta\alpha},
  \notag
  \\
  M_{\beta\alpha} = & - \frac{e^{\mathrm{i} \mathbf{p}_\beta \mathbf{x}_1 -
  \mathrm{i} \mathbf{p}_\alpha \mathbf{x}_2}}{2 \pi L}
  \notag
  \\
  & \times
  \sum_a m_a U_{\alpha a}^{*{}} U_{\beta a}^{*{}}
  \bar{u}(p_\alpha) F_a u(p_\beta),
\end{align}
where
\begin{align}\label{Fafinal}
  F_a = & \frac{1}{2 |g_a| p_a}
  \bigg[
    e^{\mathrm{i}k_1 L}
    \left(
      \frac{g_a^2}{2} + |g_a| p_a - g_a k_1 (\bm{\alpha}\mathbf{n})
    \right)
    \notag
    \\
    & -
    e^{\mathrm{i}k_2 L}
    \left(
      \frac{g_a^2}{2} - |g_a| p_a - g_a k_2 (\bm{\alpha}\mathbf{n})
    \right)
  \bigg]
  \notag
  \\
  & \times
  (1-\gamma^5),
\end{align}
and $k_{1,2}^2 = (p_a \pm |g_a|/2)^2$, $p_a = \sqrt{E^2 - m_a^2}$
is the neutrino ``momentum", $E = E_\alpha = E_\beta$ is the
energy of leptons which is conserved since nucleons are supposed
to be at rest.

Let us discuss the case of high energy leptons with $E_\lambda \gg
m_\lambda$, where $m_\lambda$ is the lepton mass. In this case the
biggest contribution to the matrix element of the total process
arises from the channel in which the helicity of leptons changes
from $-1/2$ to $1/2$. Thus the matrix element in
Eq.~\eqref{Smatrixfinal} takes the form,
\begin{align}\label{SmatrixhighE}
  M(l_\beta^{-{}} \to l_\alpha^{+{}}) \approx &
  - \frac{2 e^{\mathrm{i} \mathbf{p}_\beta \mathbf{x}_1 -
  \mathrm{i} \mathbf{p}_\alpha \mathbf{x}_2}}{\pi L}
  \sum_a E
  m_a U_{\alpha a}^{*{}} U_{\beta a}^{*{}} \langle F_a \rangle
  \notag
  \\
  & \times
  e^{-\mathrm{i}\varphi/2} \sin \frac{\vartheta}{2},
\end{align}
where the spherical angles $\varphi$ and $\vartheta$ fix the
direction of the outgoing lepton momentum with respect to the
incoming lepton momentum, which, for simplicity, is chosen to be
directed along the vector $\mathbf{L}$.

Using Eq.~\eqref{Fafinal} we can calculate the function $F_a$ in
Eq.~\eqref{SmatrixhighE} in the high energy leptons approximation,
%
%
\begin{align}\label{FahighE}
  \langle F_a \rangle = &
  \bar{u}(p_\alpha) F_a u(p_\beta)
  \\
  \notag
  & =
  \begin{cases}
    \displaystyle
    \left( 1-\frac{|g_a|}{2p_a} \right) e^{\mathrm{i}(p_a - |g_a|/2)L},
    & \displaystyle \text{if $p_a > \frac{|g_a|}{2}$}, \\
    0, & \displaystyle \text{if $p_a < \frac{|g_a|}{2}$}.
  \end{cases}
\end{align}
The details of the averaging over the lepton states in
Eqs.~\eqref{SmatrixhighE} and~\eqref{FahighE} can be found in
Ref.~\cite{Kob82}.

Finally the total cross section for the leptons scattering
$(N_1,N_2) + l_\beta^{-{}} \to (\text{neutrinos}) \to (N_1',N_2')
+ l_\alpha^{+{}}$ can be presented in the following form:
\begin{align}\label{sigmahighE}
  \sigma(l_\beta^{-{}} \to l_\alpha^{+{}}) \sim &
  \frac{G_\mathrm{F}^4 E^2}{L^2}
  \sum_{ab} m_a m_b
  U_{\alpha a}^{*{}} U_{\beta a}^{*{}} U_{\alpha b} U_{\beta b}
  \notag
  \\
  & \times
  \langle F_a \rangle \langle F_b^{*{}} \rangle
  \frac{v_\alpha}{v_\beta},
\end{align}
where $v_{\alpha,\beta}$ are the velocities of outgoing and
incoming leptons. If we study the most general
Lagrangian~\eqref{Lagrnu2antinu}, which contains both Dirac and
Majorana mass matrices, Eqs.~\eqref{FahighE}
and~\eqref{sigmahighE} involve transitions between neutrinos and
antineutrinos studied in Refs.~\cite{Kob82,SchVal81v23}  (see also
Secs.~\ref{MAJVACUUM} and~\ref{INHOMOGEN}).

Using Eqs.~\eqref{FahighE} and~\eqref{sigmahighE} one can conclude
that for light neutrinos with $m_a \ll E$ the matter contribution
is negligible since $p_a \approx E - m_a^2/2E \gg |g_a|$. However
the situation is completely different in the case of heavy
neutrinos which are not excluded
in some theoretical models~\cite{Atr09}. If we study the neutrino
propagation in the dense nuclear matter with $|g_a| \sim
\text{eV}$, the ``momentum" of such heavy neutrinos can be
comparable with the effective potential of matter interaction at
the appropriate choice of lepton energy. Nevertheless we can see
in Eq.~\eqref{FahighE} that the contribution of such heavy mass
eigenstates to neutrino-to-antineutrino transitions will be
strongly suppressed by interaction with background matter. Note
that such a behavior of the transition probability in presence of
background matter is in agreement with the results of
Ref.~\cite{EspTan97} where oscillations between neutrinos and
antineutrinos were studied in frames of the standard quantum
mechanical approach.

In the derivation of the main results~\eqref{FahighE}
and~\eqref{sigmahighE} we neglected the contribution of the
nondiagonal elements of the matrix $(g_{ab}^\mu)$. One can expect
that this contribution can be responsible for the enhancement of
the probability of neutrino-to-antineutrino transitions as it can
for neutrino flavor oscillation in matter (see Sec.~\ref{MATTER}).
If one takes into account these nondiagonal elements, the neutrino
propagator in Eq.~\eqref{Majnuprop} also acquires nondiagonal
entries. It is rather difficult to analyze this kind of
``nondiagonal" propagators analytically. Thus, the approach for
the description of neutrino oscillations based on the quantum
field theory does not seem to be applicable for the situation of
big nondiagonal elements of the matrix $(g_{ab}^\mu)$.

\section{Conclusion\label{CONCL}}

In conclusion we mention that in the present work we summarized
our resent achievements in the theoretical description of neutrino
oscillations in vacuum and external fields. Basically we studied
three approaches: relativistic quantum mechanics approach
(Secs.~\ref{VACUUM}-\ref{MATTERB}, \ref{MAJVACUUM},
and~\ref{MAJNUMATTB}), classical external sources method
(Sec.~\ref{INHOMOGEN}), and quantum field theory treatment of
neutrino oscillations (Sec.~\ref{NU2ANTINU}).

The formulation of the initial condition
problem~\eqref{inicondnulambda} for the system of flavor neutrinos
is the main feature of the relativistic quantum mechanics method.
Then one looks for wave functions of flavor neutrinos at
subsequent moments of time. In vacuum the Cauchy problem for
neutrino mass eigenstates~\eqref{matrtranslambda} can be solved
exactly [see Eq.~\eqref{SpecSolpsirvac}] giving us neutrino wave
functions for arbitrary initial condition. When we study the time
evolution of flavor neutrinos in presence of external fields
within this formalism, we can reduce the dynamics of the system to
a Schr\"{o}dinger like equation [see, e.g., Eqs.~\eqref{Schr2QFT}
and~\eqref{Ham2Maj}].

Note that, using this formalism, one solves not only the problem
of the structure of the effective Hamiltonian but also we clearly
point out what kind of space of wave functions this effective
Hamiltonian acts in. In frames of the standard quantum mechanical
approach~\cite{GriPon69,LimMar88} one studies the evolution of a
quantum mechanical neutrino ``wave function" $\nu^\mathrm{T} =
(\nu_\alpha,\nu_\beta,\dotsc)$, but the origin of the components
of this wave function is disguised. Using the relativistic quantum
mechanics method one immediately gets the answer to this
questions. If we take the coefficients $a_a^{(\zeta)}$ of the
decomposition of a four component neutrino wave function over the
basis spinors [see, e.g., Eqs.~\eqref{GenSolDirpsimattB}
and~\eqref{GenSolMajmattB}] and then make a certain
transformation, like in Eqs.~\eqref{PsiPsi'transmattB} and
\eqref{matrtransfPsi}, of these coefficients, we arrive to the
quantum mechanical neutrino ``wave function".

Note that in our approach the coefficients $a_a^{(\zeta)}$ are
$c$-numbers rather than operators as in quantum field theory. It
means that this method may be called as a classical field theory
treatment of neutrino oscillations since we do not use second
quantized neutrino wave functions.

Besides resolving the conceptual problem of neutrino oscillations,
the relativistic quantum mechanics method can be applicable for
the description of neutrino oscillations not only in vacuum
(Sec.~\ref{VACUUM}), but also in background matter
(Sec.~\ref{MATTER}), spin flavor oscillations in an external
magnetic field (Sec.~\ref{B}), and in the combination of
background matter and a magnetic field (Sec.~\ref{MATTERB}). Note
that both Dirac and Majorana neutrinos can be treated within this
formalism (see Secs.~\ref{MAJVACUUM} and~\ref{MAJNUMATTB}).

When we study the evolution of neutrinos in an external field, we
use exact solutions of wave equations in this external field. That
is why our treatment is valid for arbitrary strength of external
fields. Moreover, the dynamics of mixed neutrinos is usually
described in the mass eigenstates basis. In general case, external
fields, like interaction with background matter or with an
external electromagnetic field, are not diagonal in this basis
[see Eqs.~\eqref{gmatrmatt} and~\eqref{magmomme}], i.e. the
neutrino mass eigenstates are not independent in presence of an
external field [see, e.g., Eq.~\eqref{DireqmattB}]. Nevertheless
in frames of the relativistic quantum mechanics method we can
treat this kind of coupled mass eigenstates and receive results
which are consistent with the standard quantum mechanical
approach.

As it was mentioned above, the relativistic quantum mechanics
method is a field theory counterpart of the usual quantum
mechanical description of neutrino oscillation in a sense that it
allows one to describe the evolution of mixed massive neutrinos
from the first principles and thus it throws daylight upon some
unclear issues of the quantum mechanical approach. However, trying
to follow quantum mechanical approach as close as possible, we
also adopted some of its weaknesses. For example, as in the
quantum mechanical description, in frames of our method neutrino
``wave functions" evolve in time rather than in space. It is,
however, a contradiction with the majority of the experiments
where one measures neutrino oscillations in space rather than in
time.

Among other disadvantages of the present method we mention its
limited applicability to description of neutrinos with small
initial momentum in the case of nondiagonal external fields in the
mass eigenstates basis. In this situation the ordinary
differential equations for $a_a^{(\zeta)}(t)$ and
$b_a^{(\zeta)}(t)$ evolution [see, e.g., Eq.~\eqref{ODEDirmatt}]
start to entangle, making possible the transitions between
``positive" and ``negative" energy states. Thus, for example,
instead of $4 \times 4$ effective Schr\"{o}dinger equation for
spin flavor oscillations one gets $8 \times 8$ differential
equation, which is quite difficult for the analysis.

We should also mention the fact that to get the consistency with
the quantum mechanical approach, one should consider rather broad
initial wave packet [see Eq.~\eqref{inicondnulambdaexp}] which is
rather difficult to implement in practice. In Sec~\ref{VACUUM} we
also demonstrated, for the case of neutrino evolution in vacuum,
that spatially limited initial wave packets do not reveal flavor
oscillations since they propagate as massless particles.

As we mentioned above, the relativistic quantum mechanics approach
predicts neutrino oscillations in time rather than in space as
they are observed in experiments. To resolve this contradiction in
Sec.~\ref{INHOMOGEN} we developed a new approach to the
description of neutrino oscillations in vacuum which is based on
the consideration of flavor neutrinos emitted by classical
sources. Within this method one could study neutrino oscillations
in space of both Dirac [Eqs.~\eqref{tranprspin}
and~\eqref{survprspin}] and Majorana [Eqs.~\eqref{Pnulambda}]
neutrinos. For the case of Majorana neutrinos we could also
describe the transitions between neutrinos and antineutrinos [see
Eq.~\eqref{Pnulambdac}]. This our result is also consistent with
the previous studies~\cite{Kob82,SchVal81v23}. In frames of this
formalism we exactly take into account neutrino masses as well as
examine what kind of neutrino sources could give us a stable
oscillations picture (see also Appendix~\ref{ERROR}).

However, the external classical source approach is not free from
disadvantages. This method is based on the use of the retarded
Green functions [Eqs.~\eqref{spinretGfun} and~\eqref{SRretM}] in
the neutrino mass eigenstates basis. Thus, if one tries to apply
this formalism for the description of neutrino oscillations in an
external field, which, as we mentioned above, is typically
nondiagonal in the mass eigenstate basis, one encounters a problem
to find a nondiagonal retarded Green function. It is quite
difficult to calculate this kind of function analytically. Thus
this approach is unlikely to be applicable to the description of
neutrino oscillations in an external field, in contrast to the
relativistic quantum mechanics method with gives satisfactory
results in presence of external fields at least for
ultrarelativistic neutrinos, unless we study a special case of an
external field which is diagonal in the neutrino mass eigenstates
basis.

We made an attempt to describe transitions between neutrinos and
antineutrinos using the relativistic quantum mechanics method
[Eq.~\eqref{PtrnuanuMajVac}] and the external classical source
approach [Eq.~\eqref{Pnulambdac}] and obtained the corresponding
transition probability formulae which resemble the previously
derived ones~\cite{Kob82,SchVal81v23}. However, in case of
Majorana neutrinos, particles are identical to their
antiparticles. The concept of antiparticles is an inherent
component of quantum field theory. Thus the approaches based on
classical physics do not seem to be an appropriate tool for the
description of oscillations between neutrinos and antineutrinos.
That is why in Sec.~\ref{NU2ANTINU} we use the quantum field
theory method to study particles and antiparticles in background
matter in case of Majorana neutrinos. Note that the analogous
method for the description of neutrino oscillations in vacuum was
previously used in Refs.~\cite{Kob82,GiuKimLeeLee93,Beu02}.

It should be noted that oscillations between neutrinos and
antineutinos manifest in ($0\nu2\beta$)-decay~\cite{Bil10}. During
this process two nucleons inside a nucleus exchange a virtual
Majorana neutrino. That is why it is important to examine the
influence of dense nuclear matter on the process of Majorana
neutrinos propagation.

In our analysis we considered the situation when the interaction
with matter is diagonal in the mass eigenstates basis since the
case of a nondiagonal neutrino propagator is rather difficult to
study analytically. This difficulty is analogous to that in the
external classical source approach. Nevertheless we found that
this type of matter interaction suppresses transitions between
neutrinos and antineutrinos. Moreover, in case of hypothetical
very heavy Majorana neutrinos the transition probability vanishes
[see Eqs.~\eqref{FahighE} and~\eqref{sigmahighE}]. It means that
one cannot explore the presence of this kind of heavy neutrinos
studying ($0\nu2\beta$)-decay.

In Secs.~\ref{SPECMATT}, \ref{STERILE}, and~\ref{MAJSPECMATT} we
discussed several applications of the results obtained in frames
of the relativistic quantum mechanics method to the studies of
astrophysical neutrinos emitted during the core collapse of a
supernova. First we studied spin flavor oscillations of Dirac
(Sec.~\ref{SPECMATT}) and Majorana (Sec.~\ref{MAJSPECMATT})
neutrinos in an expanding envelope under the influence of the
magnetic field of a supernova. It was found that for Dirac
neutrinos the amplification of neutrino oscillations can be
achieved in a moderate magnetic field [see
Figs.~\ref{Dirnulowdens}(d)-\ref{Dirnuhighdens}(d)]. On the
contrary, in the Majorana neutrinos case, neutrino oscillations
can be enhanced only in a strong magnetic field [see
Fig.~\ref{Majnuenv}(b), (d) and~(f)].

In Sec.~\ref{STERILE} another channel of oscillations of
astrophysical neutrinos was considered. We studied the possibility
of spin flavor oscillations between electron and additional
quasi-degenerate in mass sterile neutrinos. We obtained that the
correction~\eqref{deltaH} to the effective
Hamiltonian~\eqref{HQMfromQFT}, found in frames of the
relativistic quantum mechanics method, results in the appearance
of the new resonance in this oscillations channel.

Finally in Appendix~\ref{DETAILSABSOL} we present the general
solution of the ordinary differential equations for the
coefficients $a_a^{(\zeta)}$ which one encounters in the
relativistic quantum mechanics method. In Appendix~\ref{ANALAPPR}
we discuss the validity of the relativistic quantum mechanics
approach to the description of neutrino spin flavor oscillations.
In particular, we analyze various factors which influence
oscillations between electron and sterile neutrinos with very
small $\delta m^2$ (see also Sec.~\ref{STERILE}). In
Appendix~\ref{ERROR} we study how neutrino wave functions converge
to the values corresponding to the stable oscillations picture in
the external classical source approach (Sec.~\ref{INHOMOGEN}). In
Appendix~\ref{NU2ANTINUINT} we present the details of the
calculation of the $S$-matrix element involving transitions
between neutrinos and antineutrinos in frames of the quantum field
theory description of neutrino oscillations
(Sec.~\ref{NU2ANTINU}).

Summarizing we can say that a theoretical approach for the
description of neutrino oscillations should satisfy the following
requirements: the evolution equation governing the dynamics of
mixed neutrinos should be derived from Lorentz invariant
Lagrangian of the system and thus it should account for the
relativistic invariance, at least implicitly; and such a method
should be applicable in presence of various external fields. Among
various theoretical approaches~\cite{BerGuzNis10,Beu03} the
relativistic quantum mechanics method is likely to be most
appropriate formalism for the description of neutrino
oscillations.

\begin{acknowledgments}
  This work has been supported by Conicyt (Chile), Programa
  Bicentenario PSD-91-2006. The author is thankful to
  S.~G.~Kovalenko, T.~Morozumi, and J.~Maalampi
  for helpful discussions.
\end{acknowledgments}

\appendix

\section{Solution to the ordinary differential equations for the functions
$a_a^{(\zeta)}$\label{DETAILSABSOL}}

In this Appendix we describe the formalism for the analysis of
ordinary differential equations for the functions $a_a^{(\zeta)}$
which one encounters in Secs.~\ref{MATTER} and~\ref{B}.

Let us study the time evolution of the two component spinor
$\mathbf{Z}^\mathrm{T}=(Z_1,Z_2)$ which is governed by the
Schr\"odinger equation of the form,
\begin{equation}\label{Zequ}
  \mathrm{i}\dot{\mathbf{Z}}=H\mathbf{Z},
\end{equation}
where the Hamiltonian has the following form:
\begin{equation}\label{ZHam}
  H= g
  \begin{pmatrix}
    0 & e^{\mathrm{i}\omega t} \\
    e^{-\mathrm{i}\omega t} & 0 \
  \end{pmatrix}.
\end{equation}
Here $g$ and $\omega$ are real parameters. Eq.~\eqref{Zequ} should
be supplied with the initial condition $\mathbf{Z}(0)$. To find
the solution of Eqs.~\eqref{Zequ} and~\eqref{ZHam} we introduce
the new spinor $\mathbf{Z}'$ by the relation,
$\mathbf{Z}=\mathcal{U}\mathbf{Z}'$, where the unitary matrix
$\mathcal{U}$ reads
\begin{equation}\label{ZZ'matr}
  \mathcal{U}=
  \begin{pmatrix}
    e^{\mathrm{i}\omega t/2} & 0 \\
    0 & e^{-\mathrm{i}\omega t/2} \
  \end{pmatrix}.
\end{equation}
Now Eq.~\eqref{Zequ} is rewritten in the following way:
\begin{equation}\label{Z'equ}
  \mathrm{i}\dot{\mathbf{Z}}'=H'\mathbf{Z}',
\end{equation}
with the new Hamiltonian $H'$ which is obtained with help of
Eqs.~\eqref{Zequ}-\eqref{ZZ'matr},
\begin{equation}\label{Z'Ham}
  H'=\mathcal{U}^\dag H \mathcal{U} -
  \mathrm{i}\mathcal{U}^\dag\dot{\mathcal{U}}=
  \begin{pmatrix}
    \omega/2 & g \\
    g & -\omega/2 \
  \end{pmatrix}.
\end{equation}
Note that the initial condition for the spinor $\mathbf{Z}'(0)$ is
the same as for $\mathbf{Z}(0)$, $\mathbf{Z}'(0)=\mathbf{Z}(0)$,
due to the special form of the matrix $\mathcal{U}$ in
Eq.~\eqref{ZZ'matr}.

Supposing that the Hamiltonian $H'$ in Eqs.~\eqref{Z'equ}
and~\eqref{Z'Ham} does not depend on time we get the solution to
Eq.~\eqref{Z'equ} as
\begin{align}\label{Z'sol}
  \mathbf{Z}'(t) = & \exp{(-\mathrm{i}H't)}\mathbf{Z}'(0)
  \notag
  \\
  & =
  \left(
    \cos\varOmega t-\mathrm{i}(\bm{\sigma}\mathbf{n})\sin\varOmega t
  \right)
  \mathbf{Z}'(0),
\end{align}
where $\mathbf{n}=(g,0,\omega/2)/\varOmega$ is the unit vector and
$\varOmega=\sqrt{g^2+(\omega/2)^2}$. Using Eqs.~\eqref{ZZ'matr}
and~\eqref{Z'sol} we arrive to the expressions for the components
of $\mathbf{Z}$ written in terms of the initial condition
$\mathbf{Z}(0)$:
\begin{align}\label{Zsol}
  Z_1(t)= &
  \left(
    \cos\varOmega t-\mathrm{i}\frac{\omega}{2\varOmega}\sin\varOmega t
  \right)e^{\mathrm{i}\omega t/2}Z_1(0)
  \notag
  \\
  & -
  \mathrm{i}\frac{g}{\varOmega}\sin(\varOmega t)
  e^{\mathrm{i}\omega t/2}Z_2(0),
  \notag
  \\
  Z_2(t)= &
  \left(
    \cos\varOmega t+\mathrm{i}\frac{\omega}{2\varOmega}\sin\varOmega t
  \right)e^{-\mathrm{i}\omega t/2}Z_2(0)
  \notag
  \\
  & -
  \mathrm{i}\frac{g}{\varOmega}\sin(\varOmega t)
  e^{-\mathrm{i}\omega t/2}Z_1(0),
\end{align}
which can be directly applied for the analysis of ordinary
differential equations from Secs.~\ref{MATTER} and~\ref{B}.

To get the solution of Eq.~\eqref{aeqmatt} we identify the
components of the spinor $\mathbf{Z}$ with $a_{1,2}^{-{}}$ and the
parameter $\omega$ with $\omega_{-{}}$ (see Sec.~\ref{MATTER}).
Finally we arrive to Eq.~\eqref{aeqsolmatt}. We can also apply
Eq.~\eqref{Zsol} to obtain the solution of Eq.~\eqref{aeqB}. For
this purpose one considers two cases:
\begin{itemize}
  \item For $\mathbf{Z}^\mathrm{T}=(a_1^{+{}},a_2^{+{}})$, $g = - \mu B$,
  $\omega=\omega_{+{}}$ and $\varOmega=\Omega_{+{}}$;
  \item For $\mathbf{Z}^\mathrm{T}=(a_1^{-{}},a_2^{-{}})$, $g = \mu B$,
  $\omega=\omega_{-{}}$ and $\varOmega=\Omega_{-{}}$.
\end{itemize}
Using these formulae together with Eqs.~\eqref{Zsol} we readily
arrive to Eqs.~\eqref{aeqsolB}-\eqref{OmegaomegaB}. Note that the
dynamics of the system~\eqref{Zequ} and~\eqref{ZHam} is analogous
to the quantum mechanical description of neutrino spin flavor
oscillation in a twisting magnetic field studied in
Ref.~\cite{Smi91}.

\section{Analysis of approximations made in the derivation of
the effective Hamiltonian\label{ANALAPPR}}

In this Appendix we analyze the validity of the relativistic
quantum mechanics approach for the description of neutrino spin
flavor oscillations used in Sec.~\ref{MATTERB}. The correction to
the effective Hamiltonian~\eqref{deltaH}
is a rather small quantity. Therefore we should evaluate other
factors which can also give the contributions, comparable with
Eq.~\eqref{deltaH}, to the effective Hamiltonian. In this section
we analyze the contributions to the effective Hamiltonian from
longitudinal magnetic field, matter polarization and possible
corrections from the new interactions.

First we should remind that we use the relativistic quantum
mechanics approach, with the external fields being independent of
spatial coordinates. If external fields depend on the spatial
coordinates, Dirac wave packets theory reveals various additional
phenomena such as particles creation by the external field
inhomogeneity~\cite{ItzZub84p80}. For the approximation of the
spatially constant external fields to be valid, the typical length
scale of the external field variation $L_\mathrm{ext}$ should be
much greater than the Compton length of a
neutrino~\cite{ItzZub84p80}: $L_\mathrm{ext} \gg \lambdabar_C =
\hbar/m_\nu c$~\cite{DvoStu02}. For a neutrino with $m_\nu \sim
1\thinspace\text{eV}$ this condition reads $L_\mathrm{ext} \gg
10^{-5}\thinspace\text{cm}$, that is fulfilled for almost all
realistic external fields.

A general remark on the ``perturbative" approach used in
Sec.~\ref{MATTERB} should be made. In the modified
Eq.~\eqref{ODEDirmatt}, which includes the interaction with the
magnetic field, the terms containing $a_a^{\pm{}}$ and
$a_a^{\mp{}}$ are coupled and the coupling terms are proportional
to $g$ and $\mu B$~\eqref{hH}. As we demonstrate in
Sec.~\ref{VACUUM}, mass eigenstates decouple in vacuum are their
values depend on the initial condition only. While solving the
modified Eq.~\eqref{ODEDirmatt}, we could just take a term linear
in $g$ and $\mu B$ (see, e.g., Eq.~\eqref{nualphaRpertB} as well
as Refs.~\cite{Dvo06,DvoMaa07}). However we take into account
these terms exactly in the further analysis [see
Eqs.~\eqref{Schr1QFT} and~\eqref{Schr2QFT}]. It is equivalent to
the summation of all terms in the perturbation series.

While deriving the effective Hamiltonian~\eqref{Schr2QFT} in
Sec.~\ref{MATTERB} we supposed that magnetic field is transverse
with respect to the neutrino motion. The effect of the
longitudinal magnetic field on neutrino oscillations was studied
in Ref.~\cite{AkhKhl88}. It was found there that diagonal entries
of the effective Hamiltonian receive additional small
contributions $\mu_a B_{\parallel{}}(m_a/k)$. In order to neglect
the longitudinal magnetic field contribution in comparison with
our corrections~\eqref{deltaH}, its strength should satisfy the
condition,
\begin{equation}\label{Blong}
  \frac{B_{\parallel{}}}{B_{\perp{}}} \ll
  \frac{1}{16  k B_{\perp{}}|\mu_a m_a - \mu_b m_b|}
  \left|
    \frac{m_a^2 g_a^3}{\mathcal{M}_a^2}-
    \frac{m_b^2 g_b^3}{\mathcal{M}_b^2}
  \right|,
\end{equation}
where $B_{\perp{}}$ is the transverse component of the magnetic
field.

A regular magnetic field of a supernova typically has both
poloidal and toriodal components. Of course, an irregular
turbulent magnetic field can be also present but its length scale
appears to be quite small. A toroidal magnetic field can be $\sim
10^{16}\thinspace\text{G}$ and is concentrated near the equator of
the star at the distance $\sim 10\thinspace\text{km}$ from the
star center~\cite{ArdBisMoi05}. Thus in our case a toroidal
magnetic field is unlikely to significantly contribute to the
dynamics of neutrino oscillations.

Let us evaluate the fraction of neutrinos for which the new
correction~\eqref{deltaH} to the effective Hamiltonian gives
bigger contribution to the resonance enhancement of oscillations
compared to that of the longitudinal component of the poloidal
magnetic field. Using Eq.~\eqref{Blong} we find that these
neutrinos should be emitted inside the solid angle near the
equatorial plane with the spread $2\vartheta$, where $\vartheta
\sim B_{\parallel{}}/B_{\perp{}}$. Assuming the radially symmetric
neutrino emission we find that about 10\% of the total neutrino
flux is affected by the new resonance~\eqref{rescond1}, i.e. the
influence of the longitudinal magnetic field is negligible for
oscillations of such particles.

The next important approximation made in the deviation of
Eq.~\eqref{Schr2QFT} was the assumption of negligible polarization
of matter which can be not true if we study rather strong magnetic
fields. The effect of matter polarization on neutrino oscillations
was previously discussed in Refs.~\cite{DvoStu02,Nun97,LobStu01}.
Matter polarization produces the following contributions to the
diagonal entries of the effective
Hamiltonian~\cite{DvoStu02,LobStu01}: $g_a (\bm{\lambda}_f
\bm{\beta}_\nu)$ (left polarized neutrinos) and $g_a
(\bm{\lambda}_f \bm{\beta}_\nu) (m_a/k)$ (right polarized
neutrinos). Here $\bm{\beta}_\nu$ is the neutrino velocity and
$\bm{\lambda}_f$ is the mean polarization vector of background
fermions and we keep only the leading order in $m_a/k$.

First we estimate the contribution to the right polarized
neutrinos effective potential. It is clear that one should take
into account only the polarization of electrons since nucleons are
much heavier. For the weakly degenerate electrons we should
discuss the case of weak field limit (see Ref.~\cite{Nun97}),
since $2eB/m_e^2 \sim 10^{-8} \ll 1$, where $m_e$ is the
electron's mass and $B \sim 10^7\thinspace\text{G}$. Since the
temperature inside the shock wave region can be about several
MeV~\cite{Sum05} the electrons are relativistic. Hence their mean
polarization can be estimated as, $|\bm{\lambda}_f| \sim
\mu_\mathrm{B} B m_e/3 T_e^2$, where $T_e$ is the temperature of
electrons. Therefore we get that the new correction to the
effective Hamiltonian~\eqref{deltaH} becomes bigger than the
contribution of matter polarization to the effective potential of
right polarized neutrinos at $T_e > 4\thinspace\text{MeV}$.

To evaluate the contribution $g_a (\bm{\lambda}_f \bm{\beta}_\nu)$
to the effective Hamiltonian, which corresponds to the left
polarized neutrinos effective potential, we notice that the vector
$\bm{\lambda}_f$ should be directed along the external magnetic
field. For this term to be much less than the new correction to
the effective Hamiltonian~\eqref{deltaH}, the angle $\vartheta$,
defined above, should be very small: $\vartheta \ll 10^{-8}$, for
$T_e \sim 4\thinspace\text{MeV}$. It means that we can neglect the
polarization effects only if neutrinos are emitted very close the
equator of a star. We should, however, remind that this kind of
matter polarization term contributes only to $(H_{QM})_{22}$ in
Eq.~\eqref{HQMfromQFT} since only this entry corresponds to
$\nu_e^\mathrm{L}$. Thus the presence of the term $g_a
(\bm{\lambda}_f \bm{\beta}_\nu)$ does not directly affect our
results since we study $\nu_e^\mathrm{R} \leftrightarrow
\nu_s^\mathrm{L}$ oscillations channel (see also
Table~\ref{oscchannels}).
%

The presence of big Dirac neutrino magnetic moments implies the
existence of new interactions, beyond the standard model, which
electromagnetically couple left and right polarized neutrinos. It
is probable that these new interactions also contribute to the
effective potential of the right polarized neutrino interaction
with background matter. Despite this additional effective
potential is likely to be small, one should evaluate it and
compare with the correction~\eqref{deltaH}.

The most generic $\mathrm{SU}(2)_L \times \mathrm{U}(1)_Y$ gauge
invariant and remormalizable interaction which produces Dirac
neutrino magnetic moment was discussed in Ref.~\cite{Bel05}. It
was found that neutrino magnetic moments arise from the effective
Lagrangian involving the dimension $n=6$ operators
$\mathcal{O}_j$,
\begin{equation}\label{effLagr1}
  \mathcal{L}_\mathrm{eff} =
  \sum_j \frac{C_j}{\Lambda^2} \mathcal{O}_j + \text{h.c.},
\end{equation}
where $\Lambda \sim 1\thinspace\text{TeV}$ is a scale of the new
physics and $C_j$ are the effective operator coupling constants.
The sum in Eq.~\eqref{effLagr1} spans all the operators of the
given dimension.

One of the operators $\mathcal{O}_j$ also contributes to the
effective potential of a right-handed neutrino in matter,
\begin{equation}\label{oper6}
  \mathcal{O} =
  \kappa \bar{L} \tau_a \tilde{\phi} \sigma^{\mu\nu} \nu_\mathrm{R}
  W_{\mu\nu}^a,
\end{equation}
where $\kappa$ is the coupling constant, $\tau_a$ are Pauli
matrices, $L^\mathrm{T} = (\nu_\mathrm{L},e_\mathrm{L})$ is the
$\mathrm{SU}(2)_L$ isodoublet, $\tilde{\phi} = \mathrm{i} \tau_2
\phi^{*{}}$, with $\phi$ being a Higgs field, and $W_{\mu\nu}^a =
\partial_\mu W_\nu^a - \partial_\nu W_\mu^a - \kappa \epsilon_{abc} W_\mu^b
W_\nu^c$ is the $\mathrm{SU}(2)_L$ field strength tensor.

Assuming the spontaneous symmetry breaking at the electroweak
scale, $\phi^\mathrm{T} \to (0, v/\sqrt{2})$, we can rewrite the
Lagrangian in the form,
\begin{equation}\label{effLagr2}
  \mathcal{L}_\mathrm{eff} =
  \frac{C \kappa v}{\sqrt{2}}
  \bar{e}_\mathrm{L} \sigma^{\mu\nu} \nu_\mathrm{R}
  (W_{\mu\nu}^1 - \mathrm{i} W_{\mu\nu}^2) + \text{h.c.},
\end{equation}
which implies that a process $e^{-{}}+\nu_\mathrm{R} \to
e^{-{}}+\nu_\mathrm{L}$ should happen in background matter.

Using the results of Ref.~\cite{Bel05} we can evaluate the
contribution of the Lagrangian~\eqref{effLagr2} to the effective
Hamiltonian~\eqref{HQMfromQFT} as
\begin{equation}\label{deltaVR}
  \delta V_\mathrm{R} \sim V_{sm}
  \left(
    \frac{\mu_\nu}{\mu_\mathrm{B}}
  \right)^2
  \left(
    \frac{E_\nu}{m_e}
  \right)^2
  \frac{|\kappa|^2}{G_\mathrm{F} M_W^2},
\end{equation}
where $V_{sm} \sim G_\mathrm{F} n_e$ is the standard model
effective potential and $M_W$ is the $W$ boson mass. Taking
$\mu_\nu \sim 10^{-12} \mu_\mathrm{B}$, $E_\nu \sim
100\thinspace\text{MeV}$ and $m_\nu \sim 1\thinspace\text{eV}$
(see Sec.~\ref{STERILE}) we can get that the ratio of the
correction to the effective potential~\eqref{deltaVR} and new
correction~\eqref{deltaH} is $\sim 10^{-4}$. It means that the
influence of new interactions, which generate neutrino magnetic
moments, are not important for neutrino spin flavor oscillations.

\begin{figure}
  \centering
  \includegraphics[scale=.35]{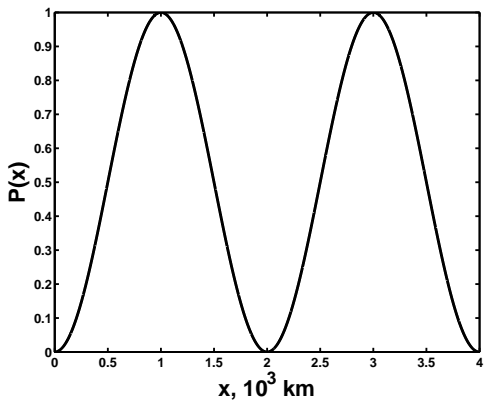}
  \caption{\label{Ptrnum}
  The numerical transition probability for neutrinos with
  $\mu = 3 \times 10^{-12}\thinspace\mu_\mathrm{B}$,
  $\mu_{\nu_e} = 10^{-13}\thinspace\mu_\mathrm{B}$,
  $E = 100\thinspace\mathrm{MeV}$,
  $m_{\nu_e} = 1\thinspace\mathrm{eV}$ and
  $\delta m^2 = 5 \times 10^{-17}\thinspace\mathrm{eV}^2$ moving in
  background matter with $\rho = 10^6\thinspace\mathrm{g/cm}^3$ and
  interacting with the magnetic field $B = 10^7\thinspace\mathrm{G}$.
  After Ref.~\cite{Dvo10}.}
\end{figure}

Now let us estimate the influence of the diagonal magnetic moments
$\mu_a$ on the dynamics of spin flavor oscillations. It was found
in Ref.~\cite{LycBli10} that to get the significant
$\nu_e^\mathrm{R}$ luminosity $\sim 10^{50}\thinspace\text{erg/s}$
the diagonal magnetic moment should be $\mu_{\nu_e} =
10^{-13}\thinspace\mu_\mathrm{B}$. Eq.~\eqref{rescond1} was
obtained under the assumption $\mu_{\nu_e} \ll \mu$. In
Eq.~\eqref{rescond3} we use $\mu = 3 \times
10^{-12}\thinspace\mu_\mathrm{B}$, i.e. the condition of the
Eq.~\eqref{rescond1} validity is satisfied. In Fig.~\ref{Ptrnum}
we present the numerical solution of the Schr\"{o}dinger
equation~\eqref{Schr2QFT}.
We remind that our simplified model with $\mu_{\nu_e} = 0$ gives
the transition probability $P(x) = \sin^2 (\mu B x)$ if the
resonance condition~\eqref{rescond1} is fulfilled. As one can see
on Fig.~\ref{Ptrnum} there is almost no difference in the dynamics
of spin flavor oscillations in our simplified model and more
realistic situation which involves non-zero diagonal magnetic
moment of an electron neutrino.

\section{Evaluation of integrals\label{ERROR}}

In this Appendix we calculate the integrals which one encounters
while studying neutrino oscillations in the model with classical
sources in Sec.~\ref{INHOMOGEN}. It is interesting to evaluate the
inexactitude which is made when we approach to the limit
$x_m\to\infty$ in Eq.~\eqref{int2}. Let us discuss two functions
\begin{align}
  \label{errorF}
  F(r,t)= & \int_0^{x_m}\mathrm{d}x
  \frac{J_1(m x)}{\sqrt{r^2+x^2}}e^{\mathrm{i}E\sqrt{r^2+x^2}}
  \notag
  \\
  & =
  \int_0^{y_m}\mathrm{d}y
  \frac{J_1(\rho y)}{\sqrt{1+y^2}}e^{\mathrm{i}\mathcal{E}\sqrt{1+y^2}},
  \\
  \label{errorF0}
  F_0(r)= & \frac{1}{\rho}(e^{\mathrm{i}\mathcal{E}}-e^{\mathrm{i}P}).
\end{align}
These functions are proportional to $\mathcal{I}_2$ in
Eqs.~\eqref{int2} and~\eqref{int2fin} respectively. In
Eqs.~\eqref{errorF} and~\eqref{errorF0} we use dimensionless
parameters $\rho=m r$, $\mathcal{E}=E r=\gamma\rho$, $\gamma=E/m$,
$y_m=\sqrt{(t/r)^2-1}$ and $P=\sqrt{\mathcal{E}^2-\rho^2}$.

In case we study neutrinos, we get that $\mathcal{E} \gg \rho \gg
1$ in almost all realistic situations. For example, suppose we
study a neutrino emitted in a supernova explosion is our Galaxy.
The typical distance is $r \sim 10\thinspace\text{kpc}$. Taking $m
\sim 1\thinspace\text{eV}$ and $E \sim 10\thinspace\text{MeV}$, we
receive that $\mathcal{E} \sim 10^{35}$ and $\rho \sim 10^{28}$.

Basing on the analysis of Sec.~\ref{DMT} we can rewrite
Eq.~\eqref{errorF} as
\begin{align}\label{deltaF}
  F(r,t) = & F_0(r)-\delta F,
  \notag
  \\
  \delta F = & \int_{y_m}^{+\infty}\mathrm{d}y
  \frac{J_1(\rho y)}{\sqrt{1+y^2}}e^{\mathrm{i}\mathcal{E}\sqrt{1+y^2}}.
\end{align}
Using the fact that $\rho \gg 1$, $y_m \gg 1$ and the
representation for the Bessel function,
\begin{equation}
  J_1(z) \approx
  \sqrt{\frac{2}{\pi z}}\cos
  \left(
    z-\frac{3\pi}{4}
  \right)
  \quad
  \text{at}
  \quad
  z \to +\infty,
\end{equation}
we obtain for the function $\delta F$ in Eq.~\eqref{deltaF} the
following expression:
\begin{align}
  \delta F \approx &
  -\frac{1}{\sqrt{2\pi\rho}}
  \big\{
    \mathrm{ci}([\gamma+1]\rho y_m)+\mathrm{ci}([\gamma-1]\rho y_m)
    \notag
    \\
    & +
    \mathrm{i}
    [\mathrm{si}([\gamma+1]\rho y_m)+\mathrm{si}([\gamma-1]\rho y_m)]
  \big\},
\end{align}
where $\mathrm{ci}(z)$ and $\mathrm{si}(z)$ are cosine and sine
integrals. Using the asymptotic expression,
\begin{equation}
  |\mathrm{ci}(z)| \sim |\mathrm{si}(z)| \sim \frac{1}{z}
  \quad
  \text{at}
  \quad
  z \to +\infty,
\end{equation}
we obtain that the function $\delta F$ approaches to zero as
$1/(y_m \rho^{3/2})$ at great values of $y_m$ and $\rho$. Note
that this result remains valid for a particle with an arbitrary
$\gamma$ factor, i.e. rapid oscillations of the function $\delta
F$ will attenuate even for slow particles. This analysis
substantiates the approximations made in Sec.~\ref{DMT}.

\begin{figure}
  \centering
  \includegraphics[scale=.45]{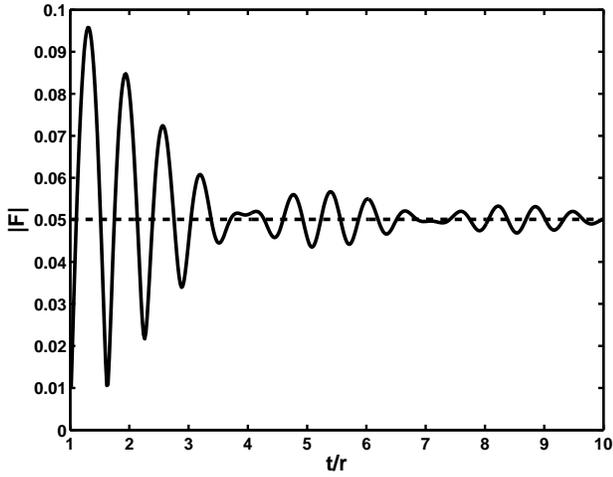}
  \caption{\label{error}
  The absolute values of the functions $F(r,t)$ and $F_0(r)$
  versus $t$.
  This figure is taken from Ref.~\cite{Dvo09}.}
\end{figure}

Finally let us illustrate the behavior of the functions $F(r,t)$
and $F_0(r)$. On Fig.~\ref{error} we present the absolute values
of these functions versus $t$.
This figure is plotted for $\rho=1$ and $\mathcal{E}=10$. The
solid line is the absolute value of the function $F(r,t)$ and the
dashed line of the function $F_0(r)$. As we mention in
Sec.~\ref{DMT} the relaxation phenomena occur when $t\gtrsim r$.
It can be also seen on Fig.~\ref{error}. It is possible to notice
that $|F(r,t)|\to|F_0(r)|$ at great values of $t$ as it is
predicted in Sec.~\ref{DMT}.

\section{Calculation of the $S$-matrix element\label{NU2ANTINUINT}}

In this Appendix we present the detailed calculation of the
$S$-matrix element in Sec.~\ref{NU2ANTINU}. Note that analogous
calculation for the case of virtual neutrinos propagating in
vacuum, i.e. when the matrix $(f_{\lambda\lambda'}^\mu)$ is absent
in Eq.~\eqref{Lagrnu2antinu}, is presented in Ref.~\cite{Kob82}.

Using Eqs.~\eqref{Tjj}-\eqref{Majnuprop} as well as after the
spatial integration and the elimination of the combinatorial
factor we can cast Eq.~\eqref{Smatrixnu} in the form,
\begin{align}\label{Smatrixx0y0}
  S = & \frac{G_\mathrm{F}^2}{\mathcal{V}\sqrt{E_\alpha E_\beta}}
  \int \mathrm{d}x_0 \mathrm{d}y_0
  e^{\mathrm{i} \mathbf{p}_\beta \mathbf{x}_1 -
  \mathrm{i} \mathbf{p}_\alpha \mathbf{x}_2}
  e^{\mathrm{i} E_\alpha x_0 -
  \mathrm{i} E_\beta y_0}
  \notag
  \\
  & \times
  \bar{u}(p_\alpha) \gamma^0 P_\mathrm{R}
  S_a(\mathbf{L}, x_0 - y_0)
  P_\mathrm{R} \gamma^0 u(p_\beta),
\end{align}
where we use the fact that $u^\mathrm{T}(-p_\alpha)C =
\bar{u}(p_\alpha)$. The integration with respect to $x_0$ and
$y_0$ can be performed with help of the new variables: $T = (x_0 +
y_0)/2$ and $t = x_0 - y_0$. After this integration one gets the
energy conservation $\delta$-function in Eq.~\eqref{Smatrixfinal}.

The Fourier transform of the neutrino propagator $S_a(x)$ in
Eq.~\eqref{Smatrixx0y0} was found in Ref.~\cite{PivStu05} and has
the form,
\begin{widetext}
\begin{equation}\label{Sa}
  S_a(k) =
  \frac{(k^2 - m_a^2 - g_a^2/4 - \mathrm{i} \sigma_{\mu\nu}\gamma^5 g_a^\mu k^\nu)
  (\gamma^\mu k_\mu + m + \gamma_\mu \gamma^5 g_a^\mu /2)}
  {(k^2 - m^2 - g_a^2/4)^2 - (g_a k)^2 + k^2 g_a^2},
\end{equation}
\end{widetext}
where $g_a^\mu \equiv g_{aa}^\mu$ are the diagonal elements of the
matrix $(g_{ab}^\mu)$ [see Eq.~\eqref{gmatrmatt}].

In case of the nonmoving and unpolarized matter the momentum
integration in Eq.~\eqref{Smatrixx0y0} can be performed using the
calculus of residues. Finally we arrive to the following result:
\begin{multline}\label{Faint}
  \int e^{\mathrm{i}\mathbf{k}\mathbf{L}}
  \frac{\mathrm{d}^3\mathbf{k}}{(2\pi)^3}
  \frac{p_a^2 - g_a^2/4 - \mathbf{k}^2 - g_a (\bm{\alpha}\mathbf{k})}
  {(\mathbf{k}^2-k_1^2)(\mathbf{k}^2-k_2^2)}(1-\gamma^5)
  \\
  \approx
  - \frac{F_a}{4 \pi L},
\end{multline}
where $F_a$ is defined in Eq.~\eqref{Fafinal}. In
Eq.~\eqref{Faint} we neglect several small terms since $k_{1,2}L
\gg 1$.

\begin{figure}
  \centering
  \includegraphics[scale=.85]{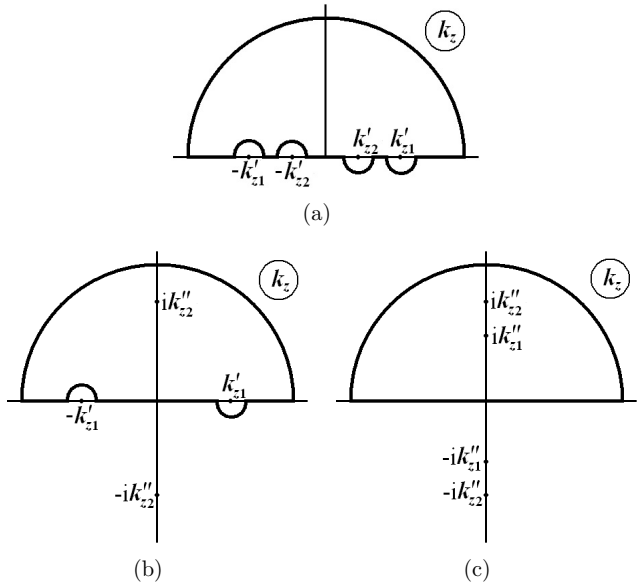}
  \caption{\label{contour}
  Integration contours for each of the integrals in Eq.~\eqref{Jabc}.
  The panel (a) corresponds to $J_a$, (b) to $J_b$, and (c) to $J_c$.}
\end{figure}

It is convenient to perform momentum integration in
Eq.~\eqref{Faint} using cylindrical coordinates pointing
$\mathbf{L}$ along the $z$-axis. Note that the integration over
$k_\phi$ is trivial and gives $2\pi$ since the terms in the
integrand [see also Eq.~\eqref{Sa}] containing $k_\phi$ just
vanish. The result of the integration over $\mathbf{k}$ can be
presented in the form,
\begin{equation}
  \int
  \frac{\mathrm{d}^2\mathbf{k}}{(2\pi)^2}
  \frac{F(k_\rho,k_z)}
  {(\mathbf{k}^2-k_1^2)(\mathbf{k}^2-k_2^2)} = J_a + J_b + J_c,
\end{equation}
where $\mathrm{d}^2\mathbf{k} = k_\rho \mathrm{d}k_\rho
\mathrm{d}k_z$ and
\begin{align}\label{Jabc}
  J_a = &
  \int_0^{k_2}
  \frac{k_\rho \mathrm{d}k_\rho}{(2\pi)^2}
  \int_{-\infty}^{+\infty} \mathrm{d}k_z
  \frac{F(k_\rho,k_z)}
  {(k_z^2 - k_{z1}'^2)(k_z^2 - k_{z2}'^2)},
  \notag
  \\
  J_b = &
  \int_{k_2}^{k_1}
  \frac{k_\rho \mathrm{d}k_\rho}{(2\pi)^2}
  \int_{-\infty}^{+\infty} \mathrm{d}k_z
  \frac{F(k_\rho,k_z)}
  {(k_z^2 - k_{z1}'^2)(k_z^2 + k_{z2}^{\prime\prime 2})}
  \notag
  \\
  J_c = &
  \int_{k_1}^{+\infty}
  \frac{k_\rho \mathrm{d}k_\rho}{(2\pi)^2}
  \int_{-\infty}^{+\infty} \mathrm{d}k_z
  \frac{F(k_\rho,k_z)}
  {(k_z^2 + k_{z1}^{\prime\prime 2})(k_z^2 + k_{z2}^{\prime\prime 2})}.
\end{align}
Here we use the notations $k_{z 1,2}' = \sqrt{k_{1,2}^2-k_\rho^2}$
and $k_{z 1,2}^{\prime\prime} = \sqrt{k_\rho^2 - k_{1,2}^2}$. The
contours of the integration for each of the integrals in
Eq.~\eqref{Jabc} are shown in Fig~\ref{contour}.
%

\end{document}